\documentclass[10pt,oneside,twocolumn,aps,pra,preprintnumbers,bibnotes10pt,longbibliography,superscriptaddress]{revtex4-2}

\usepackage[dvipsnames]{xcolor}

\usepackage{graphicx}
\usepackage{tikz}
\usepackage[absolute,overlay]{textpos}
\usepackage[utf8]{inputenc} 
\usepackage[T1]{fontenc}
\usepackage{lmodern}

\usepackage{enumitem}
\usepackage{braket}
\usepackage{booktabs}
\usepackage{amssymb}
\usepackage{amsthm}
\usepackage{float}
\usepackage{lineno}
\usepackage{wrapfig}
\usepackage{comment}
\usepackage{amsmath}
\usepackage{xcolor}
\usepackage[colorlinks=true, allcolors=blue,breaklinks=true]{hyperref}

\usepackage[font=footnotesize]{caption}
\usepackage[font=footnotesize]{subcaption}
\usepackage{placeins}

\usepackage{natbib}
\begin{document}

%\onecolumngrid
%\tableofcontents
%\clearpage
%\twocolumngrid

\title{Unveiling Energetic Advantage in Superconducting Cat-Qubits Quantum Computation}

\author{Pedro Ramos}
\affiliation{PQI -- Portuguese Quantum Institute, Portugal}

\author{Marco Pezzutto}
\affiliation{PQI -- Portuguese Quantum Institute, Portugal}
\affiliation{Physics of Information and Quantum Technologies Group, Centro de Física e Engenharia de Materiais Avançados (CeFEMA), Portugal}
\affiliation{LaPMET -- Laboratory of Physics for Materials and Emerging Technologies, Portugal}

\author{Yasser Omar}
\affiliation{Instituto Superior Técnico, Universidade de Lisboa, Portugal}
\affiliation{PQI -- Portuguese Quantum Institute, Portugal}
\affiliation{Physics of Information and Quantum Technologies Group, Centro de Física e Engenharia de Materiais Avançados (CeFEMA), Portugal}
\affiliation{LaPMET -- Laboratory of Physics for Materials and Emerging Technologies, Portugal}
\affiliation{Quantum Green Computing, Ltd.}

\date{\today}

\begin{abstract}
Quantum computers are emerging as a promising new technology due to their ability to solve complex problems that exceed the capabilities of classical systems in terms of time. Among various implementations, superconducting qubits have become the leading technology due to their scalability and compatibility with quantum error correction mechanisms. Although time has traditionally been the primary focus, energetic efficiency is becoming an increasingly important consideration, especially with the possibility of a quantum energetic advantage. In this article, the energy consumption of the Semiclassical Quantum Fourier Transform was analyzed on a superconducting quantum computing platform based on cat qubits. Quantum error correction mechanisms were studied and considered in the energy estimations.
The results show how the energy consumption scales with the number of qubits and how the most relevant parameters required for qubit stabilization, gate implementation, and error correction codes contribute to the overall energy usage. An optimization method was developed to tune these parameters with the goal of minimizing energy consumption while maintaining qubit fidelities above a given threshold. Additionally, a comparative study with state-of-the-art classical computers indicates a potential quantum energetic advantage for systems with more than 26 qubits, assuming cryogenic systems operating at Carnot efficiency, with this energetic advantage arising before any computational advantage. This behavior persists even when realistic cryogenic systems and control electronics are taken into account.

\end{abstract}

\maketitle

\section{Introduction}

Understanding the energetic performance of quantum computation is essential for determining whether this technology can realistically scale to solve real-world problems. In particular, it is important to assess whether the algorithmic advantages offered by quantum computing translate into meaningful reductions in physical resources such as energy consumption. As quantum computing emerges as a candidate to extend computational progress beyond the limits of classical technology, this question becomes central to evaluating its practical impact.

Assessing the energetic implications of quantum algorithms is particularly important given that classical computing already faces significant challenges related to energy consumption. The energy demand of data centers has increased steadily over the years~\cite{masanet2020recalibrating}, driven in part by the rapid expansion of artificial intelligence workloads~\cite{Chen2025}. According to the International Energy Agency, the electricity consumption of global data centers is projected to more than double by 2030 \cite{IEA2025EnergyAI}. Currently, these facilities already account for approximately 1.5\% of global electricity use \cite{Chen2025DataCentersAI}.

The energetic cost of quantum computation has only recently begun to attract substantial attention, particularly in contrast to the long-standing focus on algorithmic performance. From a physical standpoint, implementing quantum gates requires microwaves, lasers, or other equipment that consume energy that cannot be recovered. Moreover, isolating and precisely controlling qubits remains a significant challenge. We are currently in the noisy intermediate-scale quantum (NISQ) era, characterized by a limited number of qubits that are subject to noise effects \cite{Preskill_2018}. Decoherence poses serious engineering challenges that require sophisticated error-correction methods and innovative technologies \cite{Breuer_2006}. As a result, cryogenic systems are needed to control the qubits and isolate the system from unwanted interactions. All these technologies lead to a physically irreversible expenditure of energy. 

Recent studies have begun to explore the actual energy consumption of quantum computers \cite{Stevens_2022, Ikonen_2017, Stevens_2025, Codina2026, green, energy_initiative, Silva_Pratapsi_2023}, indicating that quantum computing may offer substantial advantages in energy efficiency compared to classical systems \cite{Moutinho2023, Meier_2025, soret2026, EERA_2026}. The observed energy advantage can be explained primarily by the algorithmic speedups. As fewer computational resources are required, the overall energy consumption is expected to decrease. However, when examining individual operations, quantum gates can consume more energy than classical logic gates due to the physical requirements involved in their implementation. Nevertheless, achieving an overall energetic advantage remains possible, depending on the hardware conditions and on how energy consumption scales when solving real-world problems with a large number of qubits. Moreover, studies on energetics should also consider the final state fidelity, since increasing the fidelity generally requires additional energy to perform the computation. As we move toward the Fault-Tolerant Quantum Computing (FTQC) era, it will be necessary to investigate the energetic performance of Quantum Error Correction (QEC) codes and assess their energetic viability~\cite{stevens_2026}.
 
A truly comprehensive study of the energetics of quantum computation is thus paramount, addressing all its
components, namely the energetic costs of the execution
of the quantum gates/circuits, of the quantum data
buses, of the baseline costs of running the experimental
setup (e.g., fields and lasers generating traps, vacuum,
cryogenics, etc.), and of the classical control of the experiment. This research agenda should include the costs
of generating non-trivial initial states, interconnecting
different quantum processors, etc., and establish benchmarks to assess the energetic performance of quantum
machines. Furthermore, the energetic costs will naturally depend on the chosen platform, requiring dedicated studies. They will also depend on the specific algorithm under consideration. In this context, this paper is part of a broader research effort aimed at understanding the energetic performance of quantum computation across different hardware platforms. In particular, besides superconducting cat qubits, such studies are being carried out for trapped ions~\cite{Gois2024}, Rydberg atoms~\cite{Oscar2024}, and semiconductor spin qubits~\cite{Santos2026}.

This paper aims to estimate the energy scaling associated with performing the Quantum Fourier Transform (QFT) on a cat-qubit superconducting platform developed by the company \textit{Alice} \& \textit{Bob}. The QFT is chosen as a benchmark due to the relative simplicity of its circuit, which facilitates implementation across different quantum platforms. As a fundamental primitive in many quantum algorithms, it provides an ideal starting point for comparing hardware architectures and establishing a theoretical estimate of energetic performance.

In addition, this study introduces an optimization strategy for implementing QFT with integrated QEC mechanisms, providing a framework to adjust the most critical parameters in order to minimize energy consumption while maintaining high qubit fidelity.

\section{Superconducting Qubits}
Superconducting qubits are among the most promising quantum computing architectures, with several companies actively developing them due to their scalability and compatibility with error-correction mechanisms~\cite{Kjaergaard2020, Ezratty2023}. In 2019, quantum supremacy was demonstrated using a programmable superconducting processor \cite{Arute2019}. Recent results highlight the leading role of such processors in quantum computation, including chips with more than $1000$ qubits~\cite{Castelvecchi2023} and bit-flip times exceeding $100\,\mathrm{s}$~\cite{Berdou_2023}. More recently, Google introduced its Willow processor, which can exponentially reduce errors as more qubits are added. On benchmark tasks, it executed a computation in under five minutes that would require an estimated $10^{25}$ years on a state-of-the-art classical supercomputer \cite{Acharya2025,GoogleWillow2024}.

Although this technological development is progressing rapidly, research on its energy consumption remains limited. As new technologies are introduced, industry efforts often prioritize development and deployment, while systematic studies of energy impact may follow later. A similar trend has been observed in the field of artificial intelligence, where advances in model size and computational intensity have driven substantial increases in energy demand \cite{Chen2025}. These developments highlight the importance of assessing energy implications alongside technological progress to ensure sustainable scaling. Understanding these energetic requirements is therefore essential to determine whether a technology can be scaled to dimensions capable of solving real-world problems.

As described, there are many types of superconducting qubits, implemented by several companies. In this article, the focus is on the implementation developed by the French company \textit{Alice} \& \textit{Bob}.

\subsection{Cat Qubits}

The Cat Qubits belong to the family of Bosonic Qubits. In this model, information is encoded in a subspace of the infinite-dimensional Hilbert space of a quantum harmonic oscillator, typically a resonator. Although several encodings exist, this study focuses on the implementation of two-component cat codes \cite{Cochrane_1999,Mirrahimi_2014,Guillaud_2023}.

This cat code uses two coherent states of the same amplitude but opposite phase, $\lvert \alpha \rangle$ and $\lvert -\alpha \rangle$, with $\alpha$ real. The cat-state amplitude is directly related to the average photon number of the coherent state, \(\alpha^2\). The qubit states \(\lvert0\rangle_\alpha\) and \(\lvert1\rangle_\alpha\) can be expressed as
\begin{align}
&\lvert 0 \rangle_{\alpha} = \frac{1}{\sqrt{2}} (\lvert +_\alpha \rangle + \lvert -_\alpha \rangle) = \lvert \alpha \rangle + \mathcal{O}(e^{-2|\alpha|^2}) \\
&\lvert 1 \rangle_{\alpha} = \frac{1}{\sqrt{2}} ((\lvert +_\alpha \rangle -\lvert -_\alpha \rangle)  = \lvert -\alpha \rangle + \mathcal{O}(e^{-2|\alpha|^2}),
\end{align} 
where \(\lvert \pm \rangle _{\alpha} = \sqrt{\frac{1}{N_{\pm}}} (\lvert\alpha\rangle \pm \lvert-\alpha\rangle) \) and \(N_\pm = 2(1\pm e^{-2 |\alpha|^2})\).

A stabilization mechanism is necessary to maintain these states in the code space, preventing all states from collapsing to the vacuum. This is achieved by implementing two-photon-driven dissipation, a method that induces the system to exchange pairs of photons with the environment, which can be described by the Lindblad operators~\cite{Lidar2019, Lescanne_2020}
\begin{equation}\label{lindblad_operator_stab}
    \hat{L}_2 = \sqrt{\kappa_2}(\hat{a}^2-\alpha^2),
\end{equation}
and $\hat{L}_2^{\dag}$, where \(\hat{a}\) is the annihilation operation of the cat qubit memory and \(\kappa_2\) the rate at which pairs of photons are exchanged with the environment.

The stabilization mechanism does not occur spontaneously. It is necessary to engineer an interaction that exchanges pairs of photons between the cat-qubit resonator, which loses photons at a rate \(\kappa_1\), and a one-photon-lossy mode, the buffer, which loses photons at a rate \(\kappa_b\). This interaction is realized using a nonlinear device responsible for coupling the storage and the buffer modes. In the rotating wave approximation, the interaction Hamiltonian is
\begin{equation}
    \frac{\hat{H}_2}{\hbar} = g_2 \hat{a}^2 \hat{b}^\dagger + g_2^* \hat{a}^{\dagger 2} \hat{b},
\end{equation}
where \(g_2\) is the interaction strength and \(\hat{b}\) is the buffer annihilation operator. 

In the physical implementation of this interaction, the nonlinearity is obtained using a device called an asymmetrically-threaded SQUID (ATS)~\cite{Lescanne_2020}. The potential energy of an ATS is proportional to \(\sin{\hat{\phi}}\), where \(\hat{\phi}\) is a linear combination of the storage and buffer mode operators, \(\hat{a}, \hat{a}^{\dagger}, \hat{b}, \hat{b}^{\dagger}\), and represents the phase difference across the ATS. The power-series expansion of $\sin(\hat{\phi})$ contains cubic and higher-order terms that give rise to nonlinear couplings between the modes. These interactions can be activated by driving the ATS with a pump at frequency $\omega_p = 2\omega_a - \omega_b$, where $\omega_a$ and $\omega_b$ are the resonance frequencies of the memory and buffer modes, respectively. This pump stimulates the conversion of two storage photons into one buffer photon and one pump photon. In addition, to engineer the reverse conversion process, a linear drive should be applied to the buffer mode at its resonance frequency, of the form \(\epsilon_d^* \hat{b} e^{-i \omega_b t} + \text{h.c.}\) \cite{Chamberland_2022}.

\begin{figure}[t]
    \centering
    \includegraphics[width=0.5\linewidth]{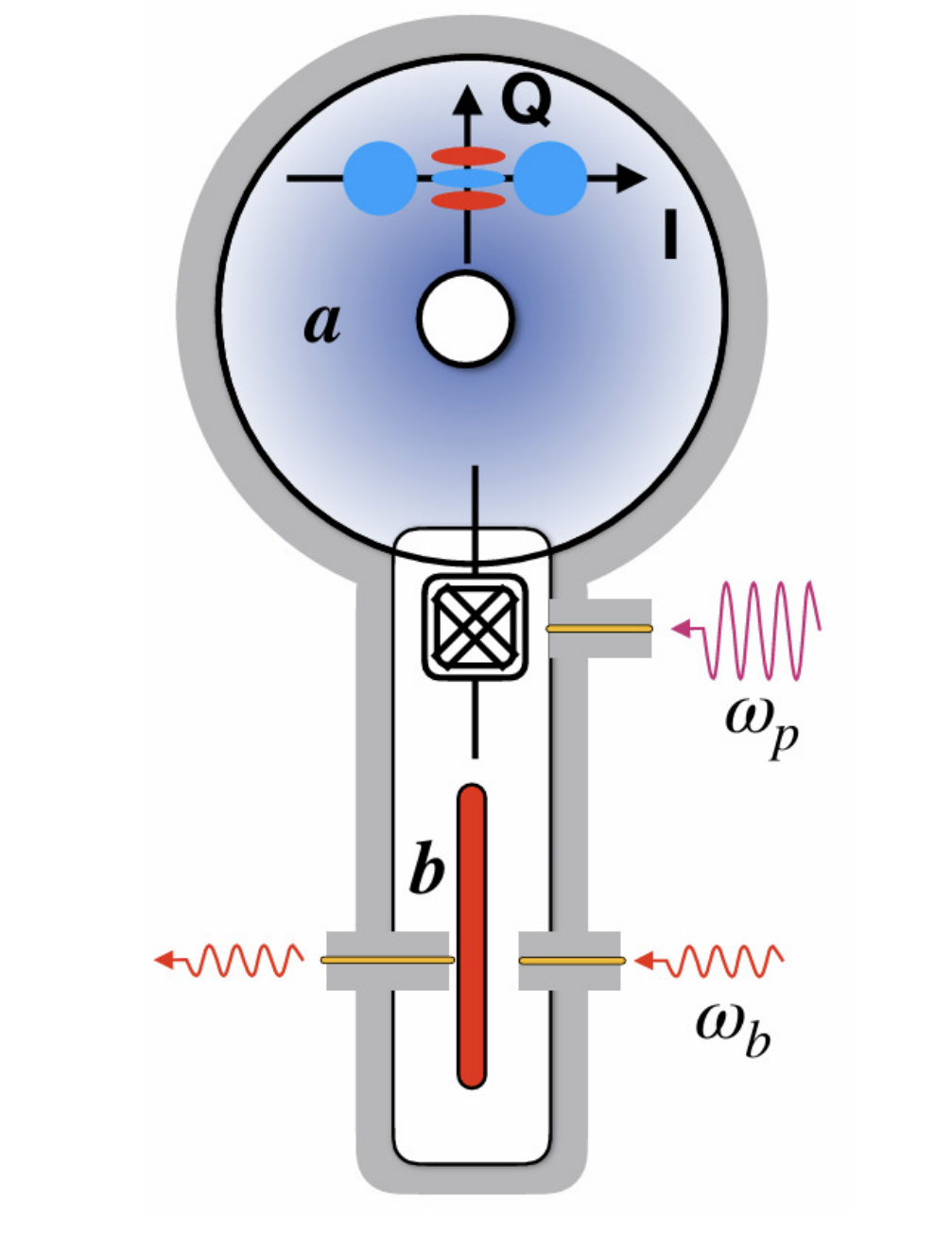}
    \caption{Cat-Qubit implementation. The memory, represented in blue, stores the state of the qubit, which is normally a quantum resonator with frequency \(\omega_a\). The buffer, in red, is pumped at its resonance frequency \(\omega_b\). The square with crosses is the ATS, which is pumped at frequency \(\omega_p=2\omega_a - \omega_b\) to create the process of qubit stabilization.  Taken from \cite{Guillaud_2023}.}
    \label{fig:cat-qubit_implementation}
\end{figure}

The implementation of a device that realizes a two-photon driven dissipative process is represented in Fig.~\ref{fig:cat-qubit_implementation}. This non-linear circuit was realized in \cite{Lescanne_2020}.

Cat qubits are promising because bit-flip errors can be exponentially suppressed with the mean number of photons in the cat state. In particular, the bit-flip error rate scales as $\gamma_X \propto e^{-2\alpha^2}$. However, the phase-flip error rate scales with \(\gamma_Z \propto \alpha^2\) \cite{Guillaud_2023,Lescanne_2020, Reglade_2024}. This property of the cat-qubits implementation has the advantage of reducing the dimensionality of the repetition code needed to prevent errors, enabling a focus on just one dimension instead of two. For example, when performing Shor's algorithm, this property allows the number of required qubits to be reduced by approximately a factor of 60 compared to superconducting qubits using a surface code \cite{Gouzien_2023}.

It has already been demonstrated that cat qubits can be controlled with bit-flip times exceeding 10 seconds \cite{Reglade_2024}. Using moon cat qubits, this lifetime can be further extended to approximately 22 seconds \cite{Rousseau2025}. More recently, \textit{Alice} \& \textit{Bob} have reported a substantial improvement, claiming bit-flip times on the order of 44 minutes in their latest experimental results \cite{AliceBobCatQubit2025}.

\subsection{Operations on Cat Qubits}
The implementation of quantum algorithms relies on the controlled manipulation of qubits. A set of gates that enables universal computation and can be implemented with the repetition cat code used in quantum error correction is \cite{Gouzien_2023}:
\begin{equation}
    \mathcal{S = \{ P _{\lvert\pm\rangle}, M_{\text{x}}, } X,Z, \text{CNOT}, \text{Toffoli} \}.
\end{equation}
This section describes the implementation of these gates.

\textbf{Preparation of the Coherent State} \(\lvert \alpha\rangle\):\\
The eigenstates of the \(Z\) operator are exponentially close to the coherent state \(\lvert \alpha\rangle\). The preparation of the cat qubit in this state begins by initializing it in the vacuum state, followed by applying a strong microwave pulse to the oscillator, which induces a displacement \(D[\pm \alpha]\). The two-photon-driven dissipation is turned on to immediately stabilize the cat qubit after the displacement \cite{Guillaud_2023, Gouzien_2023}.

\textbf{Preparation of a cat state} $\lvert \pm_{\alpha} \rangle$:\\
The preparation of the \(\lvert +_{\alpha} \rangle\) state is achieved by initializing the memory in the vacuum state and activating the two-photon-driven dissipation, as this process preserves photon-number parity.

To prepare the odd-parity state \(\lvert -_\alpha \rangle\), the system is first initialized in the even-parity state \(\lvert +_\alpha \rangle\), followed by the application of a \(Z\) gate \cite{Guillaud_2023, Gouzien_2023}.

\textbf{Measurement of the X operator:}\\
The eigenstates of the \(X\) operator, \(\lvert \pm_\alpha \rangle\), possess well-defined photon-number parity: \(\lvert +_\alpha \rangle\) has even parity, while \(\lvert -_{\alpha} \rangle\) has odd parity. Thus, measuring the photon-number parity effectively implements an \(X\)-basis measurement.

\begin{figure*}[t]
    \centering
    \includegraphics[width=0.9\textwidth]{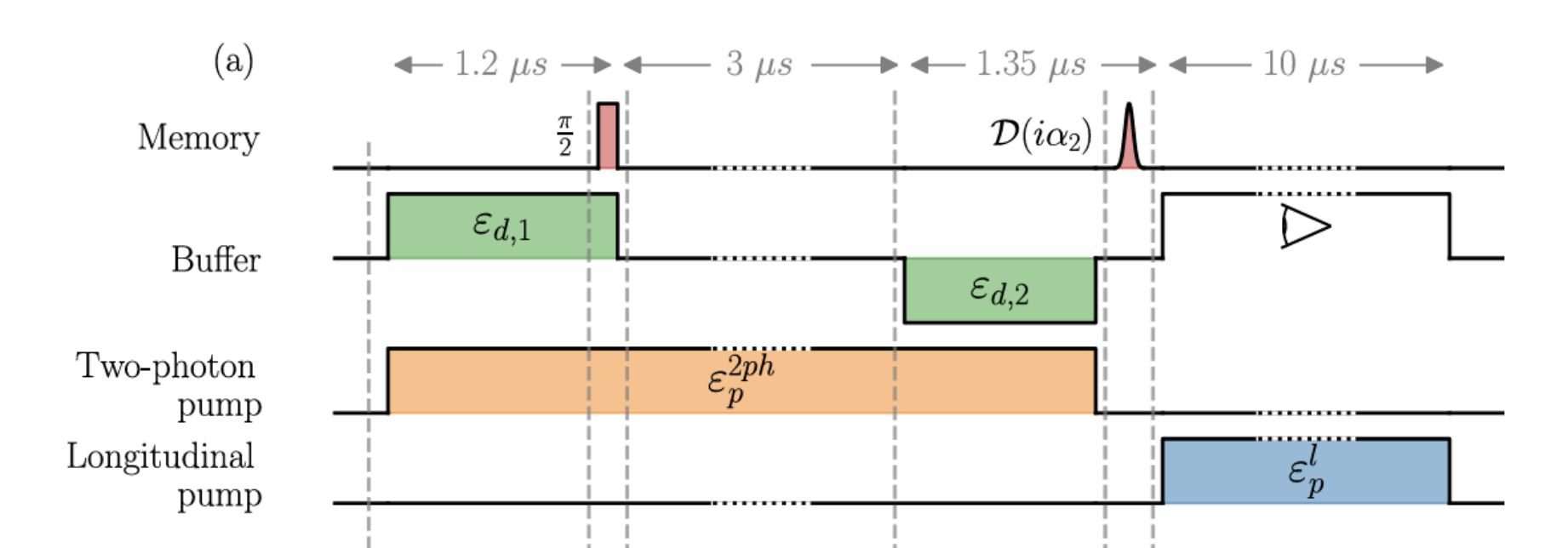}
    \caption{Pulse sequence of the quantum tomography protocol. Taken from \cite{Reglade_2024}.}
    \label{fig:fig_pulse_sequence_x_measurement}
\end{figure*}

The protocol used to perform the measurement is illustrated in Fig.~\ref{fig:fig_pulse_sequence_x_measurement}. The measurement sequence is divided into five steps \cite{Report2_confidential}. First, a \(\pi/2\) rotation about the \(Z\) axis is applied. This is followed by a deflation step, obtained by turning off the buffer drive while simultaneously turning on the two-photon pump. Next, an inflation step is performed by applying a buffer drive, which results in the following Hamiltonian:
\begin{equation}
    \hat{H}_{\mathrm{inf}} = \hbar g_2 \left( \hat{a}^{\dagger 2} \hat{b} + \hat{a}^2 \hat{b}^\dagger \right)
    + \hbar \epsilon_{d,\mathrm{inf}} \left( \hat{b} + \hat{b}^\dagger \right).
\end{equation}

Subsequently, a displacement operation is applied to the memory mode. Finally, to complete the gate sequence, a longitudinal readout is performed, described by the Hamiltonian
\begin{equation}
    \hat{H}_{\mathrm{lr}} = \hbar g_l \hat{a}^\dagger \hat{a} \left( \hat{b} + \hat{b}^\dagger \right).
\end{equation}
%where \(\hbar g_l = E_J \varphi_a^2 \varphi_b \epsilon_p^l\).

\textbf{Z(\(\theta\)) gate:}\\
The \(Z(\theta)\) gate is realized by applying a weak resonant single-photon drive in the presence of a two-photon driven dissipation, which stabilizes the cat-qubit manifold. The system Hamiltonian, in the rotating frame of the memory mode, is given by~\cite{Guillaud_2023, Gouzien_2023}
\begin{equation}
    \hat{H} = \epsilon_z \hat{a} + \epsilon_z^* \hat{a}^\dagger,
\end{equation}
where \(\epsilon_z\) denotes the amplitude of the resonant drive and \(\hat{a}\) is the annihilation operator of the cavity mode. 

The two-photon driven dissipation confines the dynamics to the stabilized cat-qubit subspace spanned by the even and odd cat states \(\{\lvert + \rangle_\alpha, \lvert - \rangle_\alpha\}\). Within this subspace, the weak single-photon drive induces transitions that change the photon-number parity, leading to Rabi oscillations between the even and odd cat states.

The effective Hamiltonian projected onto the cat-qubit manifold can be derived using first-order perturbation theory~\cite{Guillaud_2023}:
\begin{align}
    &(\lvert + \rangle_\alpha \langle + \lvert_\alpha + \lvert - \rangle_\alpha \langle - \lvert_\alpha)
    (\epsilon_z \hat{a} + \epsilon_z^* \hat{a}^\dagger)
    (\lvert + \rangle_\alpha \langle + \lvert_\alpha + \lvert - \rangle_\alpha \langle - \lvert_\alpha) \nonumber \\
    &= 2 \, \mathrm{Re}[\alpha \epsilon_z] \, \hat{Z} + \mathcal{O}(e^{-2|\alpha|^2}).
\end{align}

The resulting Rabi oscillations correspond to rotations around the \(Z\) axis of the Bloch sphere, realizing a  \(Z(\theta)\) gate. A rotation by an angle \(\theta\) is achieved by applying the drive for a duration
\begin{equation}
    T = \frac{\theta}{4 \alpha |\epsilon_z|}.
\end{equation}
To implement the \(Z(\theta)\) gate experimentally, one can simply apply the resonant drive on the memory mode while maintaining the two-photon driven dissipation active, as demonstrated in~\cite{Marquet2024}. The weak-drive condition \(|\epsilon_z| \ll \kappa_2\) ensures that the dynamics remain confined within the cat subspace, preventing leakage and enabling high-fidelity gate operations \cite{Guillaud_2023}.

\textbf{X gate:}\\
The implementation of the \(X\) gate is achieved by making the two-photon stabilization time dependent \cite{Gouzien_2023}. As a result, the Lindblad operator described in Eq.~\ref{lindblad_operator_stab} becomes
\begin{equation}
    \hat{L} = \sqrt{\kappa_2}\!\left(\hat{a}^2 - \bigl(\alpha e^{i \pi t / T}\bigr)^2\right),
\end{equation}
which allows for the adiabatic swapping of the states \(\lvert \pm \alpha \rangle\).

In an ideal scenario of a lossless oscillator, where the gate time could be taken as \(T \to \infty\), the fidelity of the \(X\) gate would be equal to \(1\). However, for finite gate times, a nonzero probability of a phase-flip error appears. To reduce this probability, the following Hamiltonian is added:
\begin{equation}
    \hat{H} = -\frac{\pi}{T}\,\hat{a}^\dagger \hat{a},
\end{equation}
which enables the gate to be performed in a non-adiabatic manner \cite{Guillaud_2019}.

\textbf{CNOT gate:}\\
The CNOT gate is realized by applying a longitudinal interaction to the target qubit, while the stabilization of the control qubit remains stable. This is achieved by tailoring the dissipative stabilization schemes of the control and target qubits, as described by the following equations:
\begin{align}
\hat{L}_{a_c} &= \hat{a_c}^2 - \alpha^2 , \\ \nonumber
\hat{L}_{a_t}(t) &= \hat{a}_t^2 - \frac{1}{2}\alpha(\hat{a}_c + \alpha)
+ \frac{1}{2}\alpha e^{2 i \frac{\pi}{T} t}(\hat{a}_c - \alpha).
\end{align}
The indices \(c\) and \(t\) refer to the control and target qubit, respectively. The dissipation channel \(\hat{L}_{a_c}\) implements a two-photon pumping process that stabilizes the control cat qubit.
The second channel, \(\hat{L}_{a_t}(t)\), acts on the target cat qubit while depending explicitly on the control mode \(\hat{a}_c\).
When the control qubit is in the state \(\ket{\alpha}\), \(\hat{L}_{a_t}(t)\) reduces to \(\hat{a}_t^2 - \alpha^2\), stabilizing the target qubit in the code space.
When the control qubit is in the state \(\ket{-\alpha}\), the effective pumping becomes
\(\hat{a}_t^2 - \left(\alpha e^{i\frac{\pi}{T}t}\right)^2\),
thereby implementing the \(X\) operation \cite{R_gent_2023}.

This total interaction is described by the Hamiltonian \cite{Guillaud_2019, R_gent_2023}
\begin{equation}
\hat{H}_{\text{CNOT}}= \hat{H}_{\text{c}}^{\text{stab}} + \hbar g_{\text{CNOT}} (\hat{a}_c + \hat{a}_c^\dagger)\hat{a}_t^\dagger \hat{a}_t,
\end{equation}
where \(\hat{H}_{\text{c}}^{\text{stab}} = g_2 \hat{a}^2 \hat{b}^\dagger + g_2^* \hat{a}^{\dagger 2} \hat{b} + \hbar\epsilon_d(\hat{b}+\hat{b}^\dagger)\). This longitudinal interaction should be realized during the time \(T_{\text{CNOT} = \frac{\pi}{4|\alpha|g_{\text{CNOT}}}}\) \cite{Report2_confidential}.

\iffalse
This is implemented by coupling the two buffer modes \(\hat{a}_c\) and \(\hat{a}_t\) to a buffer mode and
 driving it at three different frequencies \(\omega_1=2\omega_b-\omega_d, \omega_2=\frac{1}{2}(\omega_a-\omega_d)\) and \(\omega_3=\omega_d\) \cite{Guillaud_2023}.
\fi

\section{The Quantum Fourier Transform} \label{QFT_section}

\begin{figure*}[t]
    \centering
    \includegraphics[width=0.8\linewidth]{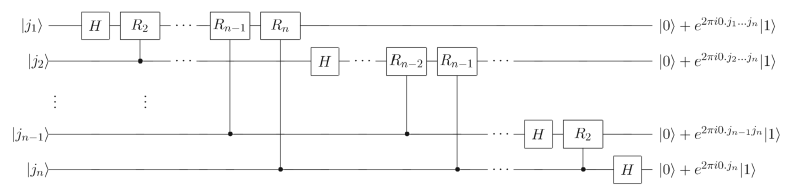}
    \caption{Quantum Fourier Transform circuit. SWAP gates at the end are omitted. Taken from \cite{Nielsen_Chuang_2010} .}
    \label{fig:quantum_fourier_transform_circuit}
\end{figure*}

\subsection{QFT circuit}

The QFT~\cite{coppersmith2002} is a quantum analogue of the Discrete Fourier Transform (DFT), which maps functions between time and frequency representations.

Consider an orthonormal basis of \(N\) states used to encode the information
\(\{ \lvert 0 \rangle, \ldots, \lvert N-1 \rangle \}\), which can be represented using \(n=\log_2(N)\) qubits. Applying the Quantum Fourier Transform (QFT) to a basis state \(\lvert k \rangle\) results in the following transformation \cite{Nielsen_Chuang_2010}:

\begin{equation}
    \lvert j \rangle \longrightarrow \frac{1}{\sqrt{N}} \sum_{k=0}^{N-1} e^{2\pi i jk/N} \lvert k \rangle.
\end{equation}
The action of the QFT on an arbitrary quantum state is given by
\begin{equation}
    \sum_{j=0}^{N-1} x_j \lvert j \rangle \longrightarrow \sum_{k=0}^{N-1} y_k \lvert k \rangle,\quad y_k = \frac{1}{\sqrt{N}} \sum_{j=0}^{N-1} x_j e^{2\pi i jk/N}.
\end{equation} 

Hence, the QFT acts as a linear transformation and can be represented by a unitary \(N \times N\) matrix. The corresponding QFT circuit is illustrated in Figure~\ref{fig:quantum_fourier_transform_circuit}. It consists of a sequence of Hadamard gates applied to each qubit, followed by controlled rotation gates. In this representation, the final swap gates are omitted. The rotation gate \(R_k\) is defined by the matrix
\begin{equation} \label{equation:R_k}
    R_k = \begin{pmatrix}
        1 & 0 \\
        0 & e^{2\pi i / 2^k}
    \end{pmatrix}.
\end{equation}

Since the circuit applies \(n\) Hadamard gates, one per qubit, \(n(n-1)/2\) controlled rotation gates, and at most \(n/2\) swap gates, the overall gate complexity of the QFT scales as \(\mathcal{O}(n^2)\).

\subsection{Semi-Classical QFT implementation}

The standard QFT circuit, composed of Hadamard and controlled rotation gates, cannot be directly implemented on physical qubits in a cat-qubit architecture. In particular, the Hadamard gates are responsible for switching the \(X\) basis of the \(Z\) basis and vice versa. Therefore, when this gate is applied, any existing phase-flip error is converted into a bit-flip error. This compromises the overall system, since bit flips are inherently protected in cat qubits.

At the logical level, however, a Hadamard operation can be implemented using fault-tolerant constructions, such as those based on Toffoli gates and magic state distillation. Despite being theoretically possible, these implementations are resource-intensive and do not provide a straightforward or efficient route to realizing the QFT.

Since our goal is to compare physical and logical qubit implementations, and given the limitations associated with Hadamard-based constructions, it is therefore natural to explore alternative formulations of the QFT that are better suited to the noise bias and operational constraints of cat-qubit architectures.

\begin{figure}[t]
    \centering
    \includegraphics[width=1\linewidth]{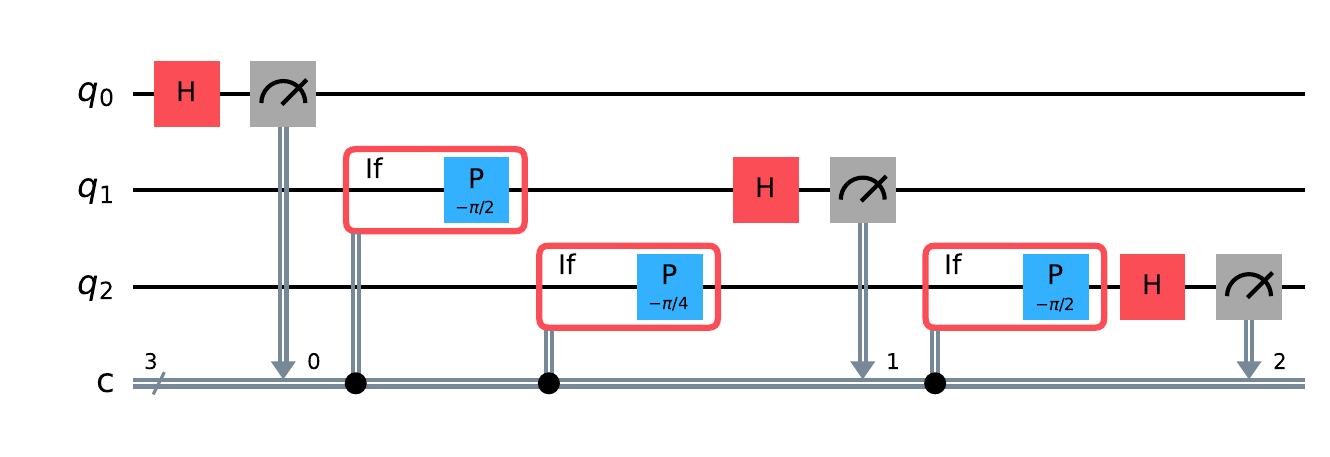}
    \caption{Semiclassical QFT circuit with three qubits.}
    \label{fig:semiclassical_QFT_circuit}
\end{figure}

The possibility is then implemented using a semiclassical approach \cite{Griffiths_1996}. The circuit for the semiclassical QFT with three qubits is shown in Figure~\ref{fig:semiclassical_QFT_circuit}. In this circuit, a Hadamard gate is applied to the first qubit, followed by a measurement along the \(Z\) axis, which effectively corresponds to a measurement along the \(X\) axis. If this measurement yields a result of 1, a phase rotation gate with angle \(-\pi / 2^i\) is applied to each subsequent qubit, where \(i \ge  1\) represents the index of the qubit where the gate is applied.  

In general, when the qubit with index \(j\) is measured along the X axis, if the result is \(1\), phase qubit gates are applied to the following qubits, with angle \(-\pi/2^{i-j}\), where \(i>j\) represents the qubit where the gate is applied. This process is repeated until all qubits have been measured.

The phase rotation gates can be written, up to a global phase, as rotations around the \(Z\) axis. Hence, the phase gate can be expressed as
\[
P(\alpha)  = 
\begin{pmatrix}
1 & 0 \\
0 & e^{i\alpha}
\end{pmatrix} = 
e^{i\frac{\alpha}{2}}
\begin{pmatrix}
e^{-i\frac{\alpha}{2}} & 0\\
0 & e^{i\frac{\alpha}{2}}
\end{pmatrix} 
= e^{-i\frac{\alpha}{2}} R_Z(\alpha)
\]

In the following energetic analysis, the whole circuit, which includes the conditional phase gates, is considered. This means that the energy consumption of the semiclassical QFT is evaluated under the assumption that all measurements, except for the last one, yield a result of \(1\), and that all phase gates are applied to the circuit.

The number of circuit repetitions is crucial for accurately characterizing the final state of the system. In general, the number of repetitions required to reliably reconstruct the output state depends on both the number of qubits and the resulting probability distribution, and may range from 1 up to $2^n$, where $n$ is the number of qubits (the number of entries in the diagonal of the density matrix) in the worst case of a nearly uniform distribution. However, this large number of repetitions is only necessary if we aim to obtain the full output state and store it classically. In general, some properties of the state can be extracted in a more efficient way. In this study, this dependence is neglected, and only a single repetition of the QFT circuit is considered. In a realistic scenario, however, the total energy cost should be multiplied by the actual number of repetitions required in each case.

\section{The Energetic Cost of the Gates}
This section describes the energy consumption and the phase-flip probability of the different gates used to implement the semiclassical QFT.

The energy can be divided into three different types according to the level at which we calculate the consumption \cite{Report1_confidential}:
\begin{enumerate}
    \item \textbf{Microscopic Level:} It refers to the qubit architecture. It is related to the energy of the drives the system receives and the implemented design.
    \item \textbf{Macroscopic Level:}  The minimal energy required when considering cryogenic mechanisms with Carnot efficiency.
    \item \textbf{Billed energy: } It is the exact energy consumed by the devices used to date. 
\end{enumerate}

Executing a given algorithm involves three distinct stages: preparation, computation, and readout. These three stages are typically estimated separately. However, the preparation of the initial state is itself a demanding problem. If the desired initial state is complex, a corresponding quantum circuit of significant depth may be required to prepare it. In~\cite{Pagni2025}, an improved quantum amplitude encoding scheme was investigated. Assuming that the quantum platform supports controlled rotation gates as native operations, the worst-case runtime scales as $\mathcal{O}(\sqrt{2^n}\, n)$, where $n$ is the number of qubits. This scaling remains more favorable than that of the classical counterpart. However, understanding the energy scaling of the initial state preparation is itself a subject of independent study, and for this reason, it lies outside the scope of this article.

Furthermore, in the semiclassical QFT, measurements play a fundamental role in the progression of the algorithm, since the measurement outcomes determine which gates are subsequently applied to other qubits. Because of this, we include the energy cost of measurements within the computation stage.

The total energy consumption should be specified for each of these categories. In this study, the cryogenics steps used at the macroscopic level, which are often treated as a separate category of energy consumption in other works \cite{Gois2024}, are instead included within each of the other categories. In other words, costs which are often referred to as baseline energy, here are not separated from the other costs. However, this baseline energy can be interpreted as the difference between macroscopic and microscopic energy. It represents the theoretical optimal baseline energy under the assumption of cryogenic systems operating at Carnot efficiency, which, in practice, is not achievable. Furthermore, for the remainder of this article, the analysis is performed under the condition \( \kappa_2>>100\kappa_1\), as required by Quantum Error Correction~\cite{Report1_confidential}.

The energy associated with each individual component of the circuit, arising from gate operations and stabilization, is now examined. Appendix~\ref{energy_parameters} presents a table summarizing all the relevant equations and the parameters associated with them.

\subsection{Energy Consumption During Stabilization}
To stabilize the cat qubit, a pump and a drive should be applied. The power required to activate the pump is given by the following expression \cite{Report1_confidential, Report2_confidential}
\begin{equation}
P_{\text{pump}} = p \frac{\kappa_b}{4}\kappa_2,
\end{equation}
where \(p\) is a constant related to the ATS parameters, for the microscopic level. For macroscopic, \(p\) should be multiplied by a constant \(M_p\) related to the attenuators at different temperatures on the drive line.

The power required to drive the buffer with the amplitude \(\epsilon_d\) is \cite{Report1_confidential, Report2_confidential}
\begin{equation}
    P_d(\epsilon_d) = d \epsilon_d^2,
\end{equation}
where \(\epsilon_d = \frac{\hbar \omega_b}{\kappa_b}\). In the macroscopic level, \(d\) should be multiplied by a constant \(M_d\).

Hence, the total power can be obtained by summing the power of the pump and the drive. 

\subsection{Energy Consumption During Preparation}

To prepare the state \(\lvert + \rangle_{\alpha}\), a stabilization drive and pump should be applied during a time \(\ T_{\text{prep}}=\frac{1}{\kappa_2}\)~\cite{Report1_confidential, Report2_confidential}.

Hence, the energy consumption during preparation can be calculated as
\begin{equation}
    E_{\text{prep \(\lvert+\rangle\)}} = (P_{\text{pump}} + P_d) \times T_{\text{prep}}
\end{equation}

To prepare a qubit in arbitrary states, one may begin with the plus state and subsequently apply appropriate operations to transform the qubit's state. Consequently, the corresponding total energy can be expressed as the sum of the preparation energy and the energy associated with the applied gates.

\subsection{Energy consumption of the Quantum Gates}

\textbf{\(Z(\theta)\) gate:}\\
The procedure for calculating the energy of each gate is analogous to that used for the preparation stage. The \( Z(\theta) \) gate is realized by applying a microwave drive \( \epsilon_z \), whose power is given by the following \cite{Report1_confidential, Report2_confidential}:
\begin{equation}
    P_z(\epsilon_z) = z \epsilon_z^2,
\end{equation}
where \(z= \frac{\hbar \omega_a}{\kappa_{\text{a}}}\) is the coupling rate between the memory line and the cavity.
Thus, the total power to perform the \(Z(\theta)\) gate is the sum of the stabilization power and \(P_z\). Again, for the macroscopic level, the constant \(z\) should be multiplied by a factor \(M_z\), related to the attenuators.

The drive should be active for a duration \(T_z = \frac{\theta}{4 |\alpha| \epsilon_z}\). To obtain the energy, one simply needs to multiply the required power by the duration. 

All the gates performed on the qubits will induce a phase-flip error probability. In particular, when performing the \(Z\) gate, this error is given by \cite{Report2_confidential}
\begin{equation} \label{phase_flip_probability_z_gate}
    p_z = \pi k_1 |\alpha|^2 T_z (1+2n_{th}^m) + \frac{ \epsilon_z^2}{ k_2  |\alpha^2|}T_z(1+2n_{th}^b),
\end{equation}
where \(n_{th}^m\) and \(n_{th}^b\) are the thermal populations of the memory and the buffer, respectively.

\textbf{CNOT gate:} \\
The energy of the CNOT gate can be obtained similarly to the \(Z\) gate.
The control qubit is always stabilized, while a longitudinal interaction is applied to the target qubit. The power of this longitudinal interaction is~\cite{Report1_confidential, Report2_confidential}
\begin{equation}
P_{\text{CNOT}} = c g_{\text{CNOT}}^2,
\end{equation}
where c is a constant related to the ATS parameters. For the macroscopic level, \(c\) should be multiplied by a constant \(M_c\). The drive should be applied during a time \(T_{\text{CNOT}} = \frac{\pi}{4|\alpha| g_{\text{cnot}}}\).

Hence, the total energy for the CNOT gate is
\begin{equation}
    E_{\text{CNOT}} = E_{\text{stab}}^{\text{control bit}} +  P_{\text{CNOT}}\times T_{\text{CNOT}}.
\end{equation}

The phase flip error induced by the CNOT gate on the target qubit is \cite{Report2_confidential}
\begin{equation}
    p_z^T = |\alpha|^2 \kappa_1 T_{\text{CNOT}}
\end{equation}

and on the control qubit is
\begin{equation}
    p_z^C = \frac{\pi}{4}( \frac{k_1 |\alpha|}{g_{\text{CNOT}}}(1+2n_{th}^m) + \frac{g_{\text{CNOT}}}{ k_2  |\alpha|}(1+2n_{th}^b)),
\end{equation}

\subsection{Energy Consumption During Readout}
According to Figure~\ref{fig:fig_pulse_sequence_x_measurement}, the protocol for measuring the qubit state in the X basis can be divided into five steps \cite{Report2_confidential}:
\begin{enumerate}
    \item \(\frac{\pi}{2}\) Rotation around the Z axis:

    This gate induces a phase-flip probability, which is given by equation~\ref{phase_flip_probability_z_gate}.

    \item Deflate Step:  Here, only the stabilization pump is turned on for a duration \(T_{\text{def}} = a_3/\kappa_2\)

    The phase-flip probability induced is given by
    \begin{equation}
        p_z^{\text{def}} = a_1 \frac{\kappa_1}{\kappa_2} + a_2 n_{th}^{m}.
    \end{equation}
    \item Inflate Step: The power considered is that of the stabilization pump, applied for a duration \(T_{\text{inf}} = 1/\kappa_2\)

    The phase-flip probability is 
    \begin{equation}
        p_z^{\text{inf}} = |\alpha|^2 \kappa_1 T_{\text{inf}} (1+2 n_{th}^m)
    \end{equation}
    \item Displacement: energy and phase-flip probability error negligible compared to other gates
    \item Fock-state longitudinal readout: The power considered was \(P_{\text{l}}= l g_l^2\), where \(l\) is a constant that depends on the ATS parameters.

    The fidelity of this longitudinal interaction is given by 
    \begin{equation}
        \mathcal{F}(t) = e^{- \kappa_1 t} \text{erf}(\text{SNR (t)/2}),
    \end{equation}
    where \(\text{SNR (t)/2}= \sqrt{4 \eta \frac{g_l}{k_b}(t+\frac{3}{k_b})}\).
    
\end{enumerate}

\section{Fidelity and Energy of the QFT without QEC}
\subsection{Fidelity of the Semiclassical QFT}
When a certain circuit is executed, the final fidelity is a measure that helps characterize whether the correct results are being obtained. Hence, it is an important factor in determining if the algorithm produces the expected outcomes.

In the semiclassical QFT, presented in Figure~\ref{fig:semiclassical_QFT_circuit}, it can be observed that only single-qubit gates are used. Therefore, a qubit does not depend on the state of other qubits, except for the measurement performed at the end of the calculations on the qubits above. Consequently, each qubit can be treated as independent, and the final fidelity of the state is the product of the final fidelity of each qubit
\begin{equation}
    F_{\text{total}} = \prod_{i=1}^{N} F_i
\end{equation}

The single-qubit fidelity can be obtained by considering the phase-flip channel, since the only error affecting the qubit, when only bias-preserving gates are chosen, is the phase-flip error. A phase-flip channel for a gate \(i\) with a given phase-flip probability \(p_i\) can be described by~\cite{Nielsen_Chuang_2010}
\begin{equation}
    \mathcal{E} = (1-p_i) \rho + p_i Z \rho Z
\end{equation}

For a sequence of N gates, the quantum channel can be represented by
\begin{equation} \label{quantum_channel_total_phase_flip_probability}
    \mathcal{E} = (1-p_{i_t}) \rho + p_{i_t} Z \rho Z
\end{equation}
where \(p_{i_t}=\frac{1-\lambda}{2}\) and \(\lambda=\prod_{i=1}^{N} (1-2 p_i)\).

The final average fidelity for each qubit is given by
\begin{equation}
    F_{i_{\text{avg}}} = 1 - \frac{2p_{i_t}}{3}
\end{equation}

\subsection{Energetics of the Semiclassical QFT}
\begin{figure*}[t]
    \centering

    \begin{minipage}{0.48\linewidth}
        \centering
        \includegraphics[width=\linewidth]{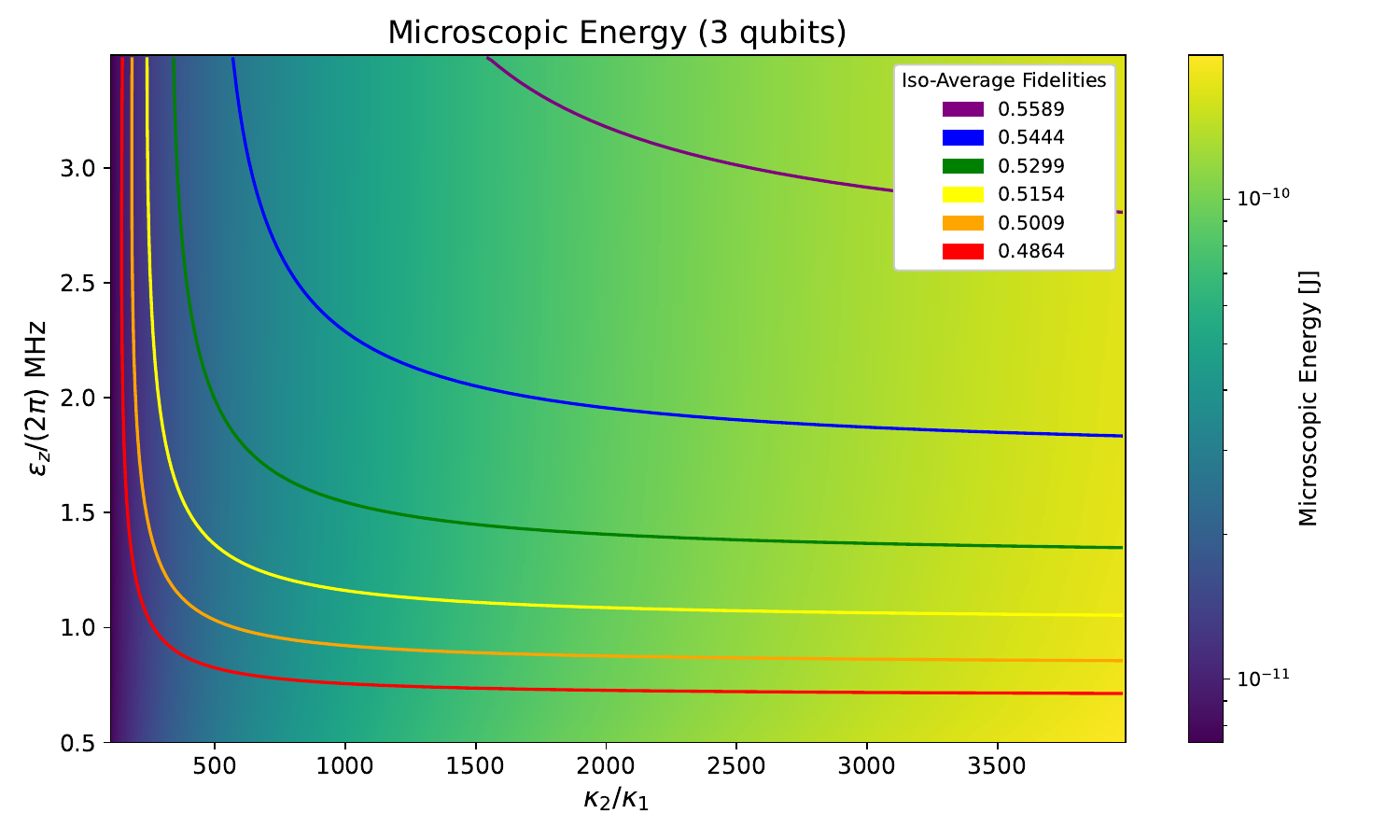}
        \caption*{(a) Microscopic Energy}
        \label{fig:microscopic_energy_physical}
    \end{minipage}
    \hfill
    \begin{minipage}{0.48\linewidth}
        \centering
        \includegraphics[width=\linewidth]{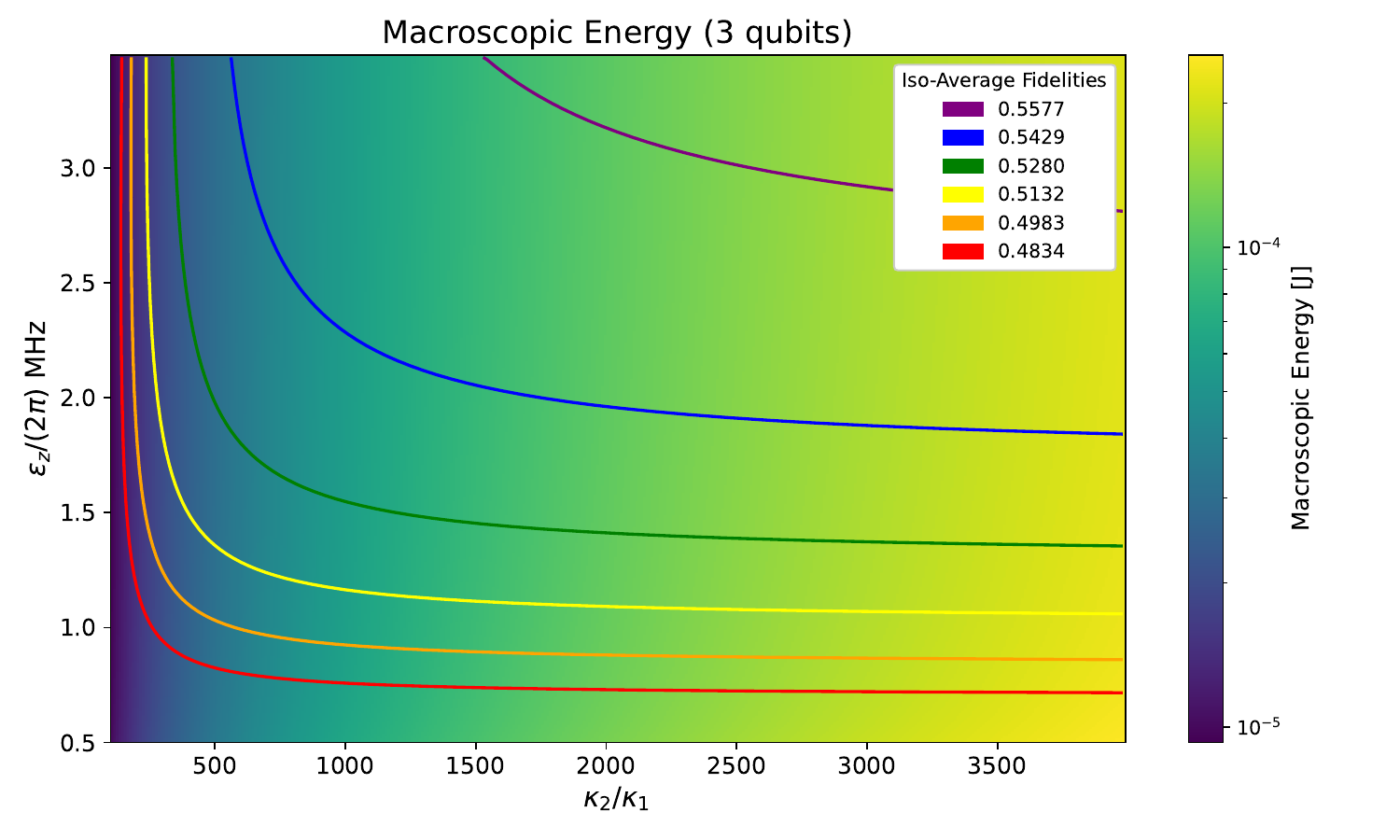}
        \caption*{(b) Macroscopic energy}
        \label{fig:macro__energy_physical}
    \end{minipage}

     \caption{Energy consumption of the semiclassical QFT applied to three qubits for both microscopic (a) and macroscopic (b) levels, for an amplitude of the coherent state \(\alpha=3\). The colored lines represent the average fidelity of the system after the QFT is performed.}
    \label{fig:semi_qft_energy_micro_macro_3_qubits}
\end{figure*}

The energetics of the QFT can be written as the sum of the energy consumption of each qubit separately, \(
E_{\text{total}} = \sum_{i=1}^N E_i,\)
where the energy associated with each qubit is computed by summing the contributions of all gates applied to it, together with the stabilization energy required. This includes both the stabilization needed during the execution of its own gates and the stabilization required while other gates are being applied to different qubits, and the qubit remains idle. Furthermore, once a qubit is measured, it becomes a classical signal. Therefore, no additional energy cost is considered after the measurement.
\begin{equation}
    E_i = E_{i_{\text{gates}}} + E_{i_{\text{stab}}},
\end{equation}
\begin{figure}[t]
    \centering
    \includegraphics[width=\linewidth]{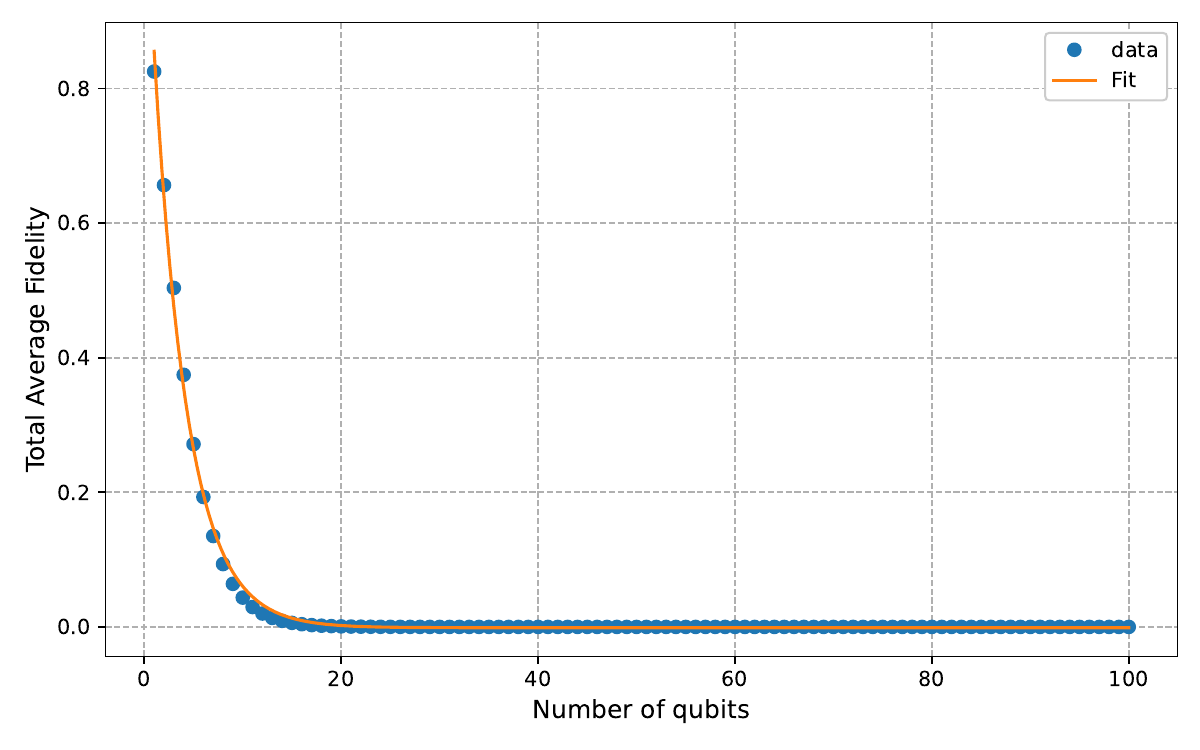}
    \caption{The Total Average Fidelity of the Semiclassical Quantum Fourier Transform decays exponentially with the number of qubits. In these estimates, \(\alpha=3\),  \(\kappa_2=1000\space \kappa_1\),  \(\frac{\epsilon_z}{2\pi} = \text{1MHz}\).}
    \label{fig:total_fidelity_macro}
\end{figure}

By summing the energy contributions of all the qubits, it is possible to estimate the total energy required to execute the algorithm. 

In the following analysis, the energy is computed under the assumption that only 
$\kappa_2$ and $\epsilon_z$ vary. These parameters are chosen because they are the 
most relevant in the calculations. The parameter $\kappa_2$ is responsible for 
stabilization, while $\epsilon_z$ controls the \(Z(\theta)\) gates, needed, in this case, to implement the phase gates, up to a global phase. All other parameters used to calculate the energy consumption of the different gates are treated as constants. 

A heatmap of the total energy as a function of \(\epsilon_z\) and \(\kappa_2/\kappa_1\) can be obtained for different numbers of qubits on which the QFT is applied. As an illustrative example, Figure~\ref{fig:semi_qft_energy_micro_macro_3_qubits} shows the energy consumption of the QFT applied to three qubits, with the microscopic and macroscopic scenarios displayed in the left and right panels, respectively. The isolines in the heatmap represent the total fidelity after the QFT is applied and are equally spaced in terms of fidelity, since the step between successive lines is constant.

In fact, when comparing both scenarios, the macroscopic energy is six orders of magnitude larger.
This difference is attributable to the cryogenics (with Carnot efficiency) required for the implementation of the algorithm. As expected, the baseline cost, directly associated with this difference, has a significant impact on the total energy required to perform the QFT.

In Fig.~\ref{fig:semi_qft_energy_micro_macro_3_qubits}, we observe that the fidelity increases with both \(\epsilon_z\) and \(\kappa_2 / \kappa_1\). However, when the range of the heatmap is extended, the growth of the fidelity begins to diminish and eventually converges. As expected, increasing the energy required to perform the QFT leads to a higher total fidelity, up to a maximum value. Beyond this point, further improvements in fidelity require reducing the probability of phase flips for each qubit involved in the computation. One possible solution is the use of QEC mechanisms that encode the information across multiple physical qubits.

Furthermore, assuming that the fidelity of each qubit is slightly less than \(1\), the total fidelity of the system can be expressed as \(\mathcal{F} = \prod_{i=1}^N \mathcal{F}_i \leq \mathcal{F}_{\text{max}}^N.\)
Hence, as the number of qubits increases, the total fidelity decreases exponentially. Figure~\ref{fig:total_fidelity_macro} shows the total fidelity when the QFT is performed on different numbers of qubits, assuming the parameters are fixed. The data were fitted with the function
\begin{equation}
    y = a\, e^{-b n} + c,
\end{equation}
which yields the values
\[
a = 1.1476, \quad b = 0.2927, \quad c = -0.0012.
\]

\section{Quantum Error Correction}\label{logical_gate_implementation}

One of the major challenges in quantum computation is the presence of noise, which reduces the lifetime of qubits. To address this issue, mechanisms for mitigating qubit decoherence have been developed. There are several approaches to encoding information using bosonic codes, such as Fock-state encoding, Binomial encoding \cite{Michael_2016}, and Cat-code encoding \cite{Cochrane_1999, Mirrahimi_2014}. The present work focuses on the latter.

Cat qubits are primarily subject to two types of errors: bit-flip errors (in which the qubit state changes from \(\lvert0\rangle\) to \(\lvert1\rangle\) or vice versa) and phase-flip errors (in which the phase of the \(\lvert1\rangle\) state is inverted). However, due to the stabilization process, the occurrence of bit-flip errors is significantly suppressed, leaving only phase-flip errors to be corrected. This enables the use of a one-dimensional quantum error correction scheme, rather than a two-dimensional one \cite{Guillaud_2019}, thereby reducing the number of physical qubits required \cite{Ruiz_2025}. A logical qubit can be encoded as a sequence of physical qubits, alternating between data and auxiliary qubits, where a repetition code is used to correct phase-flip errors. 

However, the number of required physical qubits can be reduced by a factor of five by using a low-density parity-check (LDPC) code, which allows the connectivity between distant qubits~\cite{Ruiz_2025}.

The error correction method employed in the energy estimation was a repetition code of distance \(d_c\), meaning that one logical qubit is encoded in \(d_c\) physical data qubits, along with \(d_c-1\) ancillary qubits interleaved between them. Furthermore, the quantum error correction cycle used in the energy calculation was taken from \cite{Report2_confidential}.

\begin{figure}[t]
    \centering
    \includegraphics[width=0.9\linewidth]{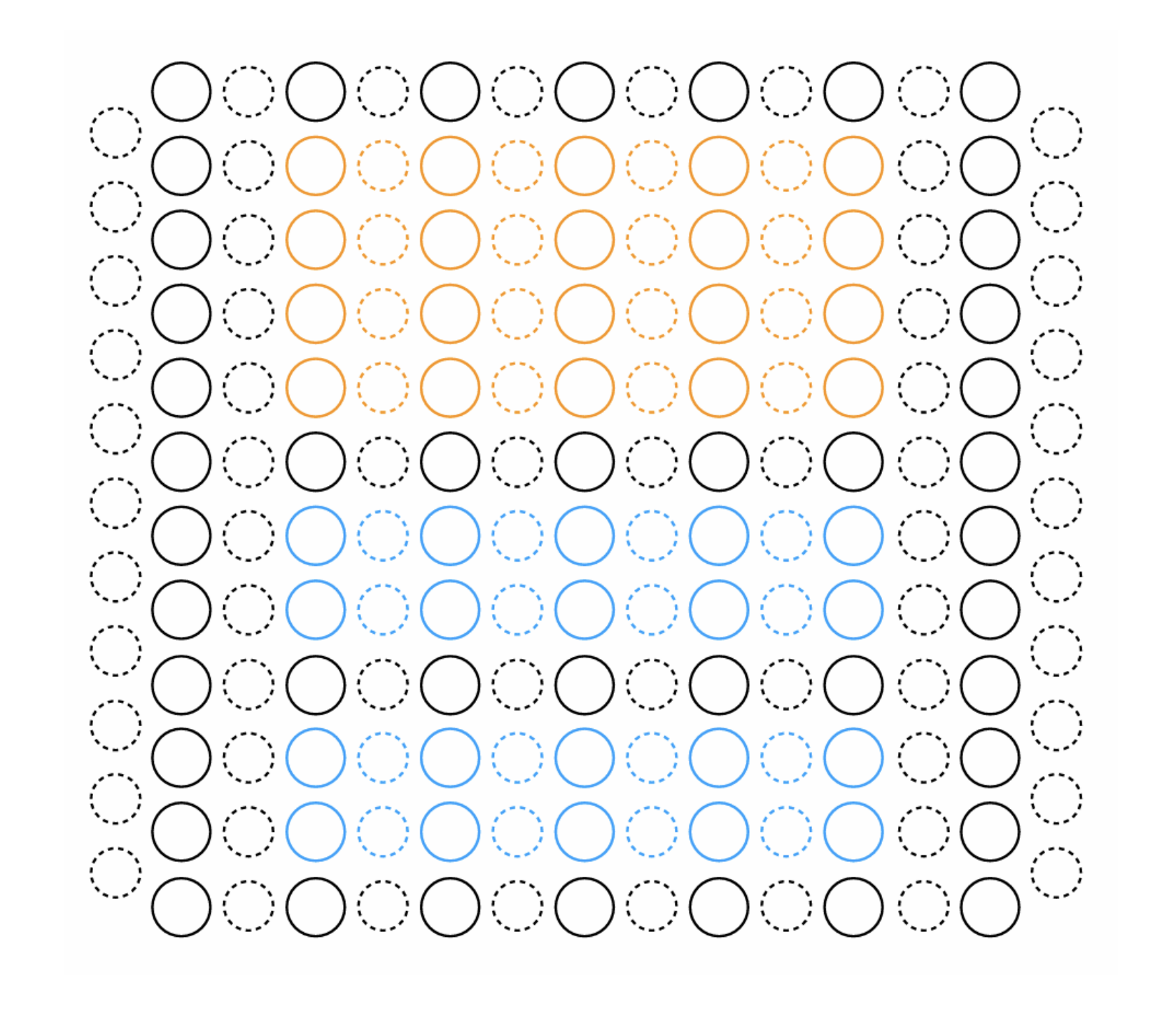}
    \caption{Layout of a quantum processor using cat qubits with four logical qubits and a code distance of~5. Each blue line represents a logical qubit: solid lines correspond to data qubits used to encode information, while dashed lines indicate ancilla qubits. The black circles represent auxiliary qubits. Taken from \cite{Gouzien_2023}. (color online)}
    \label{fig:layout_qec}
\end{figure}

In the figure \ref{fig:layout_qec}, a quantum processor is drawn. The line circles are the qubits used to encode the data. The dashed circles are ancillary qubits, allowing the measurements. The qubits used to perform computation are in blue, where each line represents a logical qubit~\cite{Gouzien_2023}.

\subsection{Implementation of Operations on Logical States}

\textbf{Preparation of the $\lvert + \rangle_L$ and $\lvert 0\rangle_L$ states}

The preparation of the $\lvert + \rangle_L$ and $\lvert 0\rangle_L$ is realized by preparing of the $\lvert + \rangle_L ^{\otimes d}$ and $\lvert 0\rangle_L ^{\otimes d}$. This preparation is proceeded by d rounds of stabilizer measurement, decoding, and correction.
\par

\textbf{Measurement of $X_L$}

This measurement is performed by measuring all the data qubits in the code on the X basis, followed by a majority vote on the measurement outcomes.

\textbf{Measurement of $Z_L$}

This measurement is performed by measuring all the data qubits in the code on the Z basis, and then taking the product of all the measurement outcomes.

\textbf{Logical $X_L$ gate}

This logical gate is implemented by performing a single physical X gate on any of the data qubits.

\textbf{Logical $Z_L$ gate}

This logical gate is implemented transversally by applying a Z gate to all the data qubits.

\textbf{Logical \(R_Z(\theta)\) gate}

For a code distance of \(d\), a logical qubit is encoded into \(d\) physical qubits, as shown in Figure~\ref{fig:layout_qec}. A logical rotation must therefore act coherently on the encoded subspace, rather than as independent single-qubit operations on each physical qubit.

For a distance-\(3\) repetition code, a logical state is
\begin{equation}
    \lvert \Psi \rangle = \alpha \lvert 000 \rangle + \beta \lvert 111 \rangle.
\end{equation}

The logical \(R_Z(\theta)\) operation is defined as
\begin{equation}
    R_Z(\theta) = e^{-i \frac{\theta}{2} Z},
\end{equation}
with action
\begin{align}
    R_Z(\theta)\lvert 0 \rangle_L &= e^{-i \frac{\theta}{2}} \lvert 0 \rangle_L, \\
    R_Z(\theta)\lvert 1 \rangle_L &= e^{+i \frac{\theta}{2}} \lvert 1 \rangle_L.
\end{align}

Thus, the desired transformation is
\begin{equation}
    \lvert \Psi' \rangle = \alpha e^{-i \frac{\theta}{2}} \lvert 000 \rangle + \beta e^{+i \frac{\theta}{2}} \lvert 111 \rangle.
\end{equation}

This logical operation can be implemented using an ancilla-assisted protocol. We introduce \(d\) auxiliary qubits initialized in \(\lvert 0 \rangle^{\otimes d}\), and apply transversal CNOT gates between data and ancilla qubits:
\begin{equation}
    \lvert \Psi_1 \rangle =
    \alpha \lvert 0 \rangle^{\otimes d}_D \lvert 0 \rangle^{\otimes d}_A
    + \beta \lvert 1 \rangle^{\otimes d}_D \lvert 1 \rangle^{\otimes d}_A.
\end{equation}

A single physical rotation \(R_Z(\theta)\) is then applied to one ancilla qubit, which yields
\begin{equation}
    \lvert \Psi_2 \rangle =
    \alpha \lvert 0 \rangle^{\otimes d}_D \lvert 0 \rangle^{\otimes d}_A
    + \beta e^{i\theta} \lvert 1 \rangle^{\otimes d}_D \lvert 1 \rangle^{\otimes d}_A.
\end{equation}

Finally, applying the inverse transversal CNOTs disentangles the ancillas:
\begin{equation}
    \lvert \Psi_3 \rangle =
    \left( \alpha \lvert 0 \rangle^{\otimes d} + \beta e^{i\theta} \lvert 1 \rangle^{\otimes d} \right)
    \otimes \lvert 0 \rangle^{\otimes d}.
\end{equation}

Up to a global phase, this corresponds exactly to the logical transformation
\begin{equation}
    \lvert \Psi' \rangle = \alpha e^{-i \frac{\theta}{2}} \lvert 000 \rangle + \beta e^{+i \frac{\theta}{2}} \lvert 111 \rangle.
\end{equation}
\begin{figure*}[t]
    \centering
    \includegraphics[width=0.8\linewidth]{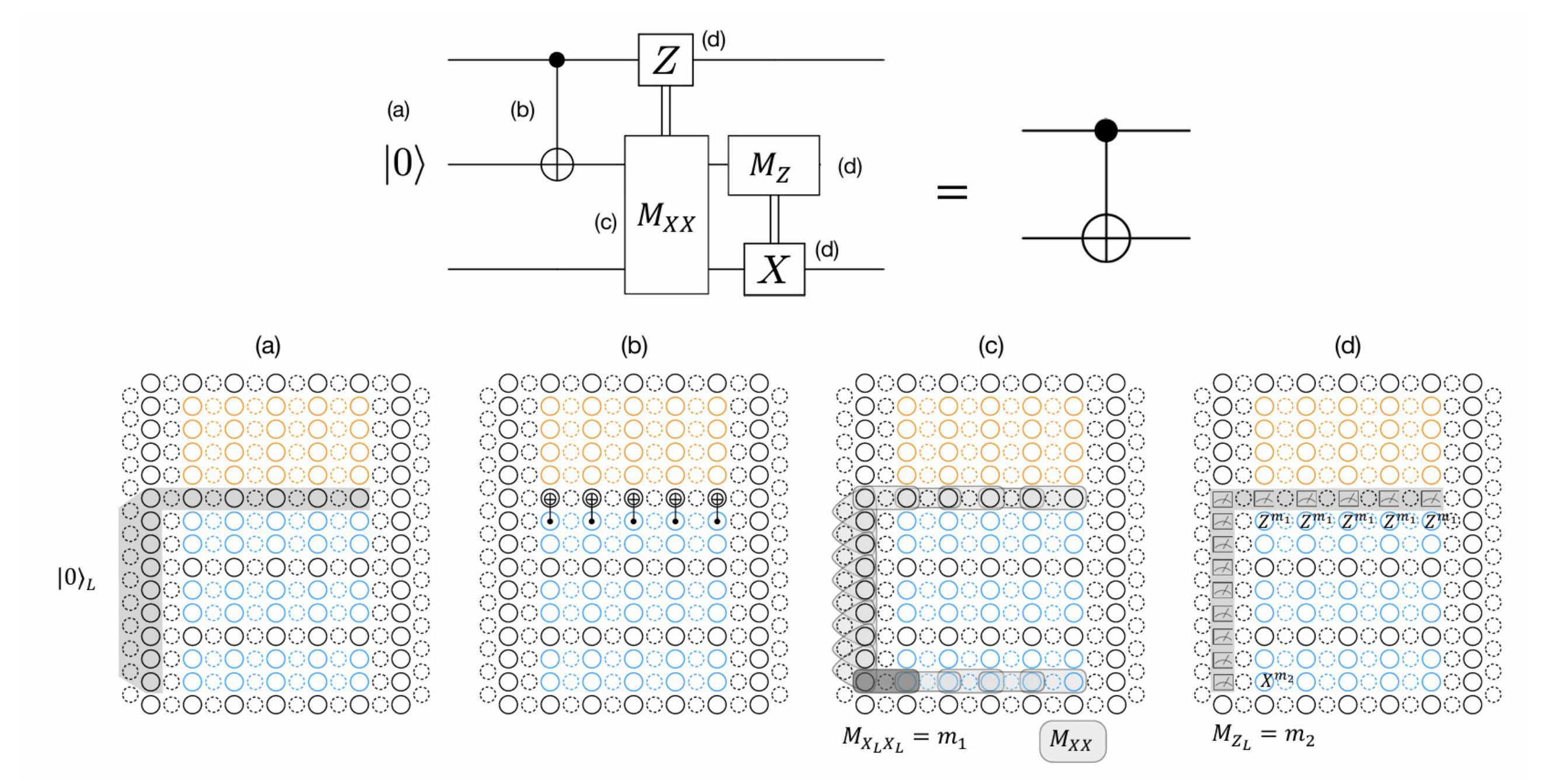}
    \caption{\textbf{Top:} Circuit used to perform a CNOT gate. \textbf{Bottom:} Steps for realizing a CNOT gate on the processor. Taken from \cite{Gouzien_2023}.}
    \label{fig:Logical_CNOT_gate}
\end{figure*}
\textbf{Logical CNOT gate}

The steps to perform a logical CNOT gate are the following:
\begin{enumerate}
    \item The ancillary routing physical qubits are used to prepare a logical ancilla state $\lvert0\rangle_L$. This ancilla forms an L over the processor, beginning in the logical control qubit and ending in the logical target qubit.
    \item Logical CNOT gate is implemented by performing a CNOT between the logical control qubit and the adjacent logical ancilla
    \item Logical $X_LX_L$ measurement is implemented by measuring d times the XX operator of the two physical qubits at the border between the logical qubits.
    \item Logical $Z_L$ is measured on the logical ancilla qubit
    \item Logical $Z_L$ is applied to the control qubit if the logical $X_L \otimes X_L$ measurement produces the output -1, and a logical $X_L$ is applied to the logical target if the logical $Z_L$ on the ancilla qubit produces the value -1. 
\end{enumerate}

\subsection{Repetition Code}
\begin{figure*}[t]
    \centering
    \includegraphics[width=0.6\linewidth]{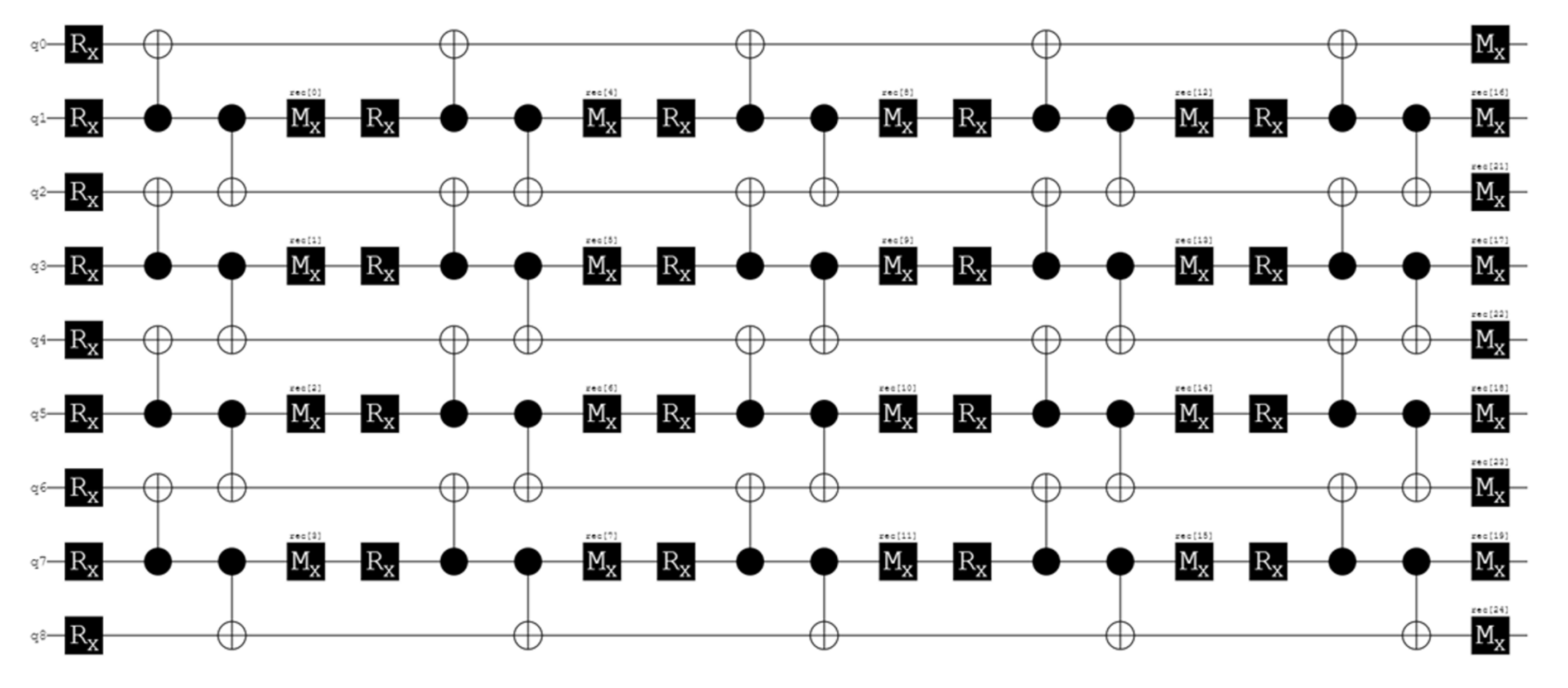}
    \caption{One round of the repetition code with distance d=5 with d cycles. The \(R_X\) is the preparation of the plus state. Taken from \cite{Report2_confidential}.}
    \label{fig:QEC_repetition_code_circuit}
\end{figure*}
To correct phase-flip errors after applying a certain number of gates, a repetition code must be implemented. The state of a single logical qubit is encoded in \(d\) physical qubits. The circuit consists of CNOT gates, measurements of the \(X\) operator, and preparations of the \(\lvert + \rangle\) state. Figure~\ref{fig:QEC_repetition_code_circuit} shows one round of the repetition code with \(5\) cycles and a code distance of \(5\). This is the specific repetition code used as the basis for the energetics analysis.

\begin{figure*}[t]
    \centering

    \begin{minipage}{0.48\linewidth}
        \centering
        \includegraphics[width=\linewidth]{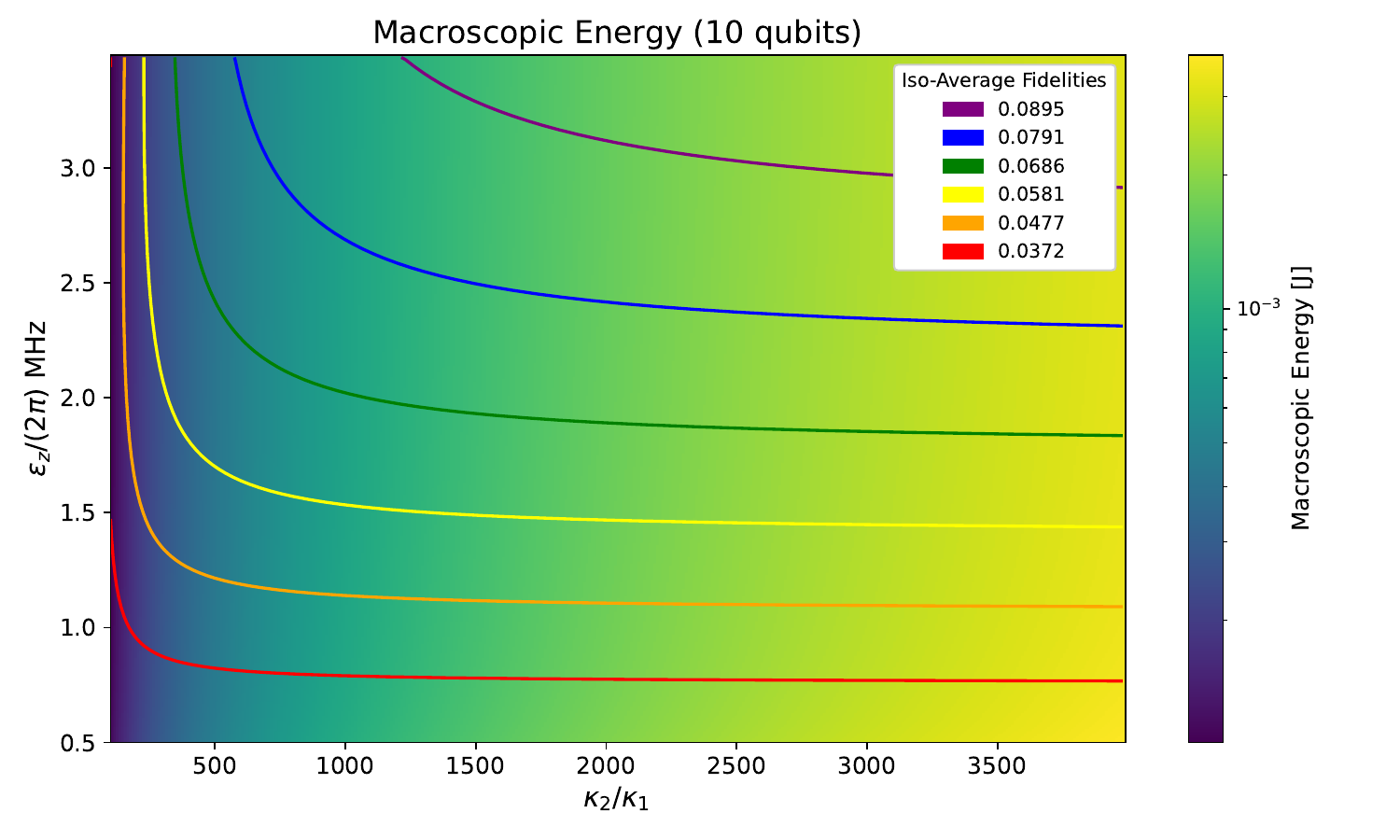}
        \caption*{(a) Physical Semiclassical QFT}
        \label{fig:physical_energy}
    \end{minipage}
    \hfill
    \begin{minipage}{0.48\linewidth}
        \centering
        \includegraphics[width=\linewidth]{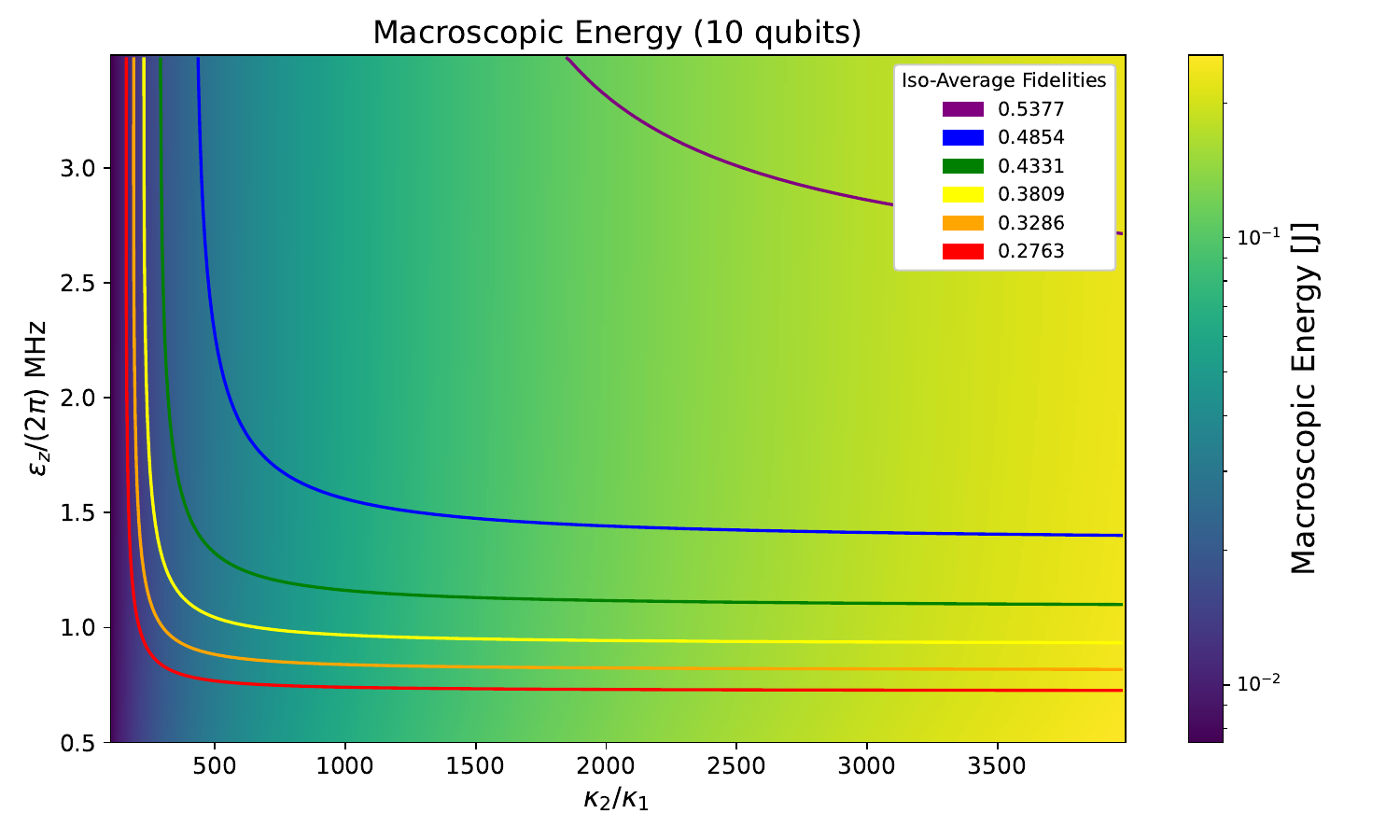}
        \caption*{(b) Logical Semiclassical QFT for \(N_b=1\) and \(d_c=5\)}
        \label{fig:logical_energy}
    \end{minipage}

    \caption{Comparison between the physical and logical energy and total average fidelity estimates for the semiclassical QFT on 10 qubits and for \(\alpha=3\). The total average fidelity increases significantly within the same heatmap ranges.}
    \label{fig:physical_vs_logical}
\end{figure*}

\subsection{Logical Fidelity}

A logical gate is implemented by applying the same physical gate to each of the \(d\) physical qubits encoding a logical qubit. Assuming that the application of a physical gate induces a phase-flip error on a single physical qubit with probability \(p\), the resulting phase-flip error probability of the logical gate is given by

\begin{equation}
    P_{\text{logical}} = \sum_{k=t+1}^d \binom{d}{k} p^k (1-p)^{d-k},
\end{equation}
where \(t = (d-1)/2\) is the maximum number of physical phase-flip errors that the error correction code can detect.

Hence, for a channel containing \(N\) gates, the circuit should be divided into \(M\) segments, each with a total number of gates equal to \(N_b\). For each segment, we compute the phase-flip probability \(p_i\), given by Eq.~\ref{quantum_channel_total_phase_flip_probability}. Since the state of the qubit is corrected after every segment, we then compute the corrected phase-flip probability \(P_i\) for each segment, which is given by
\begin{equation}
    P_{i} = \sum_{k=t+1}^d \binom{d}{k} p_i^k (1-p_i)^{d-k},
\end{equation}
The total phase-flip probability of the quantum channel for this qubit is
\begin{equation}
    P_{\text{channel i}}= \frac{1-\lambda_{\text{channel}}}{2},
\end{equation}
where \(\lambda_{\text{channel}}= \prod_{i=1}^M(1-2P_i)\) 

The total fidelity for each channel can be computed as \(\mathcal{F}_{\text{channel } i} = 1 - \frac{2}{3} P_{\text{channel } i}.\) Since in the semiclassical QFT the qubits evolve independently, the total fidelity of the system is obtained by multiplying all individual fidelities.

\subsection{Energetics of the Logical Semiclassical QFT with QEC}

\begin{figure*}[t]
    \centering

    \begin{minipage}{0.48\linewidth}
        \centering
        \includegraphics[width=\linewidth]{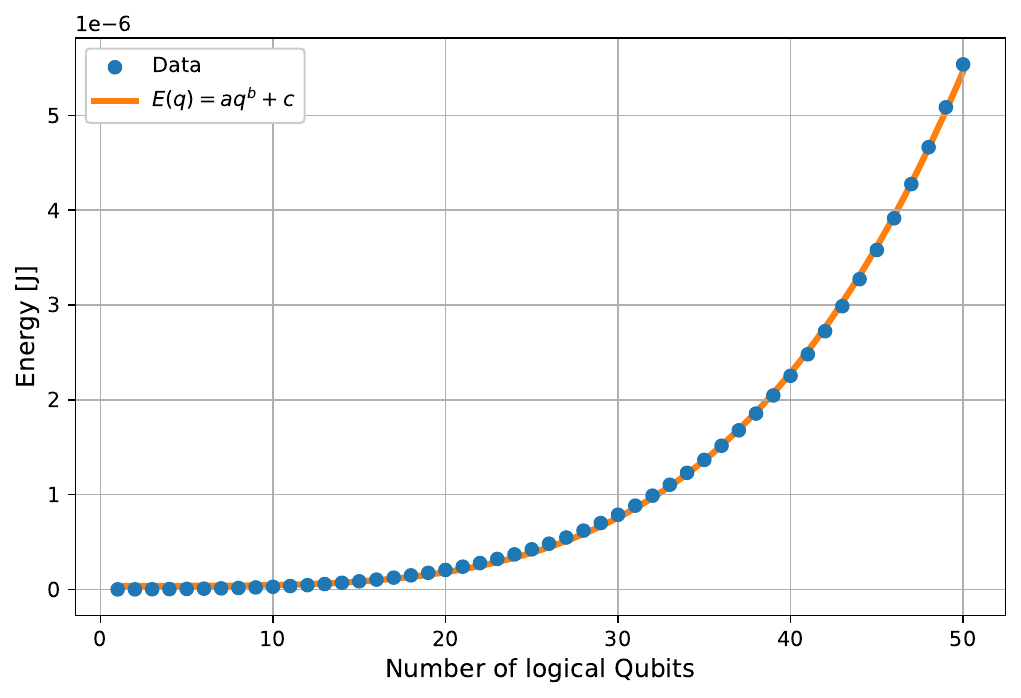}
        \caption*{(a) Microscopic energy (\(dc=5, N_b=1\))}
        \label{fig:micro}
    \end{minipage}
    \hfill
    \begin{minipage}{0.48\linewidth}
        \centering
        \includegraphics[width=\linewidth]{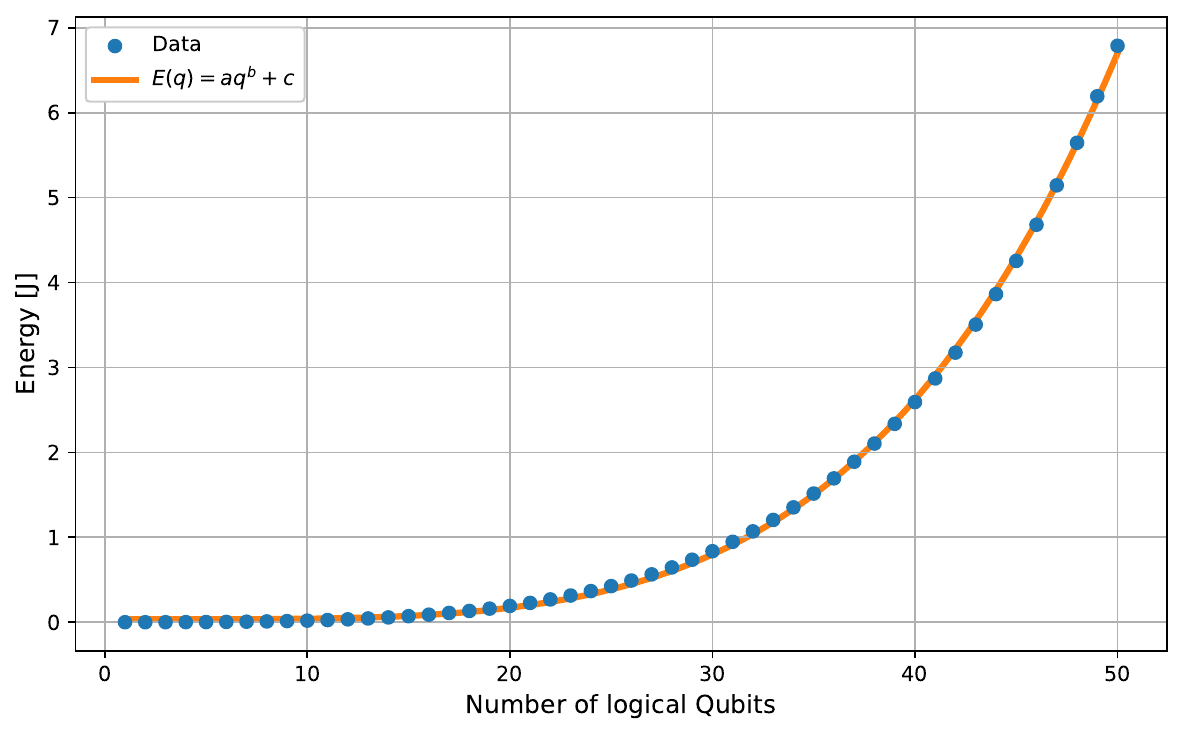}
        \caption*{(b) Macroscopic energy (\(dc=5, N_b=1\))}
        \label{fig:macro}
    \end{minipage}

    \caption{Energy estimates for microscopic and macroscopic scenarios for the QFT applied to different numbers of qubits, where the parameters \(\epsilon_z\) and \(\kappa_2\) were optimized to minimize the energy while maintaining a last-qubit fidelity of \(0.900\). The energy scaling follows a polynomial dependence on the number of qubits, with an approximate degree of 4.0 at the microscopic level and 4.2 at the macroscopic level.}
    \label{fig:macro_micro_comparison_50_qubits_fixing_nb_and_dc}
\end{figure*}

\begin{figure*}[t]
    \centering

    \begin{minipage}{0.48\linewidth}
        \centering
        \includegraphics[width=\linewidth]{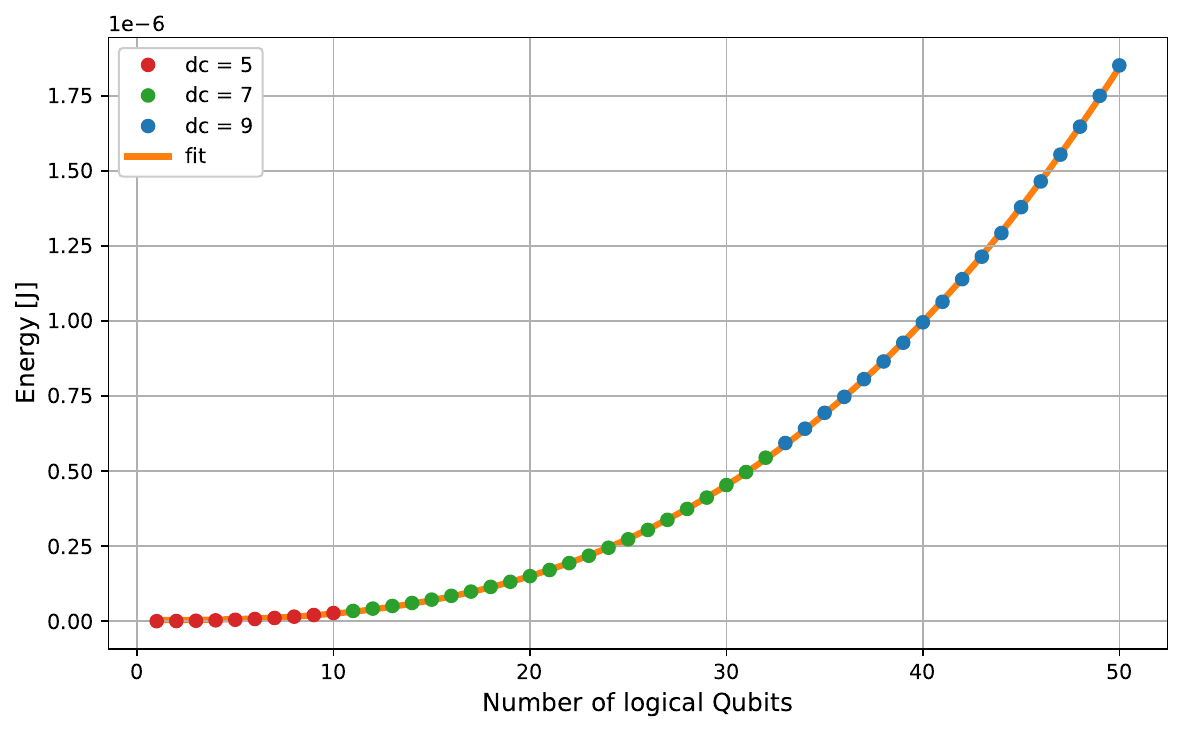}
        \caption*{(a) Microscopic energy}
        \label{fig:micro_optimized_energy}
    \end{minipage}
    \hfill
    \begin{minipage}{0.48\linewidth}
        \centering
        \includegraphics[width=\linewidth]{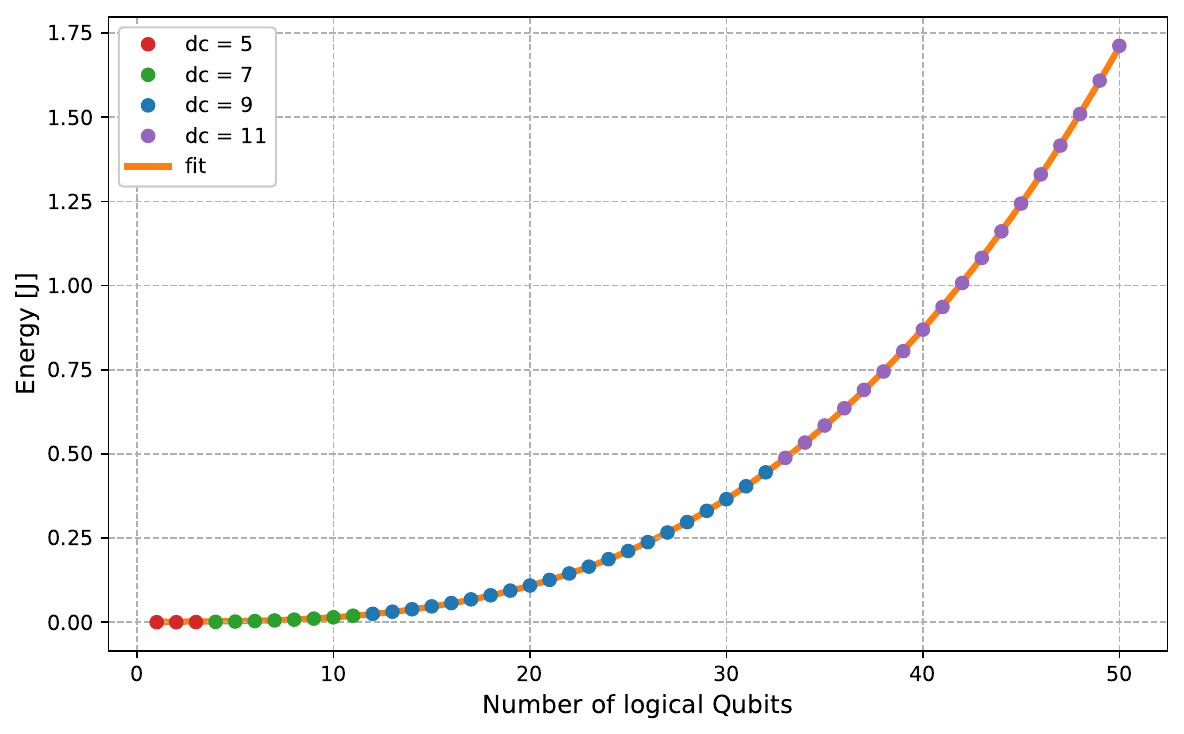}
        \caption*{(b) Macroscopic Energy }
        \label{fig:macro_optimized_energy}
    \end{minipage}

    \caption{Optimized energy for the QFT with \(\alpha = 3\), \(Nb = 1\), and a final qubit fidelity greater than $0.900$, where the different colors correspond to different distance codes. The energy scaling follows a polynomial dependence on the number of qubits of degree 2.8 at the microscopic level and 3.0 at the macroscopic level.}
    \label{fig:optimize_energy}
\end{figure*}
By following the same approach used for the physical energy, the energy consumption of the semiclassical QFT can be estimated over the ranges of $\epsilon$ and $\kappa_2/\kappa_1$ of interest. Figure~\ref{fig:physical_vs_logical} shows a comparison between the physical and logical energy estimates. The logical energy is significantly larger than the physical energy over the same heatmap frontiers, since quantum error correction mechanisms increase both the number of required operations and the time needed to perform the computation. This, in turn, leads to an increase in the stabilization energy.

The isotropic lines shown in Figure~\ref{fig:physical_vs_logical} represent the total average fidelity. As expected, the presence of QEC codes leads to a significant increase in fidelity across the range displayed in the heatmaps.

% The fidelity associated with the heatmap can also be computed. Figure~\ref{fig:physical_vs_logical} includes several contour lines, each corresponding to a fixed value of the fidelity. These fidelity contours are chosen by dividing the fidelity into equally spaced intervals.

\section{Energy scaling with number of qubits}

\subsection{Quantum Computers}
To determine how the total energy required to implement the QFT scales with the number of qubits, it is essential to select optimal values for the parameters $\epsilon_z$, $\kappa_2$, $N_b$, and $d_c$ when computing the energy cost. For a meaningful comparison across different system sizes, the fidelity of the qubits must also be taken into account. Therefore, the energy associated with performing the QFT should be evaluated under the constraint that the fidelity is kept constant throughout the comparison.

Since the qubits are independent and we operate in a regime where all gates of the semiclassical QFT are applied, the last qubit experiences the largest number of gate operations and consequently exhibits the lowest fidelity. The remaining qubits undergo fewer gates and therefore retain higher fidelities. By fixing the fidelity of the last qubit, we ensure that all other qubits achieve fidelities that are at least as high. This approach provides a fair and consistent benchmark for evaluating energy scaling.

Thus, fixing the fidelity of the last qubit allows us to tune the parameters $\epsilon_z$ and $\kappa_2/\kappa_1$ in order to minimize the total energy required to perform the QFT.

Figure~\ref{fig:macro_micro_comparison_50_qubits_fixing_nb_and_dc} shows the energy as a function of the number of qubits while maintaining a fixed fidelity for the last qubit (\(F=0.900\)). In this example, the code distance is set to $d = 5$ and $N_b = 1$. The resulting data are fitted using the function
\[
y = a\, n^{\,b} + c.
\]
For the microscopic regime, the fit parameters are
\[
a = 1.052 \times 10^{-12}, \qquad
b = 3.951, \qquad
c = 3.582 \times 10^{-8},
\]
while for the macroscopic regime, the parameters are
\[
a = 4.178 \times 10^{-7}, \qquad
b = 4.240, \qquad
c = 3.349 \times 10^{-2}.
\]

A polynomial scaling of the energy with the number of qubits was observed, with a degree of approximately \(4.240\) in the macroscopic scenario. However, the energy required to perform the QFT also depends on two additional tunable parameters, $N_b$ and $d_c$. Adjusting these parameters can lead to a reduction in total energy consumption while maintaining the desired fidelity of the final qubit. In fact, decreasing the number of gates applied between error-correction cycles may reduce the required energy for achieving a fixed last-qubit fidelity (see Appendix~\ref{Depende_on_parameters_Nb_and_d}). Although the total number of gates increases when more frequent correction cycles are implemented, the fidelity within each segment improves. This allows operation in a regime where a smaller value of $\kappa_2$ can be used, thereby reducing the overall energy.

\begin{figure*}[t]
    \centering

    % ---------- Top row: F = 0.965 ----------
    \begin{minipage}{0.48\linewidth}
        \centering
        \includegraphics[width=\linewidth]{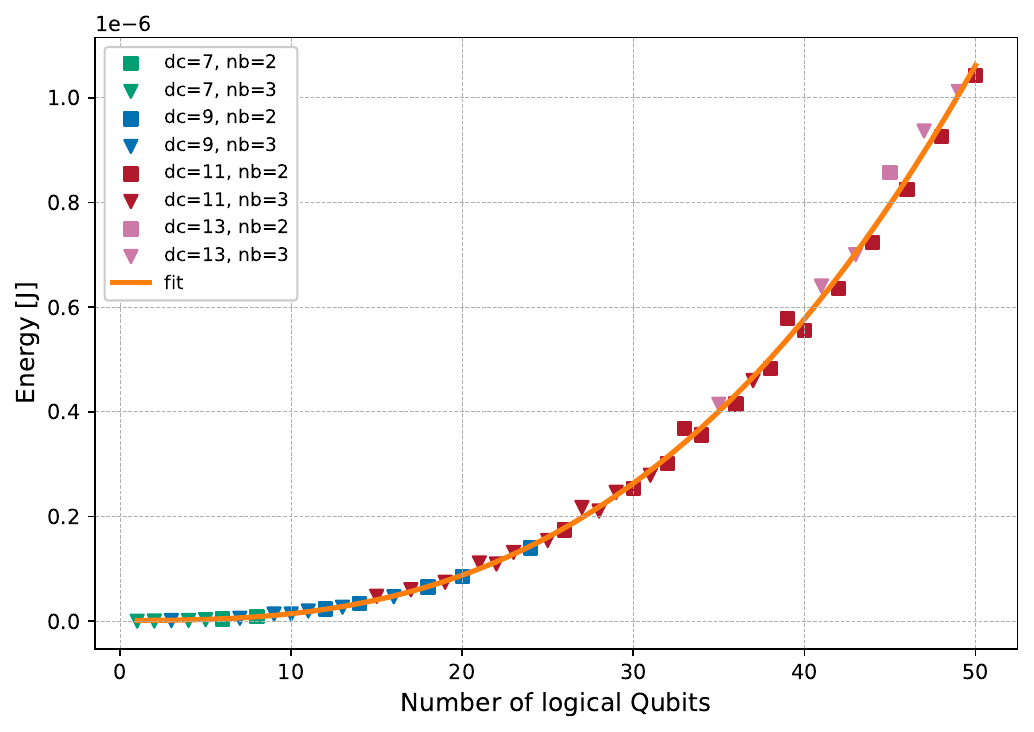}
        \caption*{(a) Microscopic energy ($F=0.900$)}
        \label{fig:energy_optimized_comparison_a}
    \end{minipage}
    \hfill
    \begin{minipage}{0.48\linewidth}
        \centering
        \includegraphics[width=\linewidth]{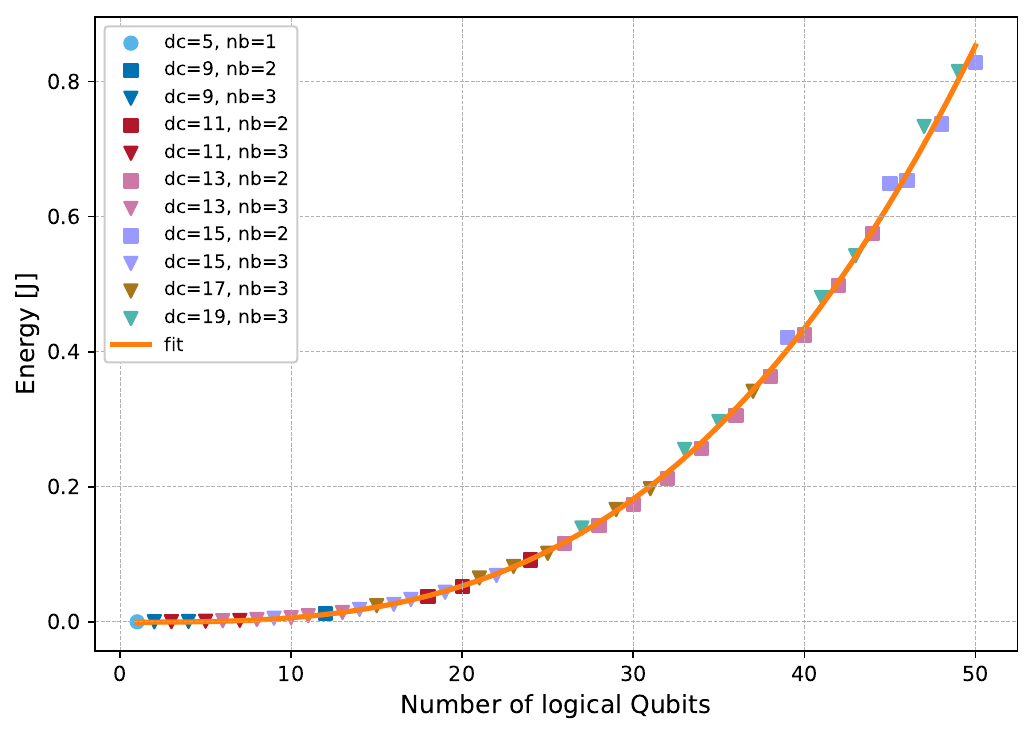}
        \caption*{(b) Macroscopic energy ($F=0.900$)}
        \label{fig:energy_optimized_comparison_b}
    \end{minipage}

    \vspace{0.5cm}

    % ---------- Bottom row: F = 0.990 ----------
    \begin{minipage}{0.48\linewidth}
        \centering
        \includegraphics[width=\linewidth]{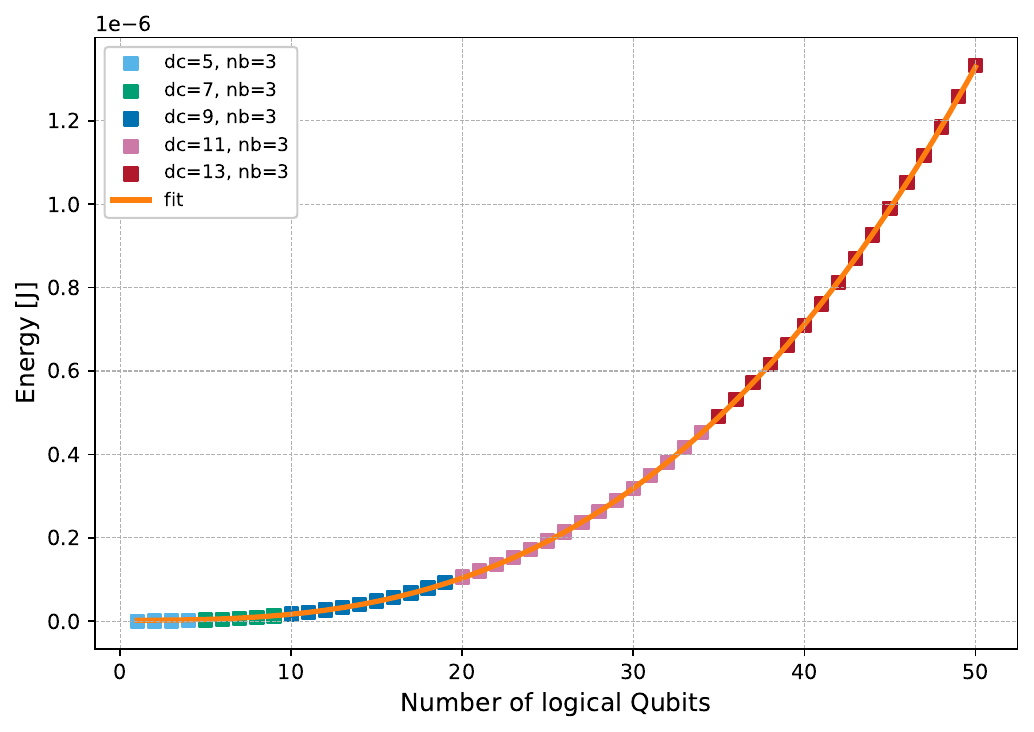}
        \caption*{(c) Microscopic energy ($F=0.990$)}
        \label{fig:energy_optimized_comparison_c}
    \end{minipage}
    \hfill
    \begin{minipage}{0.48\linewidth}
        \centering
        \includegraphics[width=\linewidth]{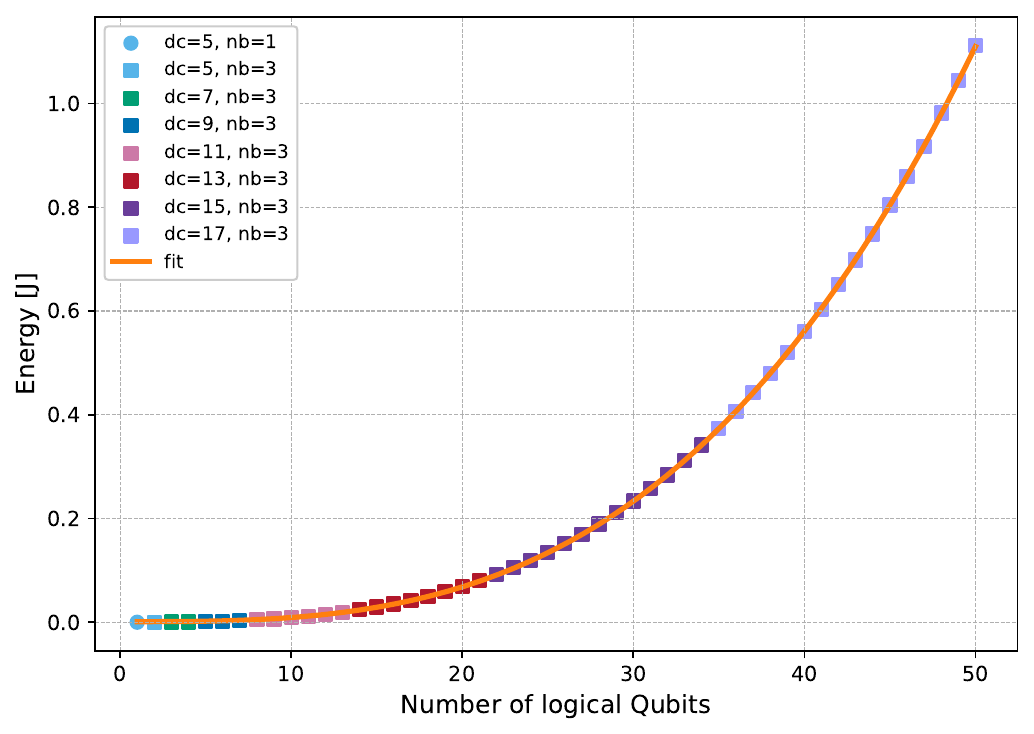}
        \caption*{(d) Macroscopic energy ($F=0.990$)}
        \label{fig:energy_optimized_comparison_d}
    \end{minipage}

    \caption{Optimized energy for microscopic and macroscopic scenarios at fidelities
    \(F=0.900\) (top row) and \(F=0.990\) (bottom row). The different colors correspond to different code distances, while the different markers denote the number of gates between repetitions of the error-correcting code. The energy scaling follows a polynomial dependence on the number of qubits of degree 2.8 (a), 3.1 (b), 2.6 (c), and 3.0 (d).}
    \label{fig:energy_optimized_comparison}
\end{figure*}

A similar optimization strategy can be applied to the code distance $d_c$, leading to the same qualitative conclusion. To systematically identify the optimal configuration, we constructed an optimization procedure that, for each system size, determines the value of the parameter that minimizes the total energy. The only constraint imposed is that the fidelity of the final qubit must exceed a defined threshold. Figure~\ref{fig:optimize_energy} illustrates a representative example of such an optimization, where $\alpha = 3$, $N_b = 1$, and the fidelity of the last qubit is constrained to be greater than $0.900$. The results indicate that, as the number of qubits increases, employing a larger code distance becomes advantageous, enabling the same target fidelity to be maintained while simultaneously lowering the total energy cost.

The obtained data were fitted using the function
\[
y = a\, n^{\,b} + c.
\]
In the microscopic regime, the best-fit parameters are
\[
a = 3.720 \times 10^{-11}, \qquad
b = 2.763, \qquad
c = 3.842 \times 10^{-9},
\]
while in the macroscopic regime, the parameters are
\[
a = 1.249 \times 10^{-5}, \qquad
b = 3.023, \qquad
c = 1.396 \times 10^{-3}.
\]

\begin{figure*}[t]
    \centering

    \begin{minipage}{0.48\linewidth}
        \centering
        \includegraphics[width=\linewidth]{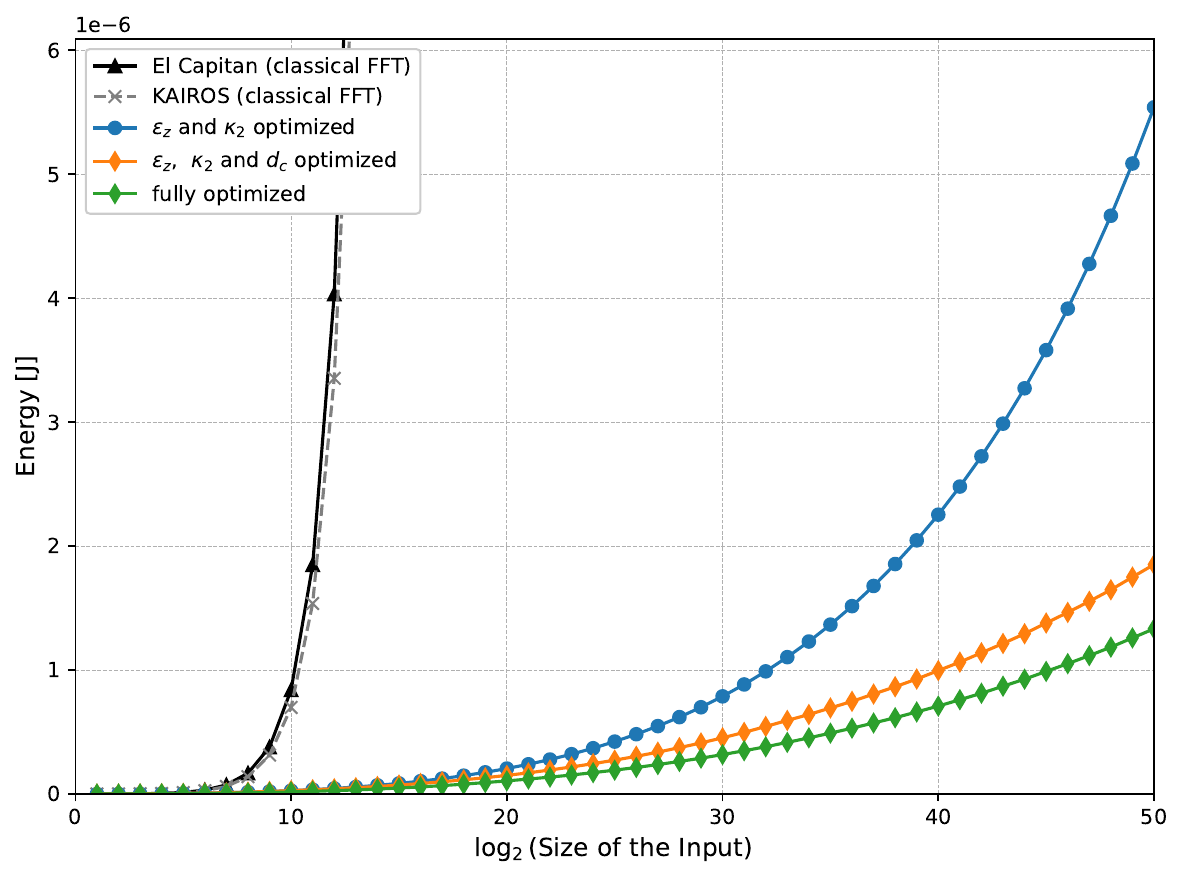}
        \caption*{(a) Microscopic energy}
        \label{fig:micro_classical_quantum}
    \end{minipage}
    \hfill
    \begin{minipage}{0.48\linewidth}
        \centering
        \includegraphics[width=\linewidth]{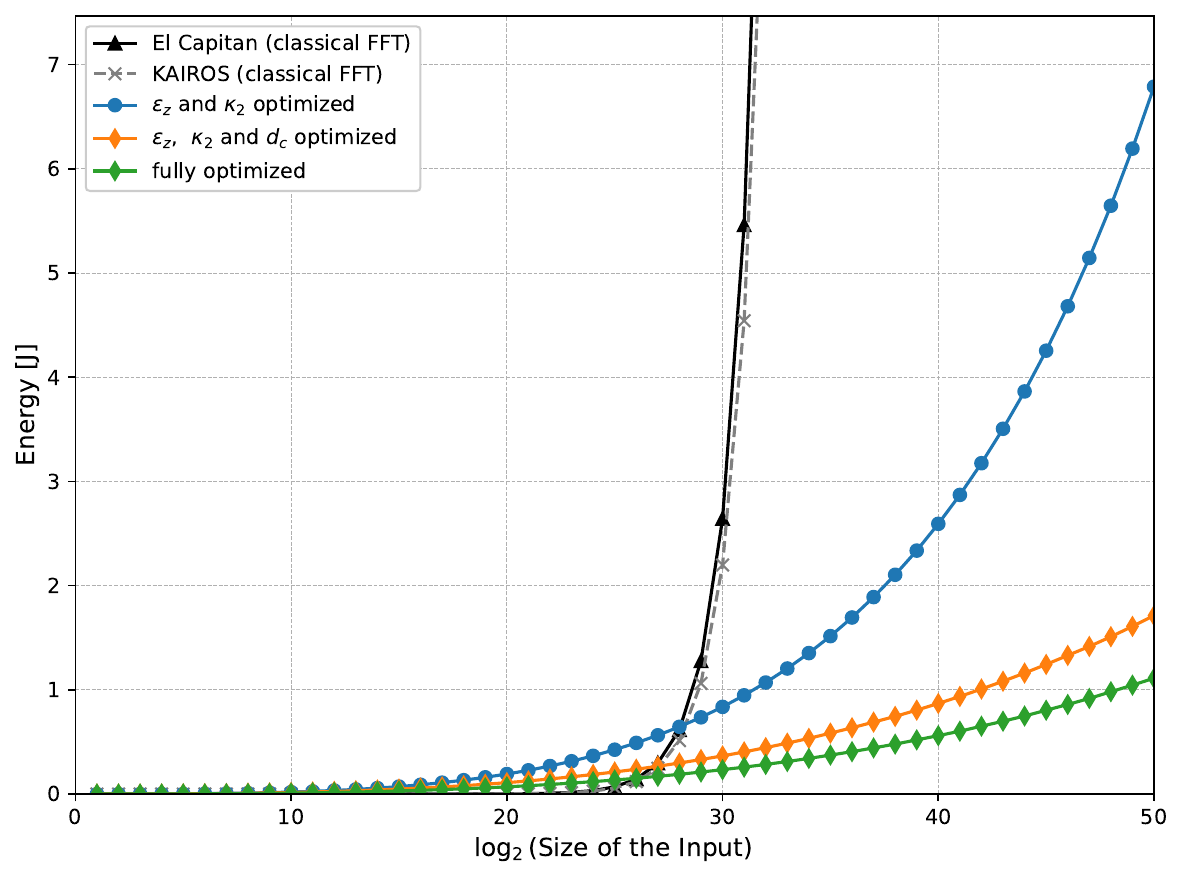}
        \caption*{(b) Macroscopic Energy }
        \label{fig:macro_classical_quantum}
    \end{minipage}

    \caption{Comparison of classical and quantum energy consumption for partially and fully optimized protocols. The blue curve corresponds to optimization of \(\kappa_2\) and \(\epsilon_z\) only, the orange curve also optimizes the code distance \(d_c\), and the green curve represents full optimization of all four parameters: \(\epsilon_z\), \(\kappa_2\), \(d_c\), and \(N_b\)}. A quantum energetic advantage is observed for systems with more than 6 qubits in the microscopic regime, and for more than 26-28 qubits (depending on the level of optimization) in the macroscopic regime. For the quantum curves, the x-axis denotes the number of logical qubits.
    \label{fig:comparison_with_optimization}
\end{figure*}

Furthermore, a fully optimized code was developed to minimize the total energy consumption subject to a fidelity constraint on the final qubit. For a given minimum required fidelity, the algorithm searches for the optimal set of parameters that yields the lowest energy while ensuring that the fidelity remains above the defined threshold. In the present implementation, only the parameter ranges are specified, and the optimization is performed within these bounds.

The Fig.~\ref{fig:energy_optimized_comparison}(a) and ~\ref{fig:energy_optimized_comparison}(b) correspond to the first example, where the minimum required fidelity of the last qubit is set to \(F = 0.900\). The optimization was carried out by varying the parameters within the following ranges:
\(\epsilon_z \in [0.5, 40.5]\)~MHz,
\(\kappa_2 / \kappa_1 \in [100, 50000]\),
\(d_c \in [5, 25]\),
and \(N_b \in [1, 3]\).
Panels (a) and (b) show the optimized microscopic and macroscopic energy, respectively. A fit to the graphs was performed using the function 
\[
y = a n^b + c.
\] 
In the microscopic regime, the parameters are 
\[
a = 2.267 \times 10^{-11}, \quad b = 2.806, \quad c =2.217 \times 10^{-9},
\] 
while in the macroscopic regime, the parameters are 
\[
a = 6.994 \times 10^{-6}, \quad b = 3.061, \quad c = 9.651 \times 10^{-4}.
\]

The Fig.~\ref{fig:energy_optimized_comparison}(c) and Fig.~\ref{fig:energy_optimized_comparison}(d) present a second example obtained using the same optimization procedure, but with a more stringent fidelity requirement of \(F = 0.990\). In this case, the parameter ranges were
\(\epsilon_z \in [0.5, 40.5]\)~MHz,
\(\kappa_2 / \kappa_1 \in [100, 50000]\),
\(d_c \in [5, 25]\),
and \(N_b \in [1, 3]\).
The resulting optimized microscopic and macroscopic energies are shown in panels (c) and (d), respectively. The fit performed yields a scaling of 2.576 and 2.989 at the microscopic and macroscopic levels.

Overall, Fig.~\ref{fig:energy_optimized_comparison} highlights how increasing the required fidelity of the last qubit impacts the optimal energy cost in both microscopic and macroscopic scenarios.

%\begin{figure}[H]

\subsection{Classical Computers}
\label{classical computers}

A comparison with the best classical supercomputers in terms of performance and efficiency was carried out~\cite{top500_nov2025, green500_nov2025}.

The performance of supercomputers is generally reported in FLOPS (Floating Point Operations per Second), which measures the number of arithmetic operations involving real numbers that a processor can perform in one second. Typically, this is determined by running a benchmark program that solves a dense system of linear equations \cite{Dongarra2003}.

However, to compare the energy consumption of classical and quantum computers, it is necessary to express performance in terms of operations per Joule (see Appendix~\ref{Appendix_best_supercomputers}). For the fastest and most energy–efficient classical computers, the energy per floating-point operation is

\begin{equation}
    E_{\text{El Capitan}} = 1.641\times10^{-11} \; [\mathrm{J/FLOP}]
\end{equation}

\begin{equation}
    E_{\text{KAIROS}} = 1.365\times10^{-11} \; [\mathrm{J/FLOP}]
\end{equation}

Considering the radix-2 Cooley-Tukey algorithm \cite{CooleyTukey1969}, a standard method for computing the FFT, the procedure requires approximately \(5N\log_2 N\) FLOPs~\cite{FFTW97,Loan1992}, where \(N\) denotes the length of the input vector. The total energy required to perform the FFT on such a system can be estimated by multiplying the energy consumed per floating-point operation by the total number of FLOPs.

\begin{equation}
    E_{\text{total}} = E_{\text{per FLOP}} \times 5N\log_2 N.
\end{equation}

\subsection{Comparison}

A comparison with classical computers is presented in Fig.~\ref{fig:comparison_with_optimization}. To ensure a fair comparison, it is assumed that both algorithms (QFT and FFT) operate on the same input size. In the quantum case, if \(N\) denotes the size of the input, it can be represented using \(n = \log_2(N)\) qubits. The black and gray curves represent the energy consumption of state-of-the-art classical supercomputers performing the FFT. The colored curves correspond to the energy consumption of quantum hardware implementing the QFT, assuming a minimum last-qubit fidelity of \(F = 0.900\). The blue curve reproduces the result shown in Fig.~\ref{fig:macro_micro_comparison_50_qubits_fixing_nb_and_dc} and corresponds to the QFT energy obtained by fixing the code distance to \(d_c = 5\) and the number of gates between repetitions of the error-correcting code to \(N_b = 1\), while optimizing \(\epsilon_z\) and \(\kappa_2\). The orange curve represents the energy obtained by additionally optimizing the code distance, as presented in Fig.~\ref{fig:optimize_energy}. Finally, the green curve corresponds to the fully optimized scenario, in which all four parameters are chosen to minimize the total energy while maintaining the fidelity of the last qubit above \(F = 0.900\), as shown in Fig.~\ref{fig:energy_optimized_comparison}.

A quantum energetic advantage is observed beyond approximately 6 qubits when considering microscopic energy, and between 26 and 28 qubits in the macroscopic energy model. This advantage arises from the fundamentally different energy scaling behaviors of classical and quantum implementations. In the classical case, the energy consumption scales exponentially with system size, whereas for the quantum implementation, it follows a polynomial scaling.

\begin{figure}[t]
    \centering
    \includegraphics[width=\linewidth]{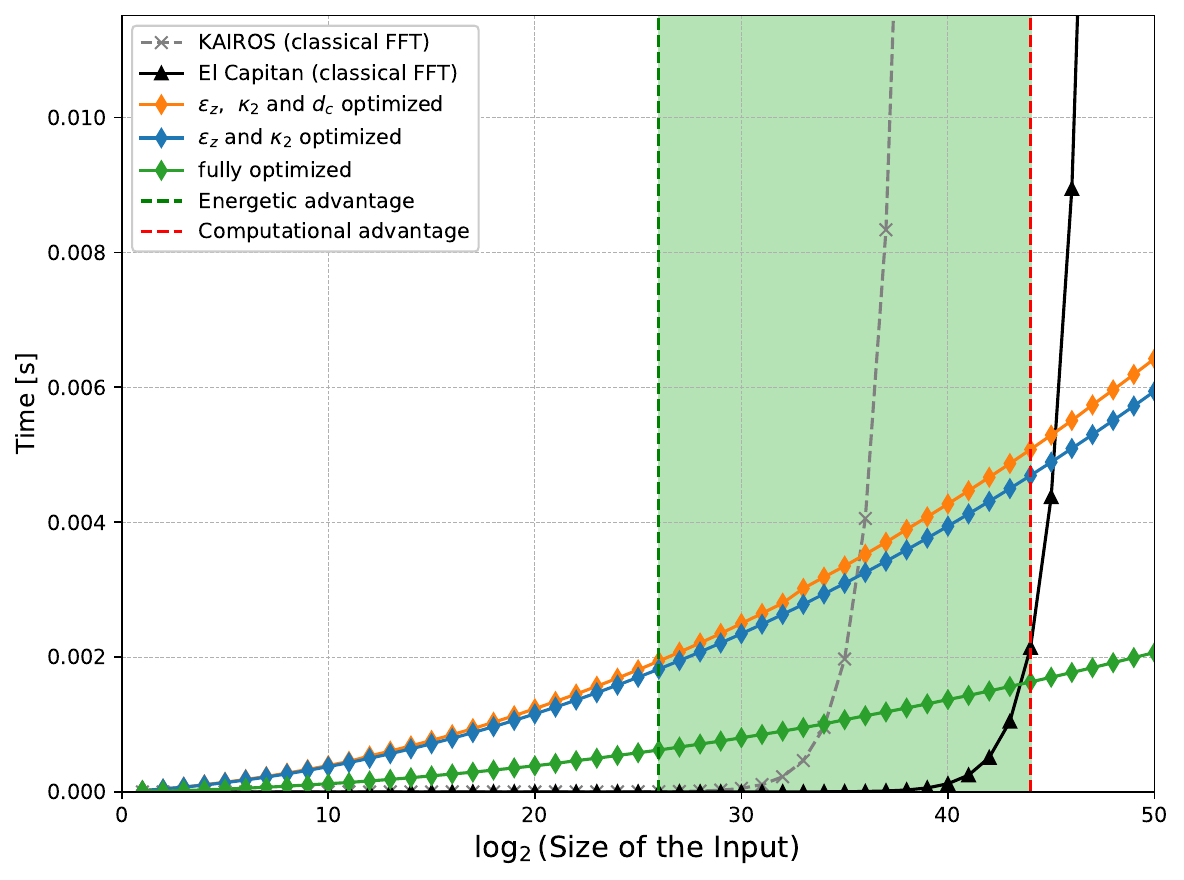}    
    \caption{Comparison of classical and quantum execution times for partially and fully optimized protocols. 
    A quantum computational time advantage is observed for systems with more than 34 qubits with respect to \textit{KAIROS} 
    (indicated by the red vertical line) and 44 qubits with respect to \textit{El Capitan}. 
    The green region represents the regime where there is an energetic advantage but no computational advantage. The execution time of the QFT was calculated based on the parameters obtained from the energy optimization for the macroscopic scenario. For the quantum curves, the x-axis denotes the number of logical qubits.}
    \label{fig:macro_classical_quantum_time}
\end{figure}

Moreover, when comparing the execution time of the FFT and the QFT, a time advantage only appears for systems with more than approximately 34 qubits for \textit{KAIROS} and approximately 44 qubits for \textit{El Capitan}, according to Fig.~\ref{fig:macro_classical_quantum_time}. The parameters to compute the time were considered using the ones resulting from the energy optimization in the macroscopic scenario. Hence, the energetic advantage arises in a regime where running the QFT does not yet provide a temporal computational advantage. This is represented by the green region defined in Fig.~\ref{fig:macro_classical_quantum_time}. The lower limit represents the point at which quantum computers spend less energy than the most efficient classical counterpart (\textit{KAIROS}), while the upper limit corresponds to the point at which quantum computers surpass the execution time of the fastest classical implementation (\textit{El Capitan}).

The energetic advantage observed in this region can be explained using Table~\ref{table:energy_time_gates_flops}. In classical computation, the computational cost of the scaling is generally expressed in terms of the number of arithmetic operations (FLOPs), whereas in quantum computation it is expressed in terms of the number of gate operations. When comparing a single FLOP with a single quantum gate, the ratio of gate time to FLOP time is significantly larger than the corresponding ratio of gate energy to FLOP energy. As a result, an energetic advantage arises in this regime. This advantage arises from the physical characteristics of current quantum hardware implementations, where gate operations typically require relatively long execution times compared to arithmetic operations in classical computers. 

\begin{table}[ht] 
\centering
\textit{KAIROS} \\[0.2cm]

\renewcommand{\arraystretch}{2.0}
\setlength{\tabcolsep}{6pt} % slightly more space for readability
\begin{tabular}{|c|c|c|}
\hline
Gates  & Z Gate & CNOT Gate \\ \hline
$\frac{\text{En. per Quantum Gate}}{\text{En. per FLOP}}$ & $5 \times 10^{3}$ & $3 \times 10^{4}$ \\ \hline
$\frac{\text{Time per Quantum Gate}}{\text{Time per FLOP}}$ & $1 \times 10^{7}$ & $8 \times 10^{7}$ \\ \hline
\end{tabular}

\caption{Ratio of the energy and the execution time of the Z and CNOT gates relative to the energy and time of a single floating-point operation (FLOP) performed on the most efficient supercomputer, \textit{KAIROS}. The ratio of the execution times is significantly larger than the ratio of the energies.}
\label{table:energy_time_gates_flops}
\end{table}

The general explanation for why the energetic advantage appears before the temporal one can be obtained by comparing the classical exponential scaling with the polynomial scaling in both the energy and time cases. An equation for the intersection point can be derived:
\begin{equation}
x=-\frac{(-1+n)\text{ProductLog}\!\left[-\frac{\left(\frac{\gamma \log(2)^{1-n}}{\beta}\right)^{\frac{1}{1-n}}}{-1+n}\right]}{\log(2)},
\end{equation}
where $n$ is the degree of the polynomial scaling and $\gamma/\beta$ is the ratio between the prefactors of the polynomial and exponential scalings. Hence, this equation relates the prefactor and the order of exponential and polynomial scaling algorithms, allowing one to understand the regions where this advantage appears (Appendix \ref{Appendix_comparison_Scaling}).

\subsection{Billed Energy}
The previous analysis was developed based on a complete theoretical study, in which Carnot efficiency was assumed for the cryogenic machines, and the control electronics required to manipulate the qubits were not considered. Only the energy required for the pumps and drives used to stabilize the qubits and perform the quantum gates was taken into account. However, this scenario is not representative of practical quantum computing systems.

To determine the total energy required to run the QFT on real quantum computers, it is necessary to account for the actual power consumption of existing hardware. This can be divided into two main components: the power required for the cryogenic systems and the power required for the control electronics.

For the cryogenic contribution, the equipment required to maintain the cryostat at low temperatures must be considered. This includes the pulse tube refrigerator operating at approximately 4~K, the dilution refrigeration stage reaching temperatures on the order of 10~mK, as well as the cold water loop required for heat extraction and system operation. On the other hand, for the control electronics, the complete set of electronic instrumentation used for qubit manipulation and readout must be considered.

\begin{figure}[]
    \centering
    \includegraphics[width=\linewidth]{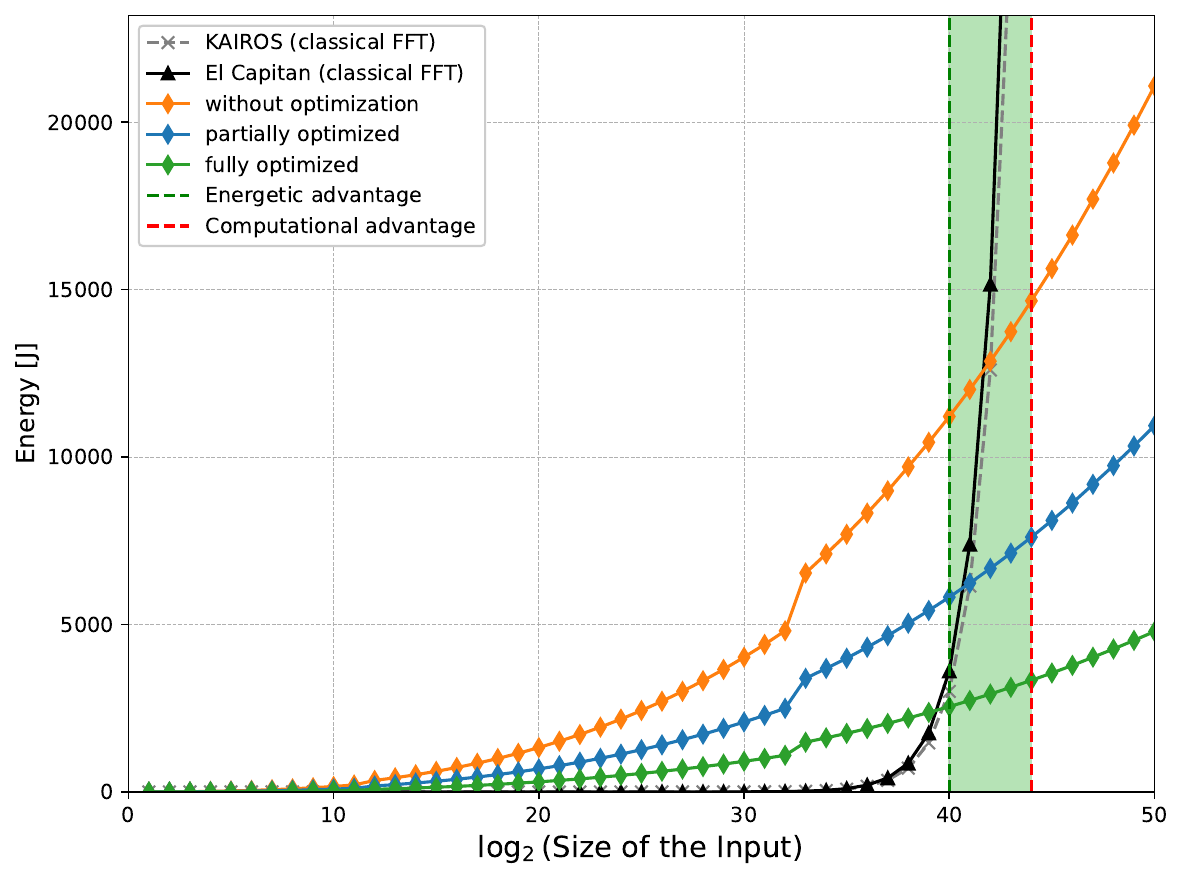}    
    \caption{This figure represents the total energy consumption of real quantum computing systems when realistic cryogenic infrastructure and control electronics are taken into account. The energy advantage is still observed to emerge before the temporal advantage. The orange line corresponds to the actual power of the current architecture, while the blue and red lines correspond to the partially and fully optimized configurations, respectively.}
    \label{fig:macro_billed_energy}
\end{figure}

According to \cite{Report1_confidential}, three total power values per qubit were considered: the current implementation, a partially optimized configuration, and a fully optimized configuration. Figure~\ref{fig:macro_billed_energy} illustrates how the total energy consumption scales, and a comparison with state-of-the-art classical computers was carried out. It is observed that the energy advantage, when realistic cryogenic and control electronic contributions are included, still emerges prior to the computational advantage.

\section{Conclusions}

In this work, we investigated the implementation of a semiclassical Quantum Fourier Transform on a cat-qubit superconducting platform to estimate the energetic cost of quantum computation in such systems. Energy consumption was analyzed at microscopic, macroscopic, and billed levels. The microscopic scale accounts for the energy directly delivered to the quantum device, while the macroscopic scale incorporates the additional energetic requirements imposed by cryogenic refrigeration operating at Carnot efficiency. Finally, the billed energy corresponds to the actual energy consumption of currently available experimental devices.

The analysis of physical qubits without quantum error correction revealed that the dominant contributor to energy consumption is the stabilization energy required to maintain the encoded states and suppress bit-flip errors. This contribution significantly exceeds the energy required to apply the different gates. Moreover, cryogenic systems introduce substantial energetic overhead at every computational stage, leading to a significant difference between microscopic and macroscopic energy consumption, even under the ideal assumption of zero losses. Further research into cryogenic technology is therefore necessary, as current systems remain far from the calculated optimal point.

Quantum Error Correction mechanisms are essential for achieving acceptable final fidelities. In this study, fidelity plays a central role, enabling a fair comparison of QFT implementations across different system sizes. By imposing a threshold fidelity for the final logical qubit, we developed an optimization framework that minimizes total energy consumption while tuning key computational parameters that determine the energy consumption of the circuit. The study concludes that employing higher distance codes and increasing the frequency at which error correction is applied can be an effective strategy for minimizing energy consumption while maintaining a constant target fidelity. Overall, this study establishes a methodology for energy optimization in quantum computation while explicitly incorporating fidelity as a fundamental constraint

A comparison with the classical Fast Fourier Transform suggests a potential energetic advantage for quantum computation. While classical energy scaling is exponential, the quantum counterpart scales polynomially, indicating a potential advantage for systems with more than approximately 6 logical qubits at the microscopic level and 26 logical qubits at the macroscopic level. In particular, this energetic advantage emerges before any computational advantage. Additionally, when real machines used for cryogenic systems and control electronics are considered, the same pattern is observed, with the energy advantage continuing to emerge before the computational advantage. In general, the existence of an energetic advantage can be explained by the computational advantage the QFT offers. However, the number of qubits required to achieve an energetic advantage is relatively small. Remarkably, this advantage arises even before any temporal advantage is observed. These results indicate that future quantum computers may be energetically viable, which would not be the case if the crossing point occurred at a much larger number of qubits.

Future work should focus on determining the number of repetitions of a given circuit, in this case the QFT, required for a given initial state and target precision. Furthermore, a rigorous assessment of the energy required for initial state preparation is necessary. Addressing these challenges will be crucial for clarifying the broader role of quantum technologies in enabling energy-efficient computation.

Future work should focus on establishing a rigorous assessment of the energy requirements for initial state preparation. Further analysis is needed to understand how this scaling behaves on a cat qubit platform, particularly when energy consumption is explicitly taken into account. Additionally, the number of repetitions of a given circuit, in this case the QFT, required to achieve a desired precision for a given initial state should be carefully studied, as it depends on the specific information one aims to extract from the circuit.

Addressing these challenges will be crucial for clarifying the broader role of quantum technologies in enabling energy-efficient computation.

\section{Acknowledgments}
We thank Jeremy Stevens and Nicholas Gialouris for the valuable and insightful discussions. Moreover, we would like to acknowledge the support of FCT — Fundação para a Ciência e a Tecnologia (Portugal), namely through project UIDB/04540/2020 and contract LA/P/0095/2020. The simulations described in this paper were carried out in the OBLIVION Supercomputer of the HPC Chair of the University of Évora, Portugal, and managed by the High-Performance Computing Centre of the same University, under project ID HPCC.250910.A1. We would like to thank the help provided by the OBLIVION Support Team and the HPC Chair.

\bibliography{bibliography}

\begin{thebibliography}{55}%
\makeatletter
\providecommand \@ifxundefined [1]{%
 \@ifx{#1\undefined}
}%
\providecommand \@ifnum [1]{%
 \ifnum #1\expandafter \@firstoftwo
 \else \expandafter \@secondoftwo
 \fi
}%
\providecommand \@ifx [1]{%
 \ifx #1\expandafter \@firstoftwo
 \else \expandafter \@secondoftwo
 \fi
}%
\providecommand \natexlab [1]{#1}%
\providecommand \enquote  [1]{``#1''}%
\providecommand \bibnamefont  [1]{#1}%
\providecommand \bibfnamefont [1]{#1}%
\providecommand \citenamefont [1]{#1}%
\providecommand \href@noop [0]{\@secondoftwo}%
\providecommand \href [0]{\begingroup \@sanitize@url \@href}%
\providecommand \@href[1]{\@@startlink{#1}\@@href}%
\providecommand \@@href[1]{\endgroup#1\@@endlink}%
\providecommand \@sanitize@url [0]{\catcode `\\12\catcode `\$12\catcode `\&12\catcode `\#12\catcode `\^12\catcode `\_12\catcode `\%12\relax}%
\providecommand \@@startlink[1]{}%
\providecommand \@@endlink[0]{}%
\providecommand \url  [0]{\begingroup\@sanitize@url \@url }%
\providecommand \@url [1]{\endgroup\@href {#1}{\urlprefix }}%
\providecommand \urlprefix  [0]{URL }%
\providecommand \Eprint [0]{\href }%
\providecommand \doibase [0]{https://doi.org/}%
\providecommand \selectlanguage [0]{\@gobble}%
\providecommand \bibinfo  [0]{\@secondoftwo}%
\providecommand \bibfield  [0]{\@secondoftwo}%
\providecommand \translation [1]{[#1]}%
\providecommand \BibitemOpen [0]{}%
\providecommand \bibitemStop [0]{}%
\providecommand \bibitemNoStop [0]{.\EOS\space}%
\providecommand \EOS [0]{\spacefactor3000\relax}%
\providecommand \BibitemShut  [1]{\csname bibitem#1\endcsname}%
\let\auto@bib@innerbib\@empty
%</preamble>
\bibitem [{\citenamefont {Masanet}\ \emph {et~al.}(2020)\citenamefont {Masanet}, \citenamefont {Shehabi}, \citenamefont {Lei}, \citenamefont {Smith},\ and\ \citenamefont {Koomey}}]{masanet2020recalibrating}%
  \BibitemOpen
  \bibfield  {author} {\bibinfo {author} {\bibfnamefont {E.}~\bibnamefont {Masanet}}, \bibinfo {author} {\bibfnamefont {A.}~\bibnamefont {Shehabi}}, \bibinfo {author} {\bibfnamefont {N.}~\bibnamefont {Lei}}, \bibinfo {author} {\bibfnamefont {S.}~\bibnamefont {Smith}},\ and\ \bibinfo {author} {\bibfnamefont {J.}~\bibnamefont {Koomey}},\ }\href@noop {} {\bibfield  {journal} {\bibinfo  {journal} {Science}\ }\textbf {\bibinfo {volume} {367}},\ \bibinfo {pages} {984} (\bibinfo {year} {2020})}\BibitemShut {NoStop}%
\bibitem [{\citenamefont {Chen}(2025{\natexlab{a}})}]{Chen2025}%
  \BibitemOpen
  \bibfield  {author} {\bibinfo {author} {\bibfnamefont {S.}~\bibnamefont {Chen}},\ }\href@noop {} {\bibfield  {journal} {\bibinfo  {journal} {Nature}\ } (\bibinfo {year} {2025}{\natexlab{a}})},\ \bibinfo {note} {published March 5, 2025}\BibitemShut {NoStop}%
\bibitem [{\citenamefont {{International Energy Agency}}(2025)}]{IEA2025EnergyAI}%
  \BibitemOpen
  \bibfield  {author} {\bibinfo {author} {\bibnamefont {{International Energy Agency}}},\ }\href {https://www.iea.org/reports/energy-and-ai} {\bibinfo {title} {{Energy and AI}}} (\bibinfo {year} {2025})\BibitemShut {NoStop}%
\bibitem [{\citenamefont {Chen}(2025{\natexlab{b}})}]{Chen2025DataCentersAI}%
  \BibitemOpen
  \bibfield  {author} {\bibinfo {author} {\bibfnamefont {S.}~\bibnamefont {Chen}},\ }\href@noop {} {\bibfield  {journal} {\bibinfo  {journal} {Nature}\ } (\bibinfo {year} {2025}{\natexlab{b}})}\BibitemShut {NoStop}%
\bibitem [{\citenamefont {Preskill}(2018)}]{Preskill_2018}%
  \BibitemOpen
  \bibfield  {author} {\bibinfo {author} {\bibfnamefont {J.}~\bibnamefont {Preskill}},\ }\href {https://doi.org/10.22331/q-2018-08-06-79} {\bibfield  {journal} {\bibinfo  {journal} {Quantum}\ }\textbf {\bibinfo {volume} {2}},\ \bibinfo {pages} {79} (\bibinfo {year} {2018})}\BibitemShut {NoStop}%
\bibitem [{\citenamefont {Breuer}\ and\ \citenamefont {Petruccione}(2007)}]{Breuer_2006}%
  \BibitemOpen
  \bibfield  {author} {\bibinfo {author} {\bibfnamefont {H.-P.}\ \bibnamefont {Breuer}}\ and\ \bibinfo {author} {\bibfnamefont {F.}~\bibnamefont {Petruccione}},\ }\href {https://doi.org/10.1093/acprof:oso/9780199213900.001.0001} {\emph {\bibinfo {title} {The Theory of Open Quantum Systems}}}\ (\bibinfo  {publisher} {Oxford University Press},\ \bibinfo {year} {2007})\BibitemShut {NoStop}%
\bibitem [{\citenamefont {Stevens}\ \emph {et~al.}(2022)\citenamefont {Stevens}, \citenamefont {Szombati}, \citenamefont {Maffei}, \citenamefont {Elouard}, \citenamefont {Assouly}, \citenamefont {Cottet}, \citenamefont {Dassonneville}, \citenamefont {Ficheux}, \citenamefont {Zeppetzauer}, \citenamefont {Bienfait}, \citenamefont {Jordan}, \citenamefont {Auff\`eves},\ and\ \citenamefont {Huard}}]{Stevens_2022}%
  \BibitemOpen
  \bibfield  {author} {\bibinfo {author} {\bibfnamefont {J.}~\bibnamefont {Stevens}}, \bibinfo {author} {\bibfnamefont {D.}~\bibnamefont {Szombati}}, \bibinfo {author} {\bibfnamefont {M.}~\bibnamefont {Maffei}}, \bibinfo {author} {\bibfnamefont {C.}~\bibnamefont {Elouard}}, \bibinfo {author} {\bibfnamefont {R.}~\bibnamefont {Assouly}}, \bibinfo {author} {\bibfnamefont {N.}~\bibnamefont {Cottet}}, \bibinfo {author} {\bibfnamefont {R.}~\bibnamefont {Dassonneville}}, \bibinfo {author} {\bibfnamefont {Q.}~\bibnamefont {Ficheux}}, \bibinfo {author} {\bibfnamefont {S.}~\bibnamefont {Zeppetzauer}}, \bibinfo {author} {\bibfnamefont {A.}~\bibnamefont {Bienfait}}, \bibinfo {author} {\bibfnamefont {A.}~\bibnamefont {Jordan}}, \bibinfo {author} {\bibfnamefont {A.}~\bibnamefont {Auff\`eves}},\ and\ \bibinfo {author} {\bibfnamefont {B.}~\bibnamefont {Huard}},\ }\bibfield  {journal} {\bibinfo  {journal} {Physical Review Letters}\ }\textbf {\bibinfo {volume} {129}},\ \href {https://doi.org/10.1103/physrevlett.129.110601} {10.1103/physrevlett.129.110601} (\bibinfo {year} {2022})\BibitemShut {NoStop}%
\bibitem [{\citenamefont {Ikonen}\ \emph {et~al.}(2017)\citenamefont {Ikonen}, \citenamefont {Salmilehto},\ and\ \citenamefont {Möttönen}}]{Ikonen_2017}%
  \BibitemOpen
  \bibfield  {author} {\bibinfo {author} {\bibfnamefont {J.}~\bibnamefont {Ikonen}}, \bibinfo {author} {\bibfnamefont {J.}~\bibnamefont {Salmilehto}},\ and\ \bibinfo {author} {\bibfnamefont {M.}~\bibnamefont {Möttönen}},\ }\bibfield  {journal} {\bibinfo  {journal} {npj Quantum Information}\ }\textbf {\bibinfo {volume} {3}},\ \href {https://doi.org/10.1038/s41534-017-0015-5} {10.1038/s41534-017-0015-5} (\bibinfo {year} {2017})\BibitemShut {NoStop}%
\bibitem [{\citenamefont {Stevens}\ and\ \citenamefont {Deffner}(2025)}]{Stevens_2025}%
  \BibitemOpen
  \bibfield  {author} {\bibinfo {author} {\bibfnamefont {J.}~\bibnamefont {Stevens}}\ and\ \bibinfo {author} {\bibfnamefont {S.}~\bibnamefont {Deffner}},\ }\href {https://doi.org/10.1088/2058-9565/ae0daf} {\bibfield  {journal} {\bibinfo  {journal} {Quantum Science and Technology}\ }\textbf {\bibinfo {volume} {10}},\ \bibinfo {pages} {04LT03} (\bibinfo {year} {2025})}\BibitemShut {NoStop}%
\bibitem [{\citenamefont {Carrasco-Codina}\ \emph {et~al.}(2026)\citenamefont {Carrasco-Codina}, \citenamefont {Escofet}, \citenamefont {Hilaire}, \citenamefont {Soret}, \citenamefont {Nerenberg}, \citenamefont {Champain}, \citenamefont {Milburn}, \citenamefont {Theophilo}, \citenamefont {Li}, \citenamefont {Bautista}, \citenamefont {Gómez}, \citenamefont {Miralles}, \citenamefont {Abadal}, \citenamefont {Almudéver}, \citenamefont {Alarcón},\ and\ \citenamefont {Yehia}}]{Codina2026}%
  \BibitemOpen
  \bibfield  {author} {\bibinfo {author} {\bibfnamefont {M.}~\bibnamefont {Carrasco-Codina}}, \bibinfo {author} {\bibfnamefont {P.}~\bibnamefont {Escofet}}, \bibinfo {author} {\bibfnamefont {P.}~\bibnamefont {Hilaire}}, \bibinfo {author} {\bibfnamefont {A.}~\bibnamefont {Soret}}, \bibinfo {author} {\bibfnamefont {S.}~\bibnamefont {Nerenberg}}, \bibinfo {author} {\bibfnamefont {V.}~\bibnamefont {Champain}}, \bibinfo {author} {\bibfnamefont {G.}~\bibnamefont {Milburn}}, \bibinfo {author} {\bibfnamefont {K.}~\bibnamefont {Theophilo}}, \bibinfo {author} {\bibfnamefont {S.~H.}\ \bibnamefont {Li}}, \bibinfo {author} {\bibfnamefont {I.}~\bibnamefont {Bautista}}, \bibinfo {author} {\bibfnamefont {A.}~\bibnamefont {Gómez}}, \bibinfo {author} {\bibfnamefont {J.}~\bibnamefont {Miralles}}, \bibinfo {author} {\bibfnamefont {S.}~\bibnamefont {Abadal}}, \bibinfo {author} {\bibfnamefont {C.~G.}\ \bibnamefont {Almudéver}}, \bibinfo {author} {\bibfnamefont {E.}~\bibnamefont {Alarcón}},\ and\ \bibinfo {author} {\bibfnamefont {R.}~\bibnamefont {Yehia}},\ }\href {https://arxiv.org/abs/2605.15090} {\bibinfo {title} {Energy efficiency of quantum computers}} (\bibinfo {year} {2026}),\ \Eprint {https://arxiv.org/abs/2605.15090} {arXiv:2605.15090 [quant-ph]} \BibitemShut {NoStop}%
\bibitem [{\citenamefont {Jaschke}\ and\ \citenamefont {Montangero}(2023)}]{green}%
  \BibitemOpen
  \bibfield  {author} {\bibinfo {author} {\bibfnamefont {D.}~\bibnamefont {Jaschke}}\ and\ \bibinfo {author} {\bibfnamefont {S.}~\bibnamefont {Montangero}},\ }\href {https://doi.org/10.1088/2058-9565/acae3e} {\bibfield  {journal} {\bibinfo  {journal} {Quantum Science and Technology}\ }\textbf {\bibinfo {volume} {8}},\ \bibinfo {pages} {025001} (\bibinfo {year} {2023})}\BibitemShut {NoStop}%
\bibitem [{\citenamefont {Auff\`eves}(2022)}]{energy_initiative}%
  \BibitemOpen
  \bibfield  {author} {\bibinfo {author} {\bibfnamefont {A.}~\bibnamefont {Auff\`eves}},\ }\bibfield  {journal} {\bibinfo  {journal} {PRX Quantum}\ }\textbf {\bibinfo {volume} {3}},\ \href {https://doi.org/10.1103/PRXQuantum.3.020101} {10.1103/PRXQuantum.3.020101} (\bibinfo {year} {2022})\BibitemShut {NoStop}%
\bibitem [{\citenamefont {Silva~Pratapsi}\ \emph {et~al.}(2023)\citenamefont {Silva~Pratapsi}, \citenamefont {Huber}, \citenamefont {Barthel}, \citenamefont {Bose}, \citenamefont {Wunderlich},\ and\ \citenamefont {Omar}}]{Silva_Pratapsi_2023}%
  \BibitemOpen
  \bibfield  {author} {\bibinfo {author} {\bibfnamefont {S.}~\bibnamefont {Silva~Pratapsi}}, \bibinfo {author} {\bibfnamefont {P.~H.}\ \bibnamefont {Huber}}, \bibinfo {author} {\bibfnamefont {P.}~\bibnamefont {Barthel}}, \bibinfo {author} {\bibfnamefont {S.}~\bibnamefont {Bose}}, \bibinfo {author} {\bibfnamefont {C.}~\bibnamefont {Wunderlich}},\ and\ \bibinfo {author} {\bibfnamefont {Y.}~\bibnamefont {Omar}},\ }\bibfield  {journal} {\bibinfo  {journal} {Applied Physics Letters}\ }\textbf {\bibinfo {volume} {123}},\ \href {https://doi.org/10.1063/5.0176719} {10.1063/5.0176719} (\bibinfo {year} {2023})\BibitemShut {NoStop}%
\bibitem [{\citenamefont {Moutinho}\ \emph {et~al.}(2023)\citenamefont {Moutinho}, \citenamefont {Pezzutto}, \citenamefont {Pratapsi}, \citenamefont {da~Silva}, \citenamefont {De~Franceschi}, \citenamefont {Bose}, \citenamefont {Costa},\ and\ \citenamefont {Omar}}]{Moutinho2023}%
  \BibitemOpen
  \bibfield  {author} {\bibinfo {author} {\bibfnamefont {J.~P.}\ \bibnamefont {Moutinho}}, \bibinfo {author} {\bibfnamefont {M.}~\bibnamefont {Pezzutto}}, \bibinfo {author} {\bibfnamefont {S.~S.}\ \bibnamefont {Pratapsi}}, \bibinfo {author} {\bibfnamefont {F.~F.}\ \bibnamefont {da~Silva}}, \bibinfo {author} {\bibfnamefont {S.}~\bibnamefont {De~Franceschi}}, \bibinfo {author} {\bibfnamefont {S.}~\bibnamefont {Bose}}, \bibinfo {author} {\bibfnamefont {A.~T.}\ \bibnamefont {Costa}},\ and\ \bibinfo {author} {\bibfnamefont {Y.}~\bibnamefont {Omar}},\ }\bibfield  {journal} {\bibinfo  {journal} {PRX Energy}\ }\textbf {\bibinfo {volume} {2}},\ \href {https://doi.org/10.1103/prxenergy.2.033002} {10.1103/prxenergy.2.033002} (\bibinfo {year} {2023})\BibitemShut {NoStop}%
\bibitem [{\citenamefont {Meier}\ and\ \citenamefont {Yamasaki}(2025)}]{Meier_2025}%
  \BibitemOpen
  \bibfield  {author} {\bibinfo {author} {\bibfnamefont {F.}~\bibnamefont {Meier}}\ and\ \bibinfo {author} {\bibfnamefont {H.}~\bibnamefont {Yamasaki}},\ }\bibfield  {journal} {\bibinfo  {journal} {PRX Energy}\ }\textbf {\bibinfo {volume} {4}},\ \href {https://doi.org/10.1103/prxenergy.4.023008} {10.1103/prxenergy.4.023008} (\bibinfo {year} {2025})\BibitemShut {NoStop}%
\bibitem [{\citenamefont {Soret}\ \emph {et~al.}(2026)\citenamefont {Soret}, \citenamefont {Dridi}, \citenamefont {Wein}, \citenamefont {Giesz}, \citenamefont {Mansfield},\ and\ \citenamefont {Emeriau}}]{soret2026}%
  \BibitemOpen
  \bibfield  {author} {\bibinfo {author} {\bibfnamefont {A.}~\bibnamefont {Soret}}, \bibinfo {author} {\bibfnamefont {N.}~\bibnamefont {Dridi}}, \bibinfo {author} {\bibfnamefont {S.~C.}\ \bibnamefont {Wein}}, \bibinfo {author} {\bibfnamefont {V.}~\bibnamefont {Giesz}}, \bibinfo {author} {\bibfnamefont {S.}~\bibnamefont {Mansfield}},\ and\ \bibinfo {author} {\bibfnamefont {P.-E.}\ \bibnamefont {Emeriau}},\ }\href {https://arxiv.org/abs/2601.08068} {\bibinfo {title} {Quantum energetic advantage before computational advantage in boson sampling}} (\bibinfo {year} {2026}),\ \Eprint {https://arxiv.org/abs/2601.08068} {arXiv:2601.08068 [quant-ph]} \BibitemShut {NoStop}%
\bibitem [{\citenamefont {Alliance}(2026)}]{EERA_2026}%
  \BibitemOpen
  \bibfield  {author} {\bibinfo {author} {\bibfnamefont {E.~E.~R.}\ \bibnamefont {Alliance}},\ }\href {https://doi.org/10.5281/zenodo.20120314} {\bibinfo {title} {Quantum computing in the net-zero transition: energy production, management, and efficiency}} (\bibinfo {year} {2026})\BibitemShut {NoStop}%
\bibitem [{\citenamefont {Stevens}\ and\ \citenamefont {Deffner}(2026)}]{stevens_2026}%
  \BibitemOpen
  \bibfield  {author} {\bibinfo {author} {\bibfnamefont {J.}~\bibnamefont {Stevens}}\ and\ \bibinfo {author} {\bibfnamefont {S.}~\bibnamefont {Deffner}},\ }\href {https://arxiv.org/abs/2605.04329} {\bibinfo {title} {Energy-error tradeoff in encoding quantum error correction}} (\bibinfo {year} {2026}),\ \Eprint {https://arxiv.org/abs/2605.04329} {arXiv:2605.04329 [quant-ph]} \BibitemShut {NoStop}%
\bibitem [{\citenamefont {G\'ois}\ \emph {et~al.}(2024)\citenamefont {G\'ois}, \citenamefont {Pezzutto},\ and\ \citenamefont {Omar}}]{Gois2024}%
  \BibitemOpen
  \bibfield  {author} {\bibinfo {author} {\bibfnamefont {F.}~\bibnamefont {G\'ois}}, \bibinfo {author} {\bibfnamefont {M.}~\bibnamefont {Pezzutto}},\ and\ \bibinfo {author} {\bibfnamefont {Y.}~\bibnamefont {Omar}},\ }\href {https://arxiv.org/abs/2404.11572} {\bibinfo {title} {Towards energetic quantum advantage in trapped-ion quantum computation}} (\bibinfo {year} {2024}),\ \Eprint {https://arxiv.org/abs/2404.11572} {arXiv:2404.11572 [quant-ph]} \BibitemShut {NoStop}%
\bibitem [{\citenamefont {Alves}\ \emph {et~al.}(2026)\citenamefont {Alves}, \citenamefont {Pezzutto},\ and\ \citenamefont {Omar}}]{Oscar2024}%
  \BibitemOpen
  \bibfield  {author} {\bibinfo {author} {\bibfnamefont {O.}~\bibnamefont {Alves}}, \bibinfo {author} {\bibfnamefont {M.}~\bibnamefont {Pezzutto}},\ and\ \bibinfo {author} {\bibfnamefont {Y.}~\bibnamefont {Omar}},\ }\href {https://arxiv.org/abs/2601.03141} {\bibinfo {title} {{Energetics of Rydberg-atom Quantum Computing}}} (\bibinfo {year} {2026}),\ \Eprint {https://arxiv.org/abs/2601.03141} {arXiv:2601.03141 [quant-ph]} \BibitemShut {NoStop}%
\bibitem [{\citenamefont {Santos}\ \emph {et~al.}(2026)\citenamefont {Santos}, \citenamefont {Nath}, \citenamefont {Pezzutto}, \citenamefont {Meunier},\ and\ \citenamefont {Omar}}]{Santos2026}%
  \BibitemOpen
  \bibfield  {author} {\bibinfo {author} {\bibfnamefont {J.}~\bibnamefont {Santos}}, \bibinfo {author} {\bibfnamefont {J.}~\bibnamefont {Nath}}, \bibinfo {author} {\bibfnamefont {M.}~\bibnamefont {Pezzutto}}, \bibinfo {author} {\bibfnamefont {T.}~\bibnamefont {Meunier}},\ and\ \bibinfo {author} {\bibfnamefont {Y.}~\bibnamefont {Omar}},\ }\href@noop {} {\bibinfo {title} {Unveiling energetic advantage in spin qubits quantum computation (in preparation)}} (\bibinfo {year} {2026})\BibitemShut {NoStop}%
\bibitem [{\citenamefont {Kjaergaard}\ \emph {et~al.}(2020)\citenamefont {Kjaergaard}, \citenamefont {Schwartz}, \citenamefont {Braum\"uller}, \citenamefont {Krantz}, \citenamefont {Wang}, \citenamefont {Gustavsson},\ and\ \citenamefont {Oliver}}]{Kjaergaard2020}%
  \BibitemOpen
  \bibfield  {author} {\bibinfo {author} {\bibfnamefont {M.}~\bibnamefont {Kjaergaard}}, \bibinfo {author} {\bibfnamefont {M.~E.}\ \bibnamefont {Schwartz}}, \bibinfo {author} {\bibfnamefont {J.}~\bibnamefont {Braum\"uller}}, \bibinfo {author} {\bibfnamefont {P.}~\bibnamefont {Krantz}}, \bibinfo {author} {\bibfnamefont {J.~I.-J.}\ \bibnamefont {Wang}}, \bibinfo {author} {\bibfnamefont {S.}~\bibnamefont {Gustavsson}},\ and\ \bibinfo {author} {\bibfnamefont {W.~D.}\ \bibnamefont {Oliver}},\ }\href {https://doi.org/10.1146/annurev-conmatphys-031119-050605} {\bibfield  {journal} {\bibinfo  {journal} {Annual Review of Condensed Matter Physics}\ }\textbf {\bibinfo {volume} {11}},\ \bibinfo {pages} {369} (\bibinfo {year} {2020})}\BibitemShut {NoStop}%
\bibitem [{\citenamefont {Ezratty}(2023)}]{Ezratty2023}%
  \BibitemOpen
  \bibfield  {author} {\bibinfo {author} {\bibfnamefont {O.}~\bibnamefont {Ezratty}},\ }\href {https://doi.org/10.1140/epja/s10050-023-01006-7} {\bibfield  {journal} {\bibinfo  {journal} {The European Physical Journal A}\ }\textbf {\bibinfo {volume} {59}},\ \bibinfo {pages} {94} (\bibinfo {year} {2023})}\BibitemShut {NoStop}%
\bibitem [{\citenamefont {Frank~Arute}\ \emph {et~al.}(2019)\citenamefont {Frank~Arute} \emph {et~al.}}]{Arute2019}%
  \BibitemOpen
  \bibfield  {author} {\bibinfo {author} {\bibfnamefont {R.~B.}\ \bibnamefont {Frank~Arute}, \bibfnamefont {Kunal~Arya}} \emph {et~al.},\ }\href {https://www.nature.com/articles/s41586-019-1666-5} {\bibfield  {journal} {\bibinfo  {journal} {Nature}\ }\textbf {\bibinfo {volume} {574}},\ \bibinfo {pages} {505–510} (\bibinfo {year} {2019})}\BibitemShut {NoStop}%
\bibitem [{\citenamefont {Castelvecchi}(2023)}]{Castelvecchi2023}%
  \BibitemOpen
  \bibfield  {author} {\bibinfo {author} {\bibfnamefont {D.}~\bibnamefont {Castelvecchi}},\ }\href {https://doi.org/10.1038/d41586-023-03854-1} {\bibfield  {journal} {\bibinfo  {journal} {Nature}\ }\textbf {\bibinfo {volume} {624}},\ \bibinfo {pages} {238} (\bibinfo {year} {2023})}\BibitemShut {NoStop}%
\bibitem [{\citenamefont {Berdou}\ \emph {et~al.}(2023)\citenamefont {Berdou}, \citenamefont {Murani}, \citenamefont {R\'eglade}, \citenamefont {Smith}, \citenamefont {Villiers}, \citenamefont {Palomo}, \citenamefont {Rosticher}, \citenamefont {Denis}, \citenamefont {Morfin}, \citenamefont {Delbecq}, \citenamefont {Kontos}, \citenamefont {Pankratova}, \citenamefont {Rautschke}, \citenamefont {Peronnin}, \citenamefont {Sellem}, \citenamefont {Rouchon}, \citenamefont {Sarlette}, \citenamefont {Mirrahimi}, \citenamefont {Campagne-Ibarcq}, \citenamefont {Jezouin}, \citenamefont {Lescanne},\ and\ \citenamefont {Leghtas}}]{Berdou_2023}%
  \BibitemOpen
  \bibfield  {author} {\bibinfo {author} {\bibfnamefont {C.}~\bibnamefont {Berdou}}, \bibinfo {author} {\bibfnamefont {A.}~\bibnamefont {Murani}}, \bibinfo {author} {\bibfnamefont {U.}~\bibnamefont {R\'eglade}}, \bibinfo {author} {\bibfnamefont {W.}~\bibnamefont {Smith}}, \bibinfo {author} {\bibfnamefont {M.}~\bibnamefont {Villiers}}, \bibinfo {author} {\bibfnamefont {J.}~\bibnamefont {Palomo}}, \bibinfo {author} {\bibfnamefont {M.}~\bibnamefont {Rosticher}}, \bibinfo {author} {\bibfnamefont {A.}~\bibnamefont {Denis}}, \bibinfo {author} {\bibfnamefont {P.}~\bibnamefont {Morfin}}, \bibinfo {author} {\bibfnamefont {M.}~\bibnamefont {Delbecq}}, \bibinfo {author} {\bibfnamefont {T.}~\bibnamefont {Kontos}}, \bibinfo {author} {\bibfnamefont {N.}~\bibnamefont {Pankratova}}, \bibinfo {author} {\bibfnamefont {F.}~\bibnamefont {Rautschke}}, \bibinfo {author} {\bibfnamefont {T.}~\bibnamefont {Peronnin}}, \bibinfo {author} {\bibfnamefont {L.-A.}\ \bibnamefont {Sellem}}, \bibinfo {author} {\bibfnamefont {P.}~\bibnamefont {Rouchon}}, \bibinfo {author} {\bibfnamefont {A.}~\bibnamefont {Sarlette}}, \bibinfo {author} {\bibfnamefont {M.}~\bibnamefont {Mirrahimi}}, \bibinfo {author} {\bibfnamefont {P.}~\bibnamefont {Campagne-Ibarcq}}, \bibinfo {author} {\bibfnamefont {S.}~\bibnamefont {Jezouin}}, \bibinfo {author} {\bibfnamefont {R.}~\bibnamefont {Lescanne}},\ and\ \bibinfo {author} {\bibfnamefont {Z.}~\bibnamefont {Leghtas}},\ }\bibfield  {journal} {\bibinfo  {journal} {PRX Quantum}\ }\textbf {\bibinfo {volume} {4}},\ \href {https://doi.org/10.1103/prxquantum.4.020350} {10.1103/prxquantum.4.020350} (\bibinfo {year} {2023})\BibitemShut {NoStop}%
\bibitem [{\citenamefont {Acharya}\ \emph {et~al.}(2025)\citenamefont {Acharya}, \citenamefont {Abanin}, \citenamefont {Aghababaie-Beni}, \citenamefont {Aleiner}, \citenamefont {Andersen}, \citenamefont {Ansmann}, \citenamefont {Arute}, \citenamefont {Arya}, \citenamefont {Asfaw}, \citenamefont {Astrakhantsev}, \citenamefont {Atalaya}, \citenamefont {Babbush}, \citenamefont {Bacon}, \citenamefont {Ballard}, \citenamefont {Bardin},\ and\ \citenamefont {et~al.}}]{Acharya2025}%
  \BibitemOpen
  \bibfield  {author} {\bibinfo {author} {\bibfnamefont {R.}~\bibnamefont {Acharya}}, \bibinfo {author} {\bibfnamefont {D.~A.}\ \bibnamefont {Abanin}}, \bibinfo {author} {\bibfnamefont {L.}~\bibnamefont {Aghababaie-Beni}}, \bibinfo {author} {\bibfnamefont {I.}~\bibnamefont {Aleiner}}, \bibinfo {author} {\bibfnamefont {T.~I.}\ \bibnamefont {Andersen}}, \bibinfo {author} {\bibfnamefont {M.}~\bibnamefont {Ansmann}}, \bibinfo {author} {\bibfnamefont {F.}~\bibnamefont {Arute}}, \bibinfo {author} {\bibfnamefont {K.}~\bibnamefont {Arya}}, \bibinfo {author} {\bibfnamefont {A.}~\bibnamefont {Asfaw}}, \bibinfo {author} {\bibfnamefont {N.}~\bibnamefont {Astrakhantsev}}, \bibinfo {author} {\bibfnamefont {J.}~\bibnamefont {Atalaya}}, \bibinfo {author} {\bibfnamefont {R.}~\bibnamefont {Babbush}}, \bibinfo {author} {\bibfnamefont {D.}~\bibnamefont {Bacon}}, \bibinfo {author} {\bibfnamefont {B.}~\bibnamefont {Ballard}}, \bibinfo {author} {\bibfnamefont {J.~C.}\ \bibnamefont {Bardin}},\ and\ \bibinfo {author} {\bibnamefont {et~al.}},\ }\href {https://doi.org/10.1038/s41586-024-08449-y} {\bibfield  {journal} {\bibinfo  {journal} {Nature}\ }\textbf {\bibinfo {volume} {638}},\ \bibinfo {pages} {920} (\bibinfo {year} {2025})}\BibitemShut {NoStop}%
\bibitem [{\citenamefont {Neven}(2024)}]{GoogleWillow2024}%
  \BibitemOpen
  \bibfield  {author} {\bibinfo {author} {\bibfnamefont {H.}~\bibnamefont {Neven}},\ }\href {https://blog.google/innovation-and-ai/technology/research/google-willow-quantum-chip/} {\bibinfo {title} {Meet willow, our state-of-the-art quantum chip}},\ \bibinfo {howpublished} {Blog post, Google Innovation \& AI} (\bibinfo {year} {2024}),\ \bibinfo {note} {accessed: 2026-03-07}\BibitemShut {NoStop}%
\bibitem [{\citenamefont {Cochrane}\ \emph {et~al.}(1999)\citenamefont {Cochrane}, \citenamefont {Milburn},\ and\ \citenamefont {Munro}}]{Cochrane_1999}%
  \BibitemOpen
  \bibfield  {author} {\bibinfo {author} {\bibfnamefont {P.~T.}\ \bibnamefont {Cochrane}}, \bibinfo {author} {\bibfnamefont {G.~J.}\ \bibnamefont {Milburn}},\ and\ \bibinfo {author} {\bibfnamefont {W.~J.}\ \bibnamefont {Munro}},\ }\href {https://doi.org/10.1103/physreva.59.2631} {\bibfield  {journal} {\bibinfo  {journal} {Physical Review A}\ }\textbf {\bibinfo {volume} {59}},\ \bibinfo {pages} {2631–2634} (\bibinfo {year} {1999})}\BibitemShut {NoStop}%
\bibitem [{\citenamefont {Mirrahimi}\ \emph {et~al.}(2014)\citenamefont {Mirrahimi}, \citenamefont {Leghtas}, \citenamefont {Albert}, \citenamefont {Touzard}, \citenamefont {Schoelkopf}, \citenamefont {Jiang},\ and\ \citenamefont {Devoret}}]{Mirrahimi_2014}%
  \BibitemOpen
  \bibfield  {author} {\bibinfo {author} {\bibfnamefont {M.}~\bibnamefont {Mirrahimi}}, \bibinfo {author} {\bibfnamefont {Z.}~\bibnamefont {Leghtas}}, \bibinfo {author} {\bibfnamefont {V.~V.}\ \bibnamefont {Albert}}, \bibinfo {author} {\bibfnamefont {S.}~\bibnamefont {Touzard}}, \bibinfo {author} {\bibfnamefont {R.~J.}\ \bibnamefont {Schoelkopf}}, \bibinfo {author} {\bibfnamefont {L.}~\bibnamefont {Jiang}},\ and\ \bibinfo {author} {\bibfnamefont {M.~H.}\ \bibnamefont {Devoret}},\ }\href {https://doi.org/10.1088/1367-2630/16/4/045014} {\bibfield  {journal} {\bibinfo  {journal} {New Journal of Physics}\ }\textbf {\bibinfo {volume} {16}},\ \bibinfo {pages} {045014} (\bibinfo {year} {2014})}\BibitemShut {NoStop}%
\bibitem [{\citenamefont {Guillaud}\ \emph {et~al.}(2023)\citenamefont {Guillaud}, \citenamefont {Cohen},\ and\ \citenamefont {Mirrahimi}}]{Guillaud_2023}%
  \BibitemOpen
  \bibfield  {author} {\bibinfo {author} {\bibfnamefont {J.}~\bibnamefont {Guillaud}}, \bibinfo {author} {\bibfnamefont {J.}~\bibnamefont {Cohen}},\ and\ \bibinfo {author} {\bibfnamefont {M.}~\bibnamefont {Mirrahimi}},\ }\bibfield  {journal} {\bibinfo  {journal} {SciPost Physics Lecture Notes}\ }\href {https://doi.org/10.21468/scipostphyslectnotes.72} {10.21468/scipostphyslectnotes.72} (\bibinfo {year} {2023})\BibitemShut {NoStop}%
\bibitem [{\citenamefont {Lidar}(2019)}]{Lidar2019}%
  \BibitemOpen
  \bibfield  {author} {\bibinfo {author} {\bibfnamefont {D.~A.}\ \bibnamefont {Lidar}},\ }\bibfield  {journal} {\bibinfo  {journal} {arXiv preprint arXiv:1902.00967}\ }\href {https://doi.org/10.48550/arXiv.1902.00967} {10.48550/arXiv.1902.00967} (\bibinfo {year} {2019})\BibitemShut {NoStop}%
\bibitem [{\citenamefont {Lescanne}\ \emph {et~al.}(2020)\citenamefont {Lescanne}, \citenamefont {Villiers}, \citenamefont {Peronnin}, \citenamefont {Sarlette}, \citenamefont {Delbecq}, \citenamefont {Huard}, \citenamefont {Kontos}, \citenamefont {Mirrahimi},\ and\ \citenamefont {Leghtas}}]{Lescanne_2020}%
  \BibitemOpen
  \bibfield  {author} {\bibinfo {author} {\bibfnamefont {R.}~\bibnamefont {Lescanne}}, \bibinfo {author} {\bibfnamefont {M.}~\bibnamefont {Villiers}}, \bibinfo {author} {\bibfnamefont {T.}~\bibnamefont {Peronnin}}, \bibinfo {author} {\bibfnamefont {A.}~\bibnamefont {Sarlette}}, \bibinfo {author} {\bibfnamefont {M.}~\bibnamefont {Delbecq}}, \bibinfo {author} {\bibfnamefont {B.}~\bibnamefont {Huard}}, \bibinfo {author} {\bibfnamefont {T.}~\bibnamefont {Kontos}}, \bibinfo {author} {\bibfnamefont {M.}~\bibnamefont {Mirrahimi}},\ and\ \bibinfo {author} {\bibfnamefont {Z.}~\bibnamefont {Leghtas}},\ }\href {https://doi.org/10.1038/s41567-020-0824-x} {\bibfield  {journal} {\bibinfo  {journal} {Nature Physics}\ }\textbf {\bibinfo {volume} {16}},\ \bibinfo {pages} {509} (\bibinfo {year} {2020})}\BibitemShut {NoStop}%
\bibitem [{\citenamefont {Chamberland}\ \emph {et~al.}(2022)\citenamefont {Chamberland}, \citenamefont {Noh}, \citenamefont {Arrangoiz-Arriola}, \citenamefont {Campbell}, \citenamefont {Hann}, \citenamefont {Iverson}, \citenamefont {Putterman}, \citenamefont {Bohdanowicz}, \citenamefont {Flammia}, \citenamefont {Keller}, \citenamefont {Refael}, \citenamefont {Preskill}, \citenamefont {Jiang}, \citenamefont {Safavi-Naeini}, \citenamefont {Painter},\ and\ \citenamefont {Brandão}}]{Chamberland_2022}%
  \BibitemOpen
  \bibfield  {author} {\bibinfo {author} {\bibfnamefont {C.}~\bibnamefont {Chamberland}}, \bibinfo {author} {\bibfnamefont {K.}~\bibnamefont {Noh}}, \bibinfo {author} {\bibfnamefont {P.}~\bibnamefont {Arrangoiz-Arriola}}, \bibinfo {author} {\bibfnamefont {E.~T.}\ \bibnamefont {Campbell}}, \bibinfo {author} {\bibfnamefont {C.~T.}\ \bibnamefont {Hann}}, \bibinfo {author} {\bibfnamefont {J.}~\bibnamefont {Iverson}}, \bibinfo {author} {\bibfnamefont {H.}~\bibnamefont {Putterman}}, \bibinfo {author} {\bibfnamefont {T.~C.}\ \bibnamefont {Bohdanowicz}}, \bibinfo {author} {\bibfnamefont {S.~T.}\ \bibnamefont {Flammia}}, \bibinfo {author} {\bibfnamefont {A.}~\bibnamefont {Keller}}, \bibinfo {author} {\bibfnamefont {G.}~\bibnamefont {Refael}}, \bibinfo {author} {\bibfnamefont {J.}~\bibnamefont {Preskill}}, \bibinfo {author} {\bibfnamefont {L.}~\bibnamefont {Jiang}}, \bibinfo {author} {\bibfnamefont {A.~H.}\ \bibnamefont {Safavi-Naeini}}, \bibinfo {author} {\bibfnamefont {O.}~\bibnamefont {Painter}},\ and\ \bibinfo {author} {\bibfnamefont {F.~G.}\ \bibnamefont {Brandão}},\ }\bibfield  {journal} {\bibinfo  {journal} {PRX Quantum}\ }\textbf {\bibinfo {volume} {3}},\ \href {https://doi.org/10.1103/prxquantum.3.010329} {10.1103/prxquantum.3.010329} (\bibinfo {year} {2022})\BibitemShut {NoStop}%
\bibitem [{\citenamefont {R\'eglade}\ \emph {et~al.}(2024)\citenamefont {R\'eglade}, \citenamefont {Bocquet}, \citenamefont {Gautier}, \citenamefont {Cohen}, \citenamefont {Marquet}, \citenamefont {Albertinale}, \citenamefont {Pankratova}, \citenamefont {Hallén}, \citenamefont {Rautschke}, \citenamefont {Sellem}, \citenamefont {Rouchon}, \citenamefont {Sarlette}, \citenamefont {Mirrahimi}, \citenamefont {Campagne-Ibarcq}, \citenamefont {Lescanne}, \citenamefont {Jezouin},\ and\ \citenamefont {Leghtas}}]{Reglade_2024}%
  \BibitemOpen
  \bibfield  {author} {\bibinfo {author} {\bibfnamefont {U.}~\bibnamefont {R\'eglade}}, \bibinfo {author} {\bibfnamefont {A.}~\bibnamefont {Bocquet}}, \bibinfo {author} {\bibfnamefont {R.}~\bibnamefont {Gautier}}, \bibinfo {author} {\bibfnamefont {J.}~\bibnamefont {Cohen}}, \bibinfo {author} {\bibfnamefont {A.}~\bibnamefont {Marquet}}, \bibinfo {author} {\bibfnamefont {E.}~\bibnamefont {Albertinale}}, \bibinfo {author} {\bibfnamefont {N.}~\bibnamefont {Pankratova}}, \bibinfo {author} {\bibfnamefont {M.}~\bibnamefont {Hallén}}, \bibinfo {author} {\bibfnamefont {F.}~\bibnamefont {Rautschke}}, \bibinfo {author} {\bibfnamefont {L.-A.}\ \bibnamefont {Sellem}}, \bibinfo {author} {\bibfnamefont {P.}~\bibnamefont {Rouchon}}, \bibinfo {author} {\bibfnamefont {A.}~\bibnamefont {Sarlette}}, \bibinfo {author} {\bibfnamefont {M.}~\bibnamefont {Mirrahimi}}, \bibinfo {author} {\bibfnamefont {P.}~\bibnamefont {Campagne-Ibarcq}}, \bibinfo {author} {\bibfnamefont {R.}~\bibnamefont {Lescanne}}, \bibinfo {author} {\bibfnamefont {S.}~\bibnamefont {Jezouin}},\ and\ \bibinfo {author} {\bibfnamefont {Z.}~\bibnamefont {Leghtas}},\ }\href {https://doi.org/10.1038/s41586-024-07294-3} {\bibfield  {journal} {\bibinfo  {journal} {Nature}\ }\textbf {\bibinfo {volume} {629}},\ \bibinfo {pages} {778–783} (\bibinfo {year} {2024})}\BibitemShut {NoStop}%
\bibitem [{\citenamefont {Gouzien}\ \emph {et~al.}(2023)\citenamefont {Gouzien}, \citenamefont {Ruiz}, \citenamefont {Le~R\'egent}, \citenamefont {Guillaud},\ and\ \citenamefont {Sangouard}}]{Gouzien_2023}%
  \BibitemOpen
  \bibfield  {author} {\bibinfo {author} {\bibfnamefont {E.}~\bibnamefont {Gouzien}}, \bibinfo {author} {\bibfnamefont {D.}~\bibnamefont {Ruiz}}, \bibinfo {author} {\bibfnamefont {F.-M.}\ \bibnamefont {Le~R\'egent}}, \bibinfo {author} {\bibfnamefont {J.}~\bibnamefont {Guillaud}},\ and\ \bibinfo {author} {\bibfnamefont {N.}~\bibnamefont {Sangouard}},\ }\bibfield  {journal} {\bibinfo  {journal} {Physical Review Letters}\ }\textbf {\bibinfo {volume} {131}},\ \href {https://doi.org/10.1103/physrevlett.131.040602} {10.1103/physrevlett.131.040602} (\bibinfo {year} {2023})\BibitemShut {NoStop}%
\bibitem [{\citenamefont {Rousseau}\ \emph {et~al.}(2025)\citenamefont {Rousseau}, \citenamefont {Ruiz}, \citenamefont {Albertinale}, \citenamefont {d'Avezac}, \citenamefont {Banys}, \citenamefont {Blandin}, \citenamefont {Bourdaud}, \citenamefont {Campanaro}, \citenamefont {Cardoso}, \citenamefont {Cottet}, \citenamefont {Cullip}, \citenamefont {Del\'eglise}, \citenamefont {Devanz}, \citenamefont {Devulder}, \citenamefont {Essig}, \citenamefont {F\'evrier}, \citenamefont {Gicquel}, \citenamefont {Gouzien}, \citenamefont {Guillaud}, \citenamefont {G\"um\"us}, \citenamefont {Hall\'en}, \citenamefont {Jacob}, \citenamefont {Magnard}, \citenamefont {Marquet}, \citenamefont {Miklass}, \citenamefont {Peronnin}, \citenamefont {Polis}, \citenamefont {Rautschke}, \citenamefont {R\'eglade}, \citenamefont {Roul}, \citenamefont {Stevens}, \citenamefont {Solard}, \citenamefont {Thomas}, \citenamefont {Ville}, \citenamefont {Wan-Fat}, \citenamefont {Lescanne}, \citenamefont {Leghtas}, \citenamefont {Cohen}, \citenamefont {Jezouin},\ and\ \citenamefont {Murani}}]{Rousseau2025}%
  \BibitemOpen
  \bibfield  {author} {\bibinfo {author} {\bibfnamefont {R.}~\bibnamefont {Rousseau}}, \bibinfo {author} {\bibfnamefont {D.}~\bibnamefont {Ruiz}}, \bibinfo {author} {\bibfnamefont {E.}~\bibnamefont {Albertinale}}, \bibinfo {author} {\bibfnamefont {P.}~\bibnamefont {d'Avezac}}, \bibinfo {author} {\bibfnamefont {D.}~\bibnamefont {Banys}}, \bibinfo {author} {\bibfnamefont {U.}~\bibnamefont {Blandin}}, \bibinfo {author} {\bibfnamefont {N.}~\bibnamefont {Bourdaud}}, \bibinfo {author} {\bibfnamefont {G.}~\bibnamefont {Campanaro}}, \bibinfo {author} {\bibfnamefont {G.}~\bibnamefont {Cardoso}}, \bibinfo {author} {\bibfnamefont {N.}~\bibnamefont {Cottet}}, \bibinfo {author} {\bibfnamefont {C.}~\bibnamefont {Cullip}}, \bibinfo {author} {\bibfnamefont {S.}~\bibnamefont {Del\'eglise}}, \bibinfo {author} {\bibfnamefont {L.}~\bibnamefont {Devanz}}, \bibinfo {author} {\bibfnamefont {A.}~\bibnamefont {Devulder}}, \bibinfo {author} {\bibfnamefont {A.}~\bibnamefont {Essig}}, \bibinfo {author} {\bibfnamefont {P.}~\bibnamefont {F\'evrier}}, \bibinfo {author} {\bibfnamefont {A.}~\bibnamefont {Gicquel}}, \bibinfo {author} {\bibfnamefont {E.}~\bibnamefont {Gouzien}}, \bibinfo {author} {\bibfnamefont {J.}~\bibnamefont {Guillaud}}, \bibinfo {author} {\bibfnamefont {E.}~\bibnamefont {G\"um\"us}}, \bibinfo {author} {\bibfnamefont {M.}~\bibnamefont {Hall\'en}}, \bibinfo {author} {\bibfnamefont {A.}~\bibnamefont {Jacob}}, \bibinfo {author} {\bibfnamefont {P.}~\bibnamefont {Magnard}}, \bibinfo {author} {\bibfnamefont {A.}~\bibnamefont {Marquet}}, \bibinfo {author} {\bibfnamefont {S.}~\bibnamefont {Miklass}}, \bibinfo {author} {\bibfnamefont {T.}~\bibnamefont {Peronnin}}, \bibinfo {author} {\bibfnamefont {S.}~\bibnamefont {Polis}}, \bibinfo {author} {\bibfnamefont {F.}~\bibnamefont {Rautschke}}, \bibinfo {author} {\bibfnamefont {U.}~\bibnamefont {R\'eglade}}, \bibinfo {author} {\bibfnamefont {J.}~\bibnamefont {Roul}}, \bibinfo {author} {\bibfnamefont {J.}~\bibnamefont {Stevens}}, \bibinfo {author} {\bibfnamefont {J.}~\bibnamefont {Solard}}, \bibinfo {author} {\bibfnamefont {A.}~\bibnamefont {Thomas}}, \bibinfo {author} {\bibfnamefont {J.-L.}\ \bibnamefont {Ville}}, \bibinfo {author} {\bibfnamefont {P.}~\bibnamefont {Wan-Fat}}, \bibinfo {author} {\bibfnamefont {R.}~\bibnamefont {Lescanne}}, \bibinfo {author} {\bibfnamefont {Z.}~\bibnamefont {Leghtas}}, \bibinfo {author} {\bibfnamefont {J.}~\bibnamefont {Cohen}}, \bibinfo {author} {\bibfnamefont {S.}~\bibnamefont {Jezouin}},\ and\ \bibinfo {author} {\bibfnamefont {A.}~\bibnamefont {Murani}},\ }\href {https://arxiv.org/abs/2502.07892} {\bibinfo {title} {Enhancing dissipative cat qubit protection by squeezing}} (\bibinfo {year} {2025}),\ \Eprint {https://arxiv.org/abs/2502.07892} {arXiv:2502.07892 [quant-ph]} \BibitemShut {NoStop}%
\bibitem [{\citenamefont {{Alice \& Bob}}(2025)}]{AliceBobCatQubit2025}%
  \BibitemOpen
  \bibfield  {author} {\bibinfo {author} {\bibnamefont {{Alice \& Bob}}},\ }\href {https://alice-bob.com/blog/just-out-of-the-lab-a-cat-qubit-that-jumps-every-hour/} {\bibinfo {title} {Just out of the lab: A cat qubit that jumps every hour}} (\bibinfo {year} {2025}),\ \bibinfo {note} {accessed: \today}\BibitemShut {NoStop}%
\bibitem [{\citenamefont {Pappalardo}\ and\ \citenamefont {Stevens}(2025)}]{Report2_confidential}%
  \BibitemOpen
  \bibfield  {author} {\bibinfo {author} {\bibfnamefont {A.}~\bibnamefont {Pappalardo}}\ and\ \bibinfo {author} {\bibfnamefont {J.}~\bibnamefont {Stevens}},\ }\href@noop {} {\bibfield  {journal} {\bibinfo  {journal} {Report, Alice and Bob}\ } (\bibinfo {year} {2025})}\BibitemShut {NoStop}%
\bibitem [{\citenamefont {Marquet}\ \emph {et~al.}(2024)\citenamefont {Marquet}, \citenamefont {Dupouy}, \citenamefont {R\'eglade}, \citenamefont {Essig}, \citenamefont {Cohen}, \citenamefont {Albertinale}, \citenamefont {Bienfait}, \citenamefont {Peronnin}, \citenamefont {Jezouin}, \citenamefont {Lescanne},\ and\ \citenamefont {Huard}}]{Marquet2024}%
  \BibitemOpen
  \bibfield  {author} {\bibinfo {author} {\bibfnamefont {A.}~\bibnamefont {Marquet}}, \bibinfo {author} {\bibfnamefont {S.}~\bibnamefont {Dupouy}}, \bibinfo {author} {\bibfnamefont {U.}~\bibnamefont {R\'eglade}}, \bibinfo {author} {\bibfnamefont {A.}~\bibnamefont {Essig}}, \bibinfo {author} {\bibfnamefont {J.}~\bibnamefont {Cohen}}, \bibinfo {author} {\bibfnamefont {E.}~\bibnamefont {Albertinale}}, \bibinfo {author} {\bibfnamefont {A.}~\bibnamefont {Bienfait}}, \bibinfo {author} {\bibfnamefont {T.}~\bibnamefont {Peronnin}}, \bibinfo {author} {\bibfnamefont {S.}~\bibnamefont {Jezouin}}, \bibinfo {author} {\bibfnamefont {R.}~\bibnamefont {Lescanne}},\ and\ \bibinfo {author} {\bibfnamefont {B.}~\bibnamefont {Huard}},\ }\bibfield  {journal} {\bibinfo  {journal} {Physical Review Applied}\ }\textbf {\bibinfo {volume} {22}},\ \href {https://doi.org/10.1103/physrevapplied.22.034053} {10.1103/physrevapplied.22.034053} (\bibinfo {year} {2024})\BibitemShut {NoStop}%
\bibitem [{\citenamefont {Guillaud}\ and\ \citenamefont {Mirrahimi}(2019)}]{Guillaud_2019}%
  \BibitemOpen
  \bibfield  {author} {\bibinfo {author} {\bibfnamefont {J.}~\bibnamefont {Guillaud}}\ and\ \bibinfo {author} {\bibfnamefont {M.}~\bibnamefont {Mirrahimi}},\ }\bibfield  {journal} {\bibinfo  {journal} {Physical Review X}\ }\textbf {\bibinfo {volume} {9}},\ \href {https://doi.org/10.1103/physrevx.9.041053} {10.1103/physrevx.9.041053} (\bibinfo {year} {2019})\BibitemShut {NoStop}%
\bibitem [{\citenamefont {R\'egent}\ \emph {et~al.}(2023)\citenamefont {R\'egent}, \citenamefont {Berdou}, \citenamefont {Leghtas}, \citenamefont {Guillaud},\ and\ \citenamefont {Mirrahimi}}]{R_gent_2023}%
  \BibitemOpen
  \bibfield  {author} {\bibinfo {author} {\bibfnamefont {F.-M.~L.}\ \bibnamefont {R\'egent}}, \bibinfo {author} {\bibfnamefont {C.}~\bibnamefont {Berdou}}, \bibinfo {author} {\bibfnamefont {Z.}~\bibnamefont {Leghtas}}, \bibinfo {author} {\bibfnamefont {J.}~\bibnamefont {Guillaud}},\ and\ \bibinfo {author} {\bibfnamefont {M.}~\bibnamefont {Mirrahimi}},\ }\href {https://doi.org/10.22331/q-2023-12-06-1198} {\bibfield  {journal} {\bibinfo  {journal} {Quantum}\ }\textbf {\bibinfo {volume} {7}},\ \bibinfo {pages} {1198} (\bibinfo {year} {2023})}\BibitemShut {NoStop}%
\bibitem [{\citenamefont {Nielsen}\ and\ \citenamefont {Chuang}(2010)}]{Nielsen_Chuang_2010}%
  \BibitemOpen
  \bibfield  {author} {\bibinfo {author} {\bibfnamefont {M.~A.}\ \bibnamefont {Nielsen}}\ and\ \bibinfo {author} {\bibfnamefont {I.~L.}\ \bibnamefont {Chuang}},\ }\href@noop {} {\emph {\bibinfo {title} {Quantum Computation and Quantum Information: 10th Anniversary Edition}}}\ (\bibinfo  {publisher} {Cambridge University Press},\ \bibinfo {year} {2010})\BibitemShut {NoStop}%
\bibitem [{\citenamefont {Coppersmith}(2002)}]{coppersmith2002}%
  \BibitemOpen
  \bibfield  {author} {\bibinfo {author} {\bibfnamefont {D.}~\bibnamefont {Coppersmith}},\ }\href {https://arxiv.org/abs/quant-ph/0201067} {\bibinfo {title} {An approximate fourier transform useful in quantum factoring}} (\bibinfo {year} {2002}),\ \Eprint {https://arxiv.org/abs/quant-ph/0201067} {arXiv:quant-ph/0201067 [quant-ph]} \BibitemShut {NoStop}%
\bibitem [{\citenamefont {Griffiths}\ and\ \citenamefont {Niu}(1996)}]{Griffiths_1996}%
  \BibitemOpen
  \bibfield  {author} {\bibinfo {author} {\bibfnamefont {R.~B.}\ \bibnamefont {Griffiths}}\ and\ \bibinfo {author} {\bibfnamefont {C.-S.}\ \bibnamefont {Niu}},\ }\href {https://doi.org/10.1103/physrevlett.76.3228} {\bibfield  {journal} {\bibinfo  {journal} {Physical Review Letters}\ }\textbf {\bibinfo {volume} {76}},\ \bibinfo {pages} {3228} (\bibinfo {year} {1996})}\BibitemShut {NoStop}%
\bibitem [{\citenamefont {Hoyau}\ \emph {et~al.}(2025)\citenamefont {Hoyau}, \citenamefont {Stevens},\ and\ \citenamefont {Magnard}}]{Report1_confidential}%
  \BibitemOpen
  \bibfield  {author} {\bibinfo {author} {\bibfnamefont {B.}~\bibnamefont {Hoyau}}, \bibinfo {author} {\bibfnamefont {J.}~\bibnamefont {Stevens}},\ and\ \bibinfo {author} {\bibfnamefont {P.}~\bibnamefont {Magnard}},\ }\href@noop {} {\bibfield  {journal} {\bibinfo  {journal} {Internship Report, Alice and Bob}\ } (\bibinfo {year} {2025})}\BibitemShut {NoStop}%
\bibitem [{\citenamefont {Pagni}\ \emph {et~al.}(2025)\citenamefont {Pagni}, \citenamefont {Huber}, \citenamefont {Epping},\ and\ \citenamefont {Felderer}}]{Pagni2025}%
  \BibitemOpen
  \bibfield  {author} {\bibinfo {author} {\bibfnamefont {V.}~\bibnamefont {Pagni}}, \bibinfo {author} {\bibfnamefont {S.}~\bibnamefont {Huber}}, \bibinfo {author} {\bibfnamefont {M.}~\bibnamefont {Epping}},\ and\ \bibinfo {author} {\bibfnamefont {M.}~\bibnamefont {Felderer}},\ }\href {https://arxiv.org/abs/2503.17113} {\bibinfo {title} {Fast quantum amplitude encoding of typical classical data}} (\bibinfo {year} {2025}),\ \Eprint {https://arxiv.org/abs/2503.17113} {arXiv:2503.17113 [quant-ph]} \BibitemShut {NoStop}%
\bibitem [{\citenamefont {Michael}\ \emph {et~al.}(2016)\citenamefont {Michael}, \citenamefont {Silveri}, \citenamefont {Brierley}, \citenamefont {Albert}, \citenamefont {Salmilehto}, \citenamefont {Jiang},\ and\ \citenamefont {Girvin}}]{Michael_2016}%
  \BibitemOpen
  \bibfield  {author} {\bibinfo {author} {\bibfnamefont {M.~H.}\ \bibnamefont {Michael}}, \bibinfo {author} {\bibfnamefont {M.}~\bibnamefont {Silveri}}, \bibinfo {author} {\bibfnamefont {R.~T.}\ \bibnamefont {Brierley}}, \bibinfo {author} {\bibfnamefont {V.~V.}\ \bibnamefont {Albert}}, \bibinfo {author} {\bibfnamefont {J.}~\bibnamefont {Salmilehto}}, \bibinfo {author} {\bibfnamefont {L.}~\bibnamefont {Jiang}},\ and\ \bibinfo {author} {\bibfnamefont {S.~M.}\ \bibnamefont {Girvin}},\ }\bibfield  {journal} {\bibinfo  {journal} {Physical Review X}\ }\textbf {\bibinfo {volume} {6}},\ \href {https://doi.org/10.1103/physrevx.6.031006} {10.1103/physrevx.6.031006} (\bibinfo {year} {2016})\BibitemShut {NoStop}%
\bibitem [{\citenamefont {Ruiz}\ \emph {et~al.}(2025)\citenamefont {Ruiz}, \citenamefont {Guillaud}, \citenamefont {Leverrier}, \citenamefont {Mirrahimi},\ and\ \citenamefont {Vuillot}}]{Ruiz_2025}%
  \BibitemOpen
  \bibfield  {author} {\bibinfo {author} {\bibfnamefont {D.}~\bibnamefont {Ruiz}}, \bibinfo {author} {\bibfnamefont {J.}~\bibnamefont {Guillaud}}, \bibinfo {author} {\bibfnamefont {A.}~\bibnamefont {Leverrier}}, \bibinfo {author} {\bibfnamefont {M.}~\bibnamefont {Mirrahimi}},\ and\ \bibinfo {author} {\bibfnamefont {C.}~\bibnamefont {Vuillot}},\ }\bibfield  {journal} {\bibinfo  {journal} {Nature Communications}\ }\textbf {\bibinfo {volume} {16}},\ \href {https://doi.org/10.1038/s41467-025-56298-8} {10.1038/s41467-025-56298-8} (\bibinfo {year} {2025})\BibitemShut {NoStop}%
\bibitem [{\citenamefont {{TOP500 Project}}(2025)}]{top500_nov2025}%
  \BibitemOpen
  \bibfield  {author} {\bibinfo {author} {\bibnamefont {{TOP500 Project}}},\ }\href {https://top500.org/lists/top500/2025/11/} {\bibinfo {title} {{TOP500 List -- November 2025}}} (\bibinfo {year} {2025}),\ \bibinfo {note} {accessed: 2026-03-04}\BibitemShut {NoStop}%
\bibitem [{\citenamefont {{TOP500}}(2026)}]{green500_nov2025}%
  \BibitemOpen
  \bibfield  {author} {\bibinfo {author} {\bibnamefont {{TOP500}}},\ }\href {https://top500.org/lists/green500/2025/06/} {\bibinfo {title} {Green500 list -- november 2025}} (\bibinfo {year} {2026}),\ \bibinfo {note} {accessed: 2026-03-04}\BibitemShut {NoStop}%
\bibitem [{\citenamefont {Dongarra}\ \emph {et~al.}(2003)\citenamefont {Dongarra}, \citenamefont {Luszczek},\ and\ \citenamefont {Petitet}}]{Dongarra2003}%
  \BibitemOpen
  \bibfield  {author} {\bibinfo {author} {\bibfnamefont {J.}~\bibnamefont {Dongarra}}, \bibinfo {author} {\bibfnamefont {P.}~\bibnamefont {Luszczek}},\ and\ \bibinfo {author} {\bibfnamefont {A.}~\bibnamefont {Petitet}},\ }\href {https://doi.org/10.1002/cpe.728} {\bibfield  {journal} {\bibinfo  {journal} {Concurrency and Computation: Practice and Experience}\ }\textbf {\bibinfo {volume} {15}},\ \bibinfo {pages} {803} (\bibinfo {year} {2003})}\BibitemShut {NoStop}%
\bibitem [{\citenamefont {Cooley}\ and\ \citenamefont {Tukey}(1969)}]{CooleyTukey1969}%
  \BibitemOpen
  \bibfield  {author} {\bibinfo {author} {\bibfnamefont {J.~W.}\ \bibnamefont {Cooley}}\ and\ \bibinfo {author} {\bibfnamefont {J.~W.}\ \bibnamefont {Tukey}},\ }in\ \href {https://direct.mit.edu/book/chapterpdf/2397157/9780262310840_cao.pdf} {\emph {\bibinfo {booktitle} {Papers on Digital Signal Processing}}}\ (\bibinfo  {publisher} {The MIT Press},\ \bibinfo {address} {Cambridge, MA},\ \bibinfo {year} {1969})\BibitemShut {NoStop}%
\bibitem [{\citenamefont {Frigo}\ and\ \citenamefont {Johnson}(1997)}]{FFTW97}%
  \BibitemOpen
  \bibfield  {author} {\bibinfo {author} {\bibfnamefont {M.}~\bibnamefont {Frigo}}\ and\ \bibinfo {author} {\bibfnamefont {S.~G.}\ \bibnamefont {Johnson}},\ }\href@noop {} {\emph {\bibinfo {title} {The Fastest {Fourier} Transform in the West}}},\ \bibinfo {type} {Tech. Rep.}\ \bibinfo {number} {MIT-LCS-TR-728}\ (\bibinfo  {institution} {Massachusetts Institute of Technology},\ \bibinfo {year} {1997})\BibitemShut {NoStop}%
\bibitem [{\citenamefont {Van~Loan}(1992)}]{Loan1992}%
  \BibitemOpen
  \bibfield  {author} {\bibinfo {author} {\bibfnamefont {C.}~\bibnamefont {Van~Loan}},\ }\href {https://doi.org/10.1137/1.9781611970999} {\emph {\bibinfo {title} {Computational Frameworks for the Fast Fourier Transform}}}\ (\bibinfo  {publisher} {Society for Industrial and Applied Mathematics},\ \bibinfo {year} {1992})\ \Eprint {https://arxiv.org/abs/https://epubs.siam.org/doi/pdf/10.1137/1.9781611970999} {https://epubs.siam.org/doi/pdf/10.1137/1.9781611970999} \BibitemShut {NoStop}%
\end{thebibliography}%
\newpage

\onecolumngrid

\appendix

\section{Formulas and Parameters}  \label{energy_parameters}

This section presents the formulas and parameters required to compute the total energy of the QFT.

\begin{table}[H]
\centering
\begin{tabular}{lccc}
\textbf{Gate} & \textbf{Power Formula} & \textbf{Duration Formula} \\
\hline
$P_+$ & $P_{P_+} = p \frac{\kappa_b}{4}\kappa_2 + d\epsilon_d^2$ & $T_{P_+} = \frac{1}{\kappa_2}$ \\

$Z(\frac{\pi}{2})$ & $P_{Z(\frac{\pi}{2})} = p \frac{\kappa_b}{4}\kappa_2 + d\epsilon_d^2 + z\epsilon_z^2$ & $T_{Z(\frac{\pi}{2})} = \frac{\pi}{8 |\alpha| \epsilon_z}$  \\

$H_{def}$ & $P_{H_{def}} = p \frac{\kappa_b}{4}\kappa_2$ & $T_{H_{def}} = \frac{a_3}{\kappa_2}$\\

$H_{inf}$ & $P_{H_{inf}} = p \frac{\kappa_b}{4}\kappa_2 + d \epsilon_d^2$ & $T_{H_{inf}} = \frac{1}{\kappa_2}$ \\

$H_{disp}$ & $P_{H_{disp}} = z \epsilon_z^2 \approx 0$ & $T_{H_{disp}} = \frac{\hbar \alpha}{\epsilon_{disp}} \approx 0$  \\

$H_{l}$ & $P_{H_{l}} = l g_l^2$ & $T_{H_{l}} = \text{argmax}(\mathcal{F}(t))$  \\

$Z(\theta)$ & $P_{Z(\theta)} = p \frac{\kappa_b}{4}\kappa_2 + d\epsilon_d^2 + z\epsilon_z^2$ & $T_{Z(\theta)} = \frac{\theta}{4 |\alpha| \epsilon_z}$\\

$\text{CNOT}$ & $P_{\text{CNOT}} = c g_{\text{CNOT}}^2 +  p \frac{\kappa_b}{4}\kappa_2 +d\epsilon_d^2$ & $T_{\text{CNOT}} = \frac{\pi}{4 |\alpha| g_{\text{CNOT}}}$  \\

\hline
\end{tabular}
\caption{Microscopic Energy calculation for different quantum gates \cite{Report1_confidential, Report2_confidential}.}
\label{tab:gate_energies_microscopic}
\end{table}

\begin{table}[H]
\centering
\begin{tabular}{lcc}
\textbf{Gate} & \textbf{Power Formula} & \textbf{Duration Formula}  \\
\hline
$P_+$ & $P_{P_+} = M_p p \frac{\kappa_b}{4}\kappa_2 + M_dd\epsilon_d^2$ & $T_{P_+} = \frac{1}{\kappa_2}$ \\

$Z(\frac{\pi}{2})$ & $P_{Z(\frac{\pi}{2})} = M_pp \frac{\kappa_b}{4}\kappa_2 + M_dd\epsilon_d^2 + M_z z\epsilon_z^2$ & $T_{Z(\frac{\pi}{2})} = \frac{\pi}{8 |\alpha| \epsilon_z}$  \\

$H_{def}$ & $P_{H_{def}} = M_pp \frac{\kappa_b}{4}\kappa_2$ & $T_{H_{def}} = \frac{a_3}{\kappa_2}$\\

$H_{inf}$ & $P_{H_{inf}} = M_p p \frac{\kappa_b}{4}\kappa_2 + M_dd \epsilon_d^2$ & $T_{H_{inf}} = \frac{1}{\kappa_2}$  \\

$H_{disp}$ & $P_{H_{disp}} = z \epsilon_z^2 \approx 0$ & $T_{H_{disp}} = \frac{\hbar \alpha}{\epsilon_{disp}} \approx 0$ \\

$H_{l}$ & $P_{H_{l}} =M_l l g_l^2$ & $T_{H_{l}} = \text{argmax}(\mathcal{F}(t))$ \\

$Z(\theta)$ & $P_{Z(\theta)} =M_p p \frac{\kappa_b}{4}\kappa_2 + M_d d\epsilon_d^2 + M_z z\epsilon_z^2$ & $T_{Z(\theta)} = \frac{\theta}{4 |\alpha| \epsilon_z}$ \\

$\text{CNOT}$ & $P_{\text{CNOT}} =M_c c g_{\text{CNOT}}^2 + M_p p \frac{\kappa_b}{4}\kappa_2 + M_dd\epsilon_d^2$ & $T_{\text{CNOT}} = \frac{\pi}{4 |\alpha| g_{\text{CNOT}}}$  \\

\hline
\end{tabular}
\caption{Macroscopic Energy calculation for different quantum gates \cite{Report1_confidential, Report2_confidential}.}
\label{tab:gate_energies_macroscopic}
\end{table}

\begin{table}[H]
\centering
\begin{tabular}{cc}
\hline
\multicolumn{1}{c}{\textbf{Parameter}} & \multicolumn{1}{c}{\textbf{Formula}} \\ \hline
\(\epsilon_d\) &  \((|\alpha^2|+\frac{\kappa_1}{\kappa_2})\frac{\sqrt{\kappa_b \kappa_2}}{2}\)\\

\(\epsilon_{z_{\text{micro}}}\) &  \(\frac{2 \kappa_2 p_z |\alpha|^3}{\theta} (1 - \sqrt{1- \frac{\theta^2 \kappa_1 }{4 \alpha^2 p_z^2 \kappa_2}})\) \\

\(\epsilon_{z_{\text{macro}}}\) &  \(\frac{2 \kappa_2 p_z |\alpha|^3}{(1+2n_{th}^b)\theta} (1 - \sqrt{1- \frac{\theta^2 \kappa_1 }{4 \alpha^2 p_z^2 \kappa_2}(1+2n_{th}^m)(1+2n_{th}^b)})\) \\

\(g_{cnot_{\text{micro}}}\) &  \(\frac{2 \kappa_2 p_z |\alpha|}{\pi} (1 - \sqrt{1- \frac{\pi^2 \kappa_1 }{4  p_z^2 \kappa_2}})\)\\

\(g_{cnot_{\text{macro}}}\) &  \(\frac{2 \kappa_2 p_z |\alpha|}{\pi(1+2n_{th}^b)} (1 - \sqrt{1- \frac{\pi^2 \kappa_1 }{4  p_z^2 \kappa_2}(1+2n_{th}^b)(1+2n_{th}^m)})\)\\
\(\mathcal{F}(t)\) &    \(e^{-\kappa_1t}\text{erf}(\sqrt{4\eta \frac{g_l^2}{\kappa_b}(t+\frac{3}{\kappa_b})}/2) \) \\
\hline
\end{tabular}
\caption{Formulas of the parameters used in the energy calculations. Here, \(p_z\) denotes the phase-flip probability of the gate \cite{Report1_confidential, Report2_confidential}.}
\label{tab:parameters_formuals}
\end{table}

\begin{table}[H]
\centering
\begin{tabular}{cc}
\hline
\multicolumn{1}{c}{\textbf{Parameter}} & \multicolumn{1}{c}{\textbf{Value}} \\ \hline
\(\kappa_1/2\pi\) &  $25$ kHz\\
\(\kappa_b/2\pi\) & $40$ kHz \\
\(n_{th}^m\)   & $0.2$ \\
\(n_{th}^b\)   & $0.02$ \\

\(p\)   & $1.35\times 10^{-21}$ \\
\(d\)   & $3.0\times 10^{-32}$ \\
\(z\)   & $1.1\times 10^{-27}$ \\
\(c\)   & $1.20\times 10^{-19}$ \\
\(l\)   & \(\frac{p}{4}\) \\
\(M_p\)   & $1.26\times 10^{6}$ \\
\(M_d\)   & $1.5\times 10^{10}$ \\
\(M_c\)   & \(M_p\) \\
\(M_c\)   & $1.5\times 10^{8}$ \\
\(M_l\)   & \(M_p\) \\
\(g_l/2\pi\) & $5\times10^{6}$ kHz\\

\(a_1\) &  $2.7$ \\
\(a_2\) &  $1$ \\ 
\(a_3\) & $2.8$ \\ \hline
\end{tabular}
\caption{Value of the parameters used in the energy calculations \cite{Marquet2024, Report2_confidential}.}
\label{tab:parameters_values}
\end{table}

\twocolumngrid

\section{Energy scaling of Quantum Computers}
This section describes the rationale for selecting the fidelity of the last qubit as the fixed comparison parameter. Two possible strategies were considered:

\begin{enumerate}
    \item Fixing the total fidelity of the system,
    \item Fixing the fidelity of the last qubit.
\end{enumerate}

\begin{figure}[t]
        \centering
        \includegraphics[width=\linewidth]{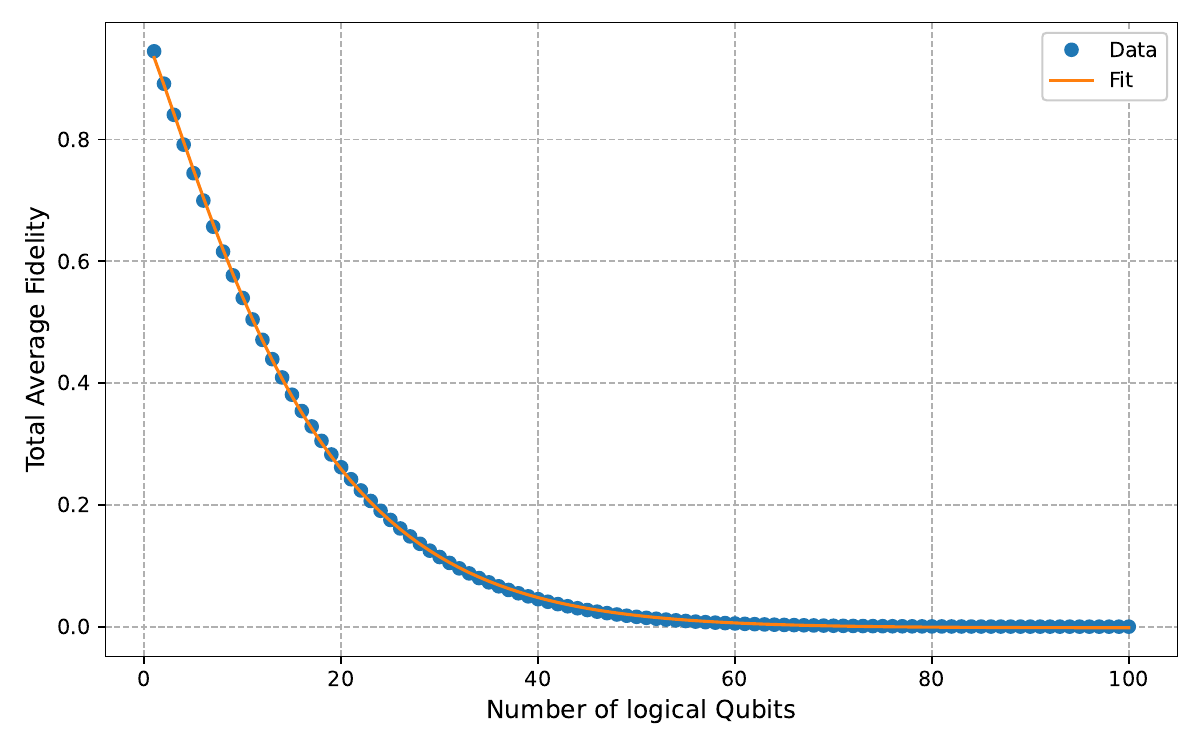}
        \caption{Total fidelity of the QFT for different numbers of qubits, given parameters \(\alpha = 3\), \(\kappa_{2} = 1000\,\kappa_{1}\), 
    \(\epsilon_{z}/(2\pi) = 10.5\,\text{MHz}\), \(d_{c} = 5\), and \(N_{b} = 1\).}
    \label{fig:total_fidelity_micro_macro}
\end{figure}

The first approach is not suitable, as the average total fidelity decreases exponentially with the number of qubits, as shown in Fig.~\ref{fig:total_fidelity_micro_macro}. Consequently, for many system sizes the fidelity achieved for a smaller number of qubits cannot be preserved when performing the QFT on a larger number of qubits. Furthermore, selecting a high total fidelity as a reference would lead to unphysical requirements, since such fidelity values would not be attainable for large systems.

To quantify this behavior, the data in Fig.~\ref{fig:total_fidelity_micro_macro} were fitted using the function
\[
y = a\, e^{-b x^{c}} + d,
\]
yielding the parameters
\[
a = 0.9737, \quad
b = 0.0391, \quad
c = 1.1740, \quad
d = -0.0019.
\]

\begin{figure}[t]
    \centering
    \includegraphics[width=\linewidth]{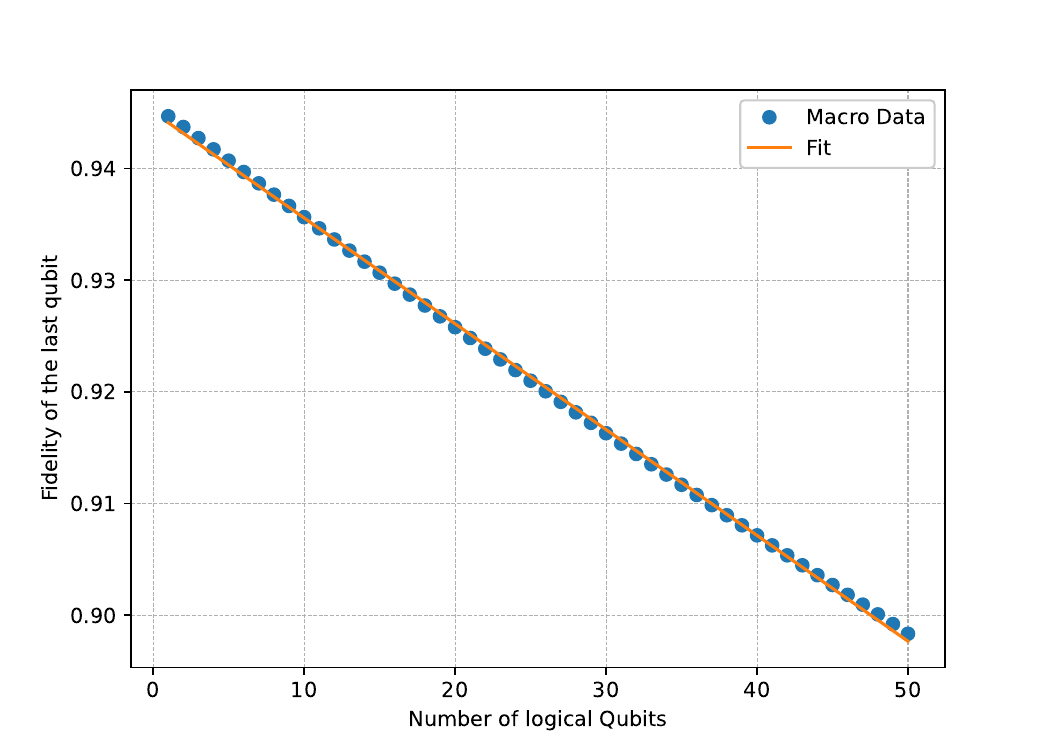}
    \caption{Final qubit fidelity as a function of the number of qubits for the microscopic and macroscopic regimes, when \(\alpha=3\), \(\kappa_2=1500 \kappa_1\), \(\frac{\epsilon_z}{2\pi}=10.5\)MHz, \(d_{c} = 5\) and \(Nb=1\).}
    \label{fig:fidelity_for_the_final_qubit_comparison}
\end{figure}

In contrast, the second option, fixing the fidelity of the last qubit, results in a behavior that depends approximately linearly on the number of qubits. Figure~\ref{fig:fidelity_for_the_final_qubit_comparison} illustrates how the fidelity of the final qubit varies as a function of system size for both the microscopic and macroscopic regimes. 

The data were fitted using a linear function of the form
\[
y = a\, x + b,
\]
yielding the parameters
\[
a = -0.000948, \qquad b = 0.9451.
\]

\section{Dependence of the QFT Energy on the Parameters \(N_b\) and \(d_c\)}
\label{Depende_on_parameters_Nb_and_d}

This section analyzes how the total energy required to perform the QFT depends on the parameters $N_b$ and $d_C$. Both parameters can be tuned to optimize energy consumption while maintaining a desired fidelity for the last qubit.

\begin{figure*}[t]
    \centering

    \begin{minipage}{0.48\linewidth}
        \centering
        \includegraphics[width=\linewidth]{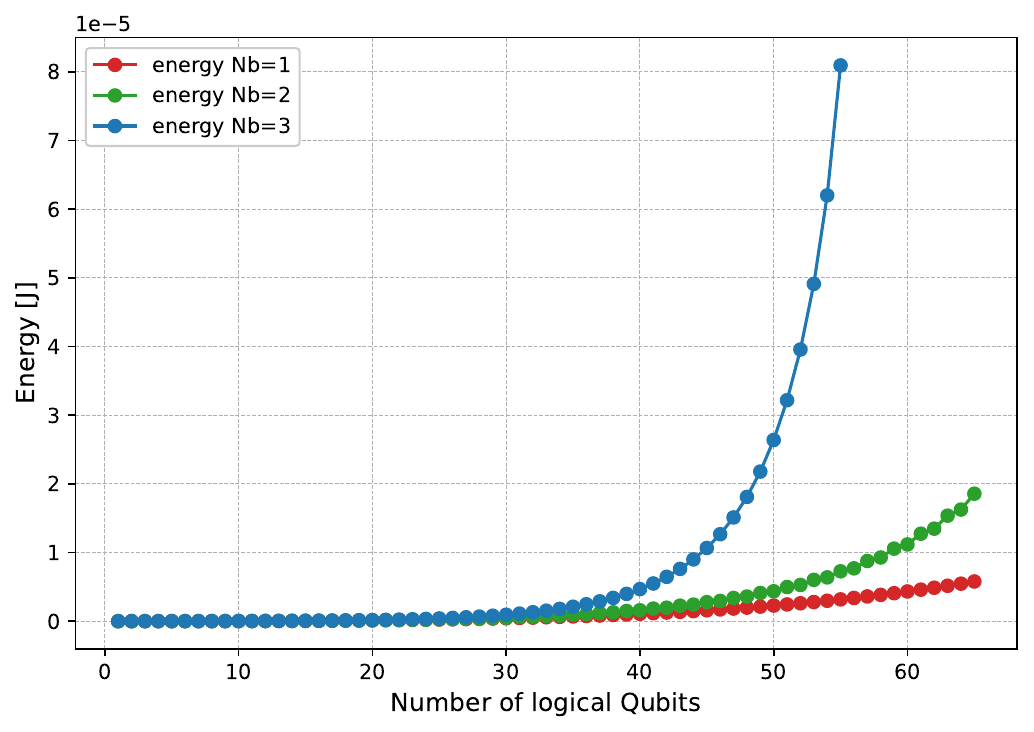}
        \caption*{(a) Microscopic energy}
        \label{fig:micro_nb}
    \end{minipage}
    \hfill
    \begin{minipage}{0.48\linewidth}
        \centering
        \includegraphics[width=\linewidth]{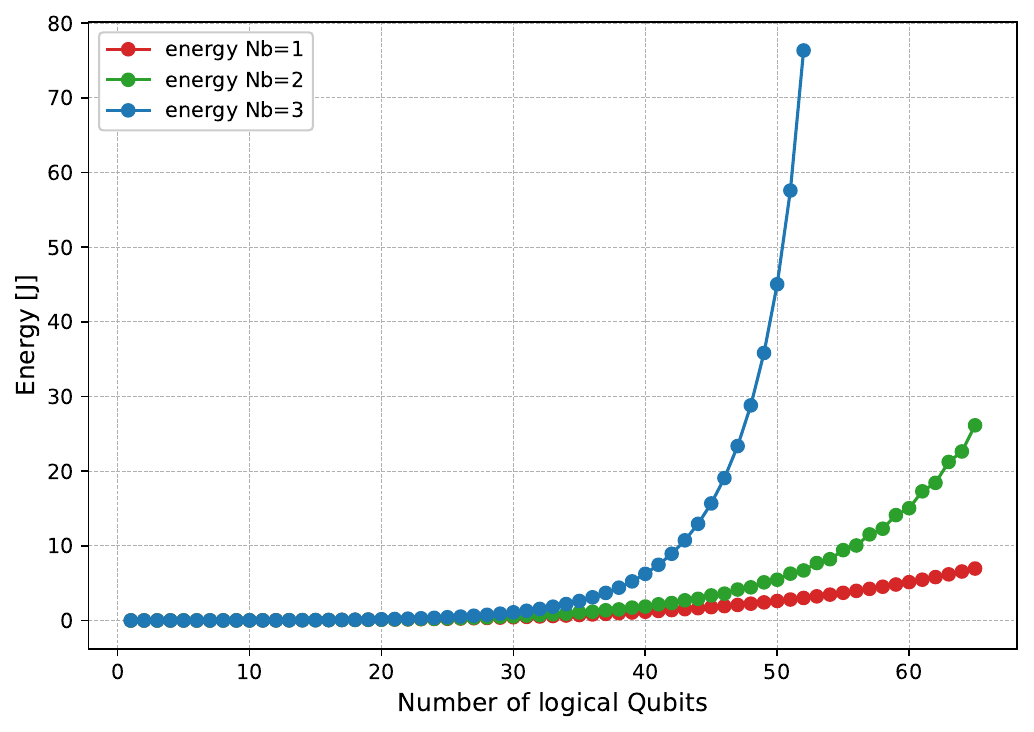}
        \caption*{(b) Macroscopic energy}
        \label{fig:macro_nb}
    \end{minipage}

    \caption{Energy consumption for different numbers of gates between repetition codes, \(N_b\) for \(d_c=5\), where the parameters \(\epsilon_z\) and \(\kappa_2\) were optimized to minimize the energy while maintaining a last-qubit fidelity of \(0.944\)}
    \label{fig:different_nb}
\end{figure*}

Figure~\ref{fig:different_nb} shows the energy as a function of the number of gates between repetition codes for a code distance of \(d_c=5\). For the same code distance, we can conclude that there are cases where it is better to use a smaller value of \(N_b\), even if this increases the total number of gates. This is because the energy required to maintain a given fidelity is higher for circuits with larger intrinsic errors. Therefore, it is often less costly to correct errors more frequently than to allow them to accumulate and increase the parameters needed to achieve the same fidelity. The oscillations in Figure~\ref{fig:different_nb}, especially for \(N_b=2\), arise from the gate block structure. For an odd number of qubits, the last qubit passes through an extra gate, altering phase-flip probabilities across segments. This reduces fidelity for odd qubit counts, requiring higher energy to reach the same target fidelity.

%\begin{figure}[H]

\begin{figure*}[t]
    \centering

    \begin{minipage}{0.48\linewidth}
        \centering
        \includegraphics[width=\linewidth]{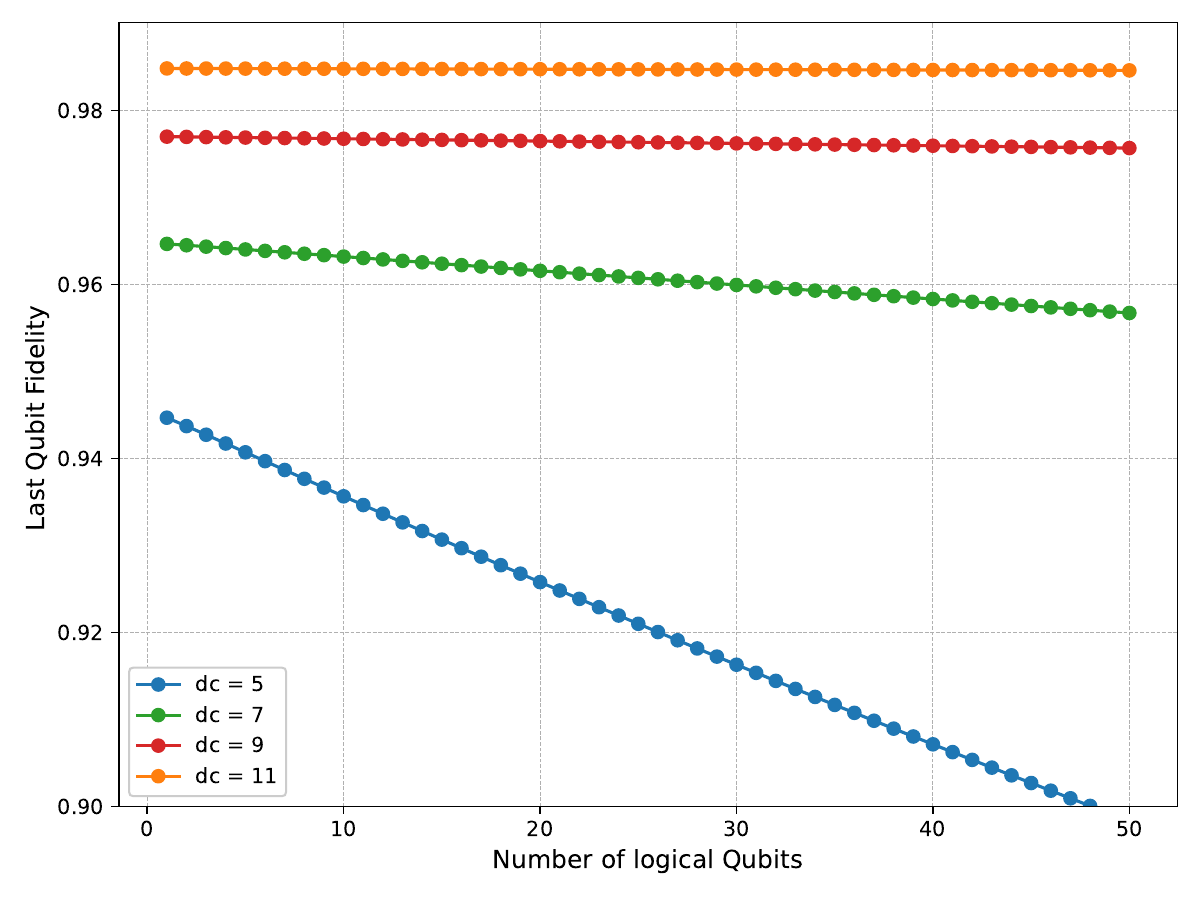}
        \caption*{(a) Last qubit fidelity}
        \label{fig:last_qubit_fidelity_different_dc}
    \end{minipage}
    \hfill
    \begin{minipage}{0.48\linewidth}
        \centering
        \includegraphics[width=\linewidth]{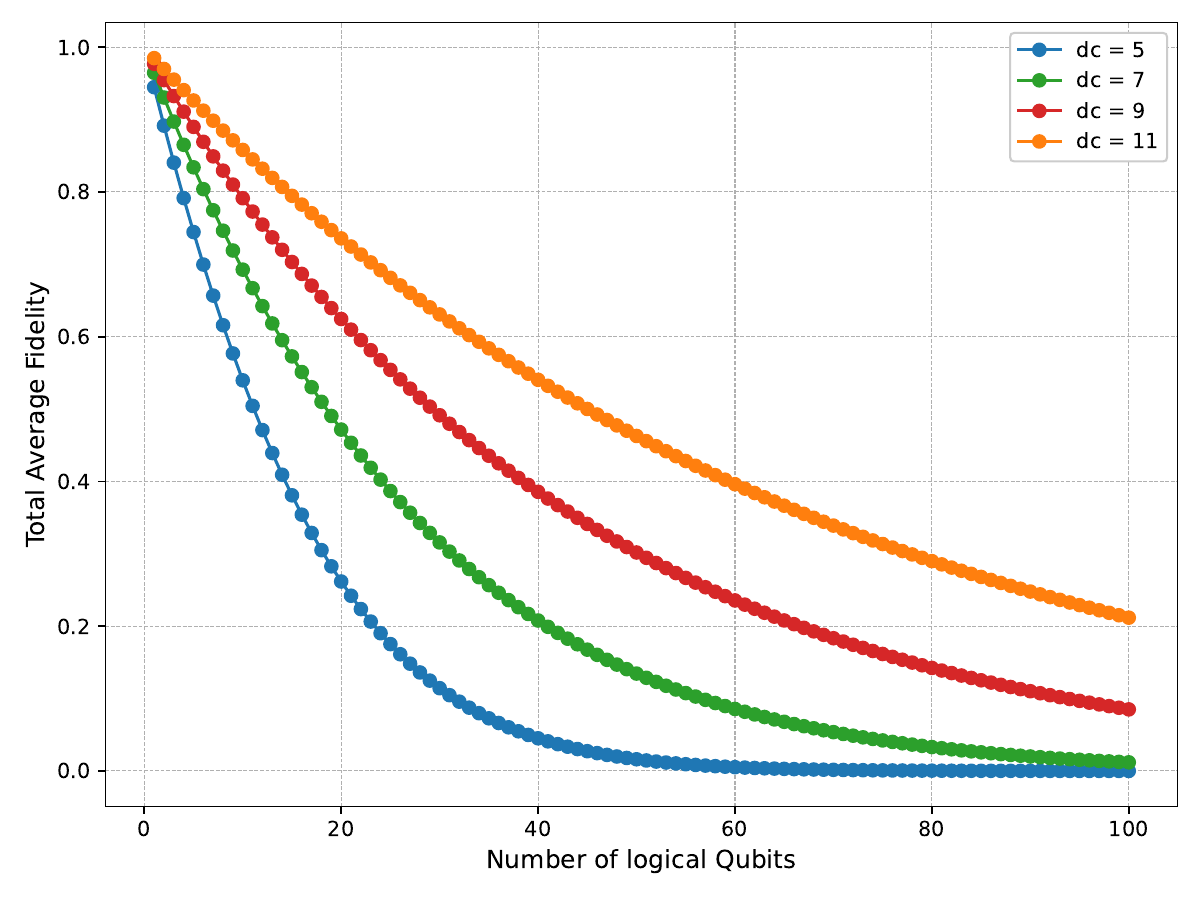}
        \caption*{(b) Total Average fidelity}
        \label{fig:total_average_fidelity_different_dc}
    \end{minipage}

    \caption{Last qubit and total average fidelity for different distance codes, when \(\alpha=3\), \(\kappa_2=1500 \kappa_1\), \(\frac{\epsilon_z}{2\pi}=10.5\)MHz and \(nb=1\). }
    \label{fig:fidelities_for_different_dc}
\end{figure*}

Furthermore, increasing the code distance generally improves both the fidelity of the last qubit and the total average fidelity of the circuit. Figure~\ref{fig:fidelities_for_different_dc} illustrates how these two fidelities vary with different code distances. If we fix the final–qubit fidelity for different code distances, choosing a different target value for each one, the total energy consumption can even decrease. This occurs because a fidelity attainable for a larger code distance may simply not be reachable for smaller distances. Therefore, when comparing codes at a fixed target fidelity, it is possible to compensate energetically by working with larger $d_c$ values.

\begin{figure*}[t]
    \centering

    \begin{minipage}{0.48\linewidth}
        \centering
        \includegraphics[width=\linewidth]{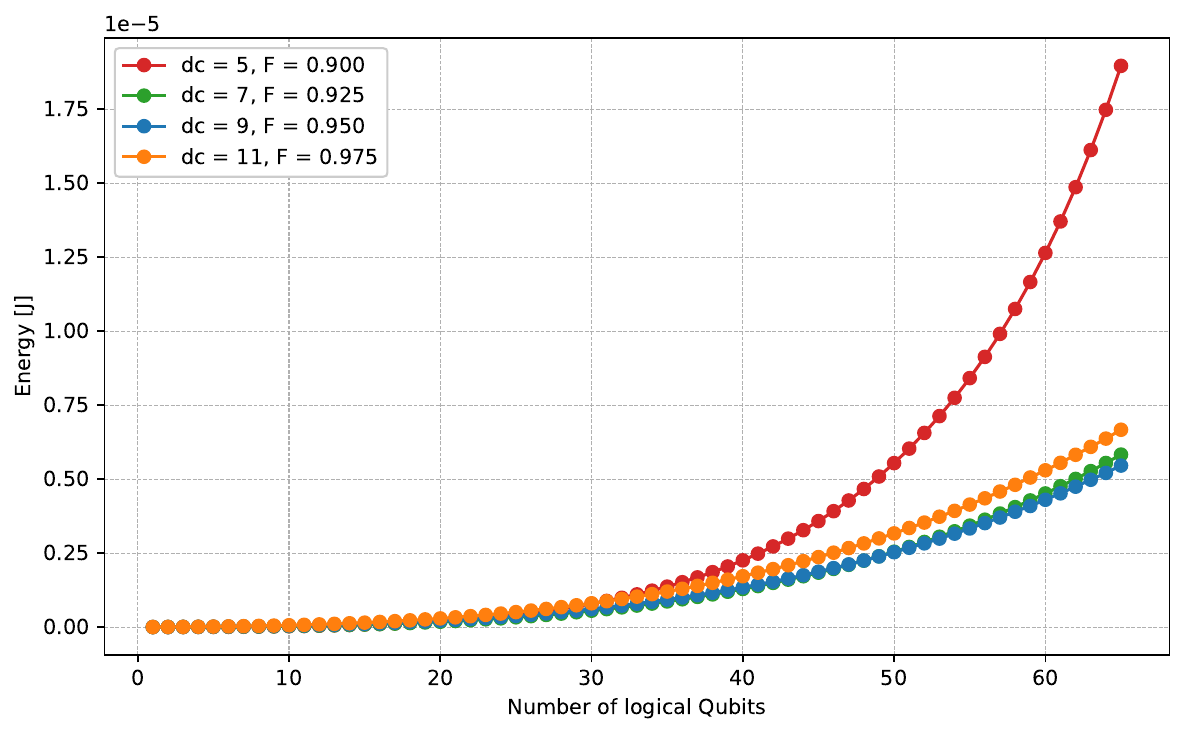}
        \caption*{(a) Microscopic energy}
        \label{fig:micro_energy_dc_comparison}
    \end{minipage}
    \hfill
    \begin{minipage}{0.48\linewidth}
        \centering
        \includegraphics[width=\linewidth]{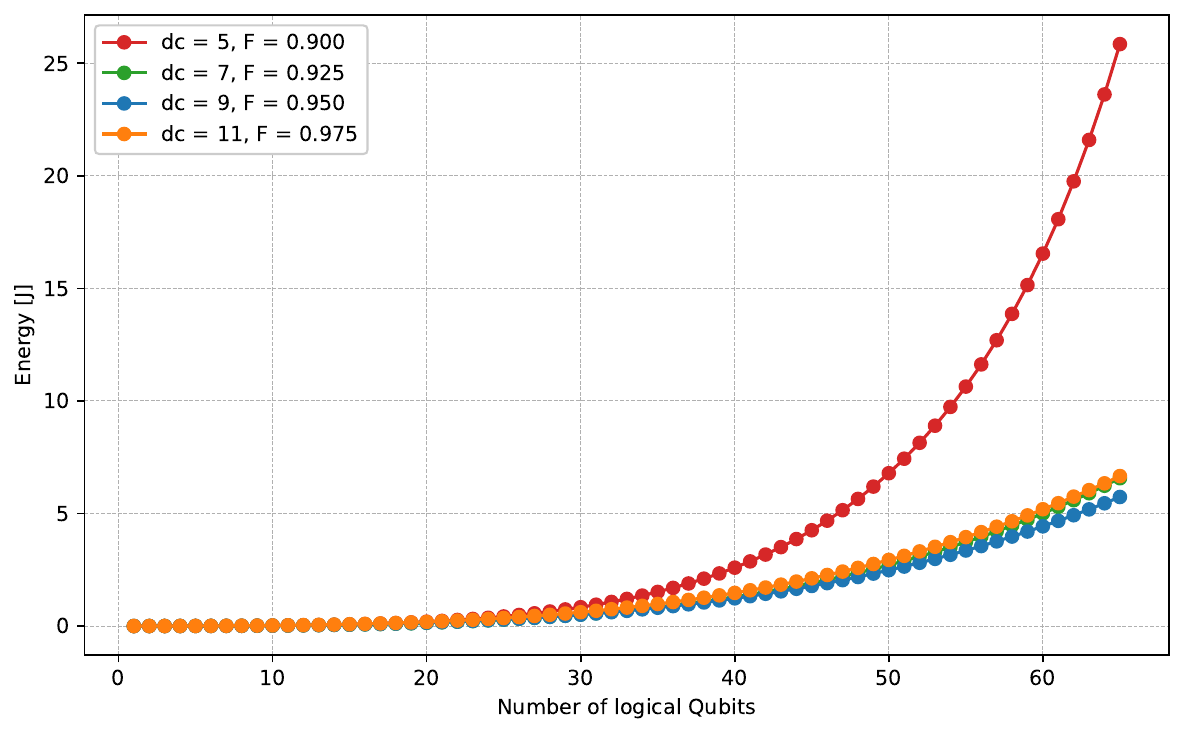}
        \caption*{(b) Macroscopic Energy }
        \label{fig:macro_energy_dc_comparison}
    \end{minipage}

    \caption{Energy consumption of the QFT for different distance codes. For each code, a specific final qubit fidelity was chosen, and the calculations were performed for \(\alpha=3\) and \(Nb=1\). Increasing the distance code allows achieving a higher final qubit fidelity without increasing the total energy required to perform the QFT.}

    \label{fig:energy_different_dc}
\end{figure*}

This effect is explained by the choice of $\kappa_2$. For larger code distances, achieving a given target fidelity allows us to work with smaller $\kappa_2$ values, which can reduce the total energy consumption even though the number of required gates increases. The reduction occurs because the stabilization energy contributes more significantly to the overall energetic cost than the gate energy. Figure~\ref{fig:energy_different_dc} presents an example illustrating this behavior.

%\begin{figure}[H]

\section{DFT consumption in Classical Computers}
\label{Appendix_best_supercomputers}
In this section, we will compute the energy consumption of the state-of-the-art classical computers.

\subsection{World's fastest Supercomputer}
According to the TOP500 \cite{top500_nov2025}, the World's fastest supercomputer in November 2025 is the El Capitan, located in the United States, with a peak performance of \(R_{peak}=2821.101809.00 \) PFlop/s, maximal performance of \(R_{max}=1809.00\) PFlop/s, and power usage of \(P=29685\) kW.

A Flop/s, Floating point operations per second, is a measurement of computer performance in computing and represents the number of floating-point arithmetic calculations that the processor can perform within a second.   

Using this, we can calculate the energy spent by a floating-point operation:
\begin{align}
    E_{\textit{per FLOPS}} &= \frac{\text{Power}}{ R_{max}} = \frac{29695\times10^{3}}{ 1809.00\times10^{15}} \nonumber \\ &= 1.641\times10^{-11} \quad [\text{J/FLOPS}]
\end{align}

\subsection{World's Green Supercomputer}
The World's Green Supercomputer in November 2025 is KAIROS, located in France, according to the Green500 list \cite{green500_nov2025}. This supercomputer has a maximal performance of \(R_{max}=3.05\) PFlop/s, a power of \(P=46\) kW and an energy efficiency of \(\eta = 73.282\) GFlops/W. Hence, we can calculate the energy spent by a bit operation
\begin{align}
    E_{\textit{per FLOPS}} &= \frac{1}{ \eta \times10^{9}} = \frac{1}{73.282\times 10^{9}}\nonumber \\
    &=1.365\times10^{-11} \quad \text{J/FLOPS}
\end{align}

\section{Comparing Exponential and Polynomial Scalings} \label{Appendix_comparison_Scaling}

\begin{figure}[t]
    \centering
    \includegraphics[width=\linewidth]{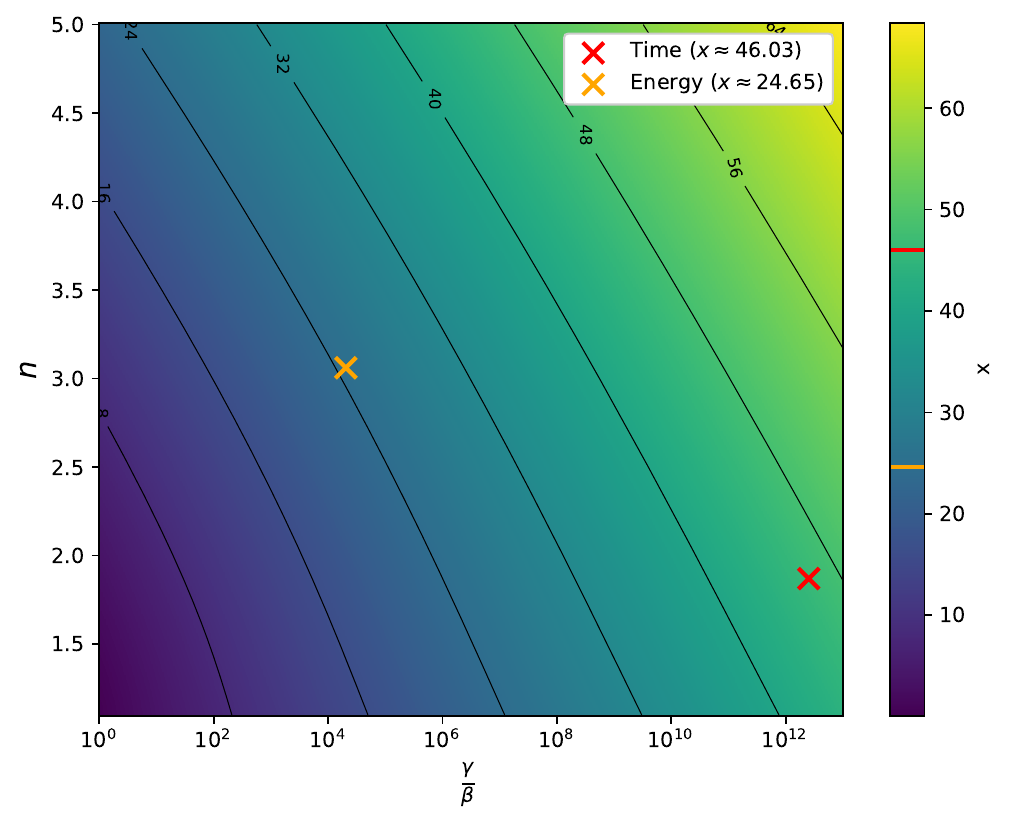}
    \caption{2D Heat Map of the values of $x$ for different values of $\gamma/\beta$ and $n$. The black lines correspond to contours where $x$ is constant. The red and orange points indicate the values obtained for the energy and time intersections, based on the obtained scaling.}
    \label{fig:comparison_scalling_exponential_polynomial}
\end{figure}

The quantum scaling, in both the energy and time scenarios, can be obtained by a polynomial fit $\gamma x^n$, where $\gamma$ is the prefactor and $n$ is the degree of the polynomial.

In the energy case,
\begin{equation}
    \gamma_{\text{energy}} = 6.994\times10^{-6}, \quad n_{\text{energy}} = 3.061
\end{equation}

For the time case,
\begin{equation}
    \gamma_{\text{time}} = 1.396\times10^{-6}, \quad n_{\text{time}} = 1.78
\end{equation}

The classical scaling, as explained in Section~\ref{classical computers}, can be written as $\beta x 2^x$, where for the energy and time cases:
\begin{equation}
    \beta_{\text{energy}} = 5E_{\text{per FLOP}}, \quad 
    \beta_{\text{time}} = \frac{5}{R_{\text{max}}}.
\end{equation}

The intersection point $x$ can be obtained by equating both expressions:
\begin{equation}
    \beta x 2^x = \gamma x^n.
\end{equation}
Solving this equation for $x$, it is possible to find
\begin{equation}
x=-\frac{(-1+n)\text{ProductLog}\!\left[-\frac{\left(\frac{\gamma \log(2)^{1-n}}{\beta}\right)^{\frac{1}{1-n}}}{-1+n}\right]}{\log(2)}.
\end{equation}

Using \textit{El Capitan} as the reference for time and \textit{KAIROS} for energy, since they are the fastest and most efficient systems, respectively, one finds
\begin{equation}
\frac{\gamma}{\beta}_{\text{energy}} = 2.1\times 10^{4}, \quad
\frac{\gamma}{\beta}_{\text{time}} = 2.5\times 10^{12}.
\end{equation}
As presented in Table~\ref{table:energy_time_gates_flops}, which compares gates and FLOPS, the energy factor is significantly smaller than the corresponding time factor. This is expected, as these factors reflect differences in scaling, consistent with the distinction between the fundamental units of computation considered: gates in the quantum case and FLOPS in the classical case.

Figure~\ref{fig:comparison_scalling_exponential_polynomial} represents the value of $x$ for different values of $\gamma/\beta$ and $n$. The black lines correspond to the cases where the same $x$ is obtained. Using the results found for energy and time, one can conclude that the intersection for the time scaling occurs at a larger $x$ than for energy, implying that a quantum energetic advantage will appear before the temporal one.

\end{document}